\documentclass[10pt,journal,compsoc]{IEEEtran}
\usepackage{hyperref}
\usepackage{xcolor}

\ifCLASSOPTIONcompsoc
  \usepackage[nocompress]{cite}
\else
  \usepackage{cite}
\fi
\ifCLASSINFOpdf
  \usepackage[pdftex]{graphicx}
  \graphicspath{{figures/}}
  \DeclareGraphicsExtensions{.pdf,.jpg,.jpeg,.png}
\else
\fi
\usepackage{amsmath}
\usepackage[caption=false, font=footnotesize, labelformat=simple]{subfig}
\begin{document}
%\title{Visualization of Model Parameter Sensitivity \\ along Trajectories in Numerical Weather \\ Predictions}
\title{Visual Analysis of Multiple Dynamic Sensitivities \\ along Ascending Trajectories in the Atmosphere}

\author{Christoph~Neuhauser,
        Maicon~Hieronymus,
        Michael~Kern,
        Marc~Rautenhaus,
        Annika~Oertel,
        and~Rüdiger~Westermann
% <-this % stops a space
\IEEEcompsocitemizethanks{\IEEEcompsocthanksitem Christoph Neuhauser and Rüdiger Westermann are with Technical University of Munich (TUM).\protect\\
E-mail: \{christoph.neuhauser\,$|$\,westermann\}@tum.de.\protect\\
\IEEEcompsocthanksitem Maicon Hieronymus is with Johannes Gutenberg University Mainz.\protect\\
E-mail: mhieronymus@uni-mainz.de. \protect\\
\IEEEcompsocthanksitem Michael Kern is with Advanced Micro Devices, Inc.\protect\\
E-mail: Michael.Kern@amd.com. \protect\\
\IEEEcompsocthanksitem Marc Rautenhaus is with Universität Hamburg.\protect\\
E-mail: marc.rautenhaus@uni-hamburg.de. \protect\\
\IEEEcompsocthanksitem Annika Oertel is with Karlsruhe Institute of Technology.\protect\\
E-mail: annika.oertel@kit.edu.}}
\IEEEtitleabstractindextext{%
\begin{abstract}
Numerical weather prediction models rely on parameterizations for subgrid-scale processes, e.g., for cloud microphysics. These parameterizations are a well-known source of uncertainty in weather forecasts that can be quantified via algorithmic differentiation, which computes the sensitivities of prognostic variables to changes in model parameters. It is particularly interesting to use sensitivities to analyze the validity of physical assumptions on which microphysical parameterizations in the numerical model source code are based. In this article, we consider the use case of strongly ascending trajectories, so-called warm conveyor belt trajectories, known to have a significant impact on intense surface precipitation rates in extratropical cyclones. We present visual analytics solutions to analyze interactively the sensitivities of a selected prognostic variable, i.e. rain mass density, to multiple model parameters along such trajectories. We propose a visual interface that enables to a) compare the values of multiple sensitivities at a single time step on multiple trajectories, b) assess the spatio-temporal relationships between sensitivities and the shape and location of trajectories, and c) a comparative analysis of the temporal development of sensitivities along multiple trajectories. We demonstrate how our approach enables atmospheric scientists to interactively analyze the uncertainty in the microphysical parameterizations, and along the trajectories, with respect to a selected prognostic variable. We apply our approach to the analysis of convective trajectories within the extratropical cyclone ``Vladiana'', which occurred between 22-25 September 2016 over the North Atlantic.
\end{abstract}

}

\maketitle

\begin{figure*}[ht]
  \centering
  \includegraphics[width=\linewidth]{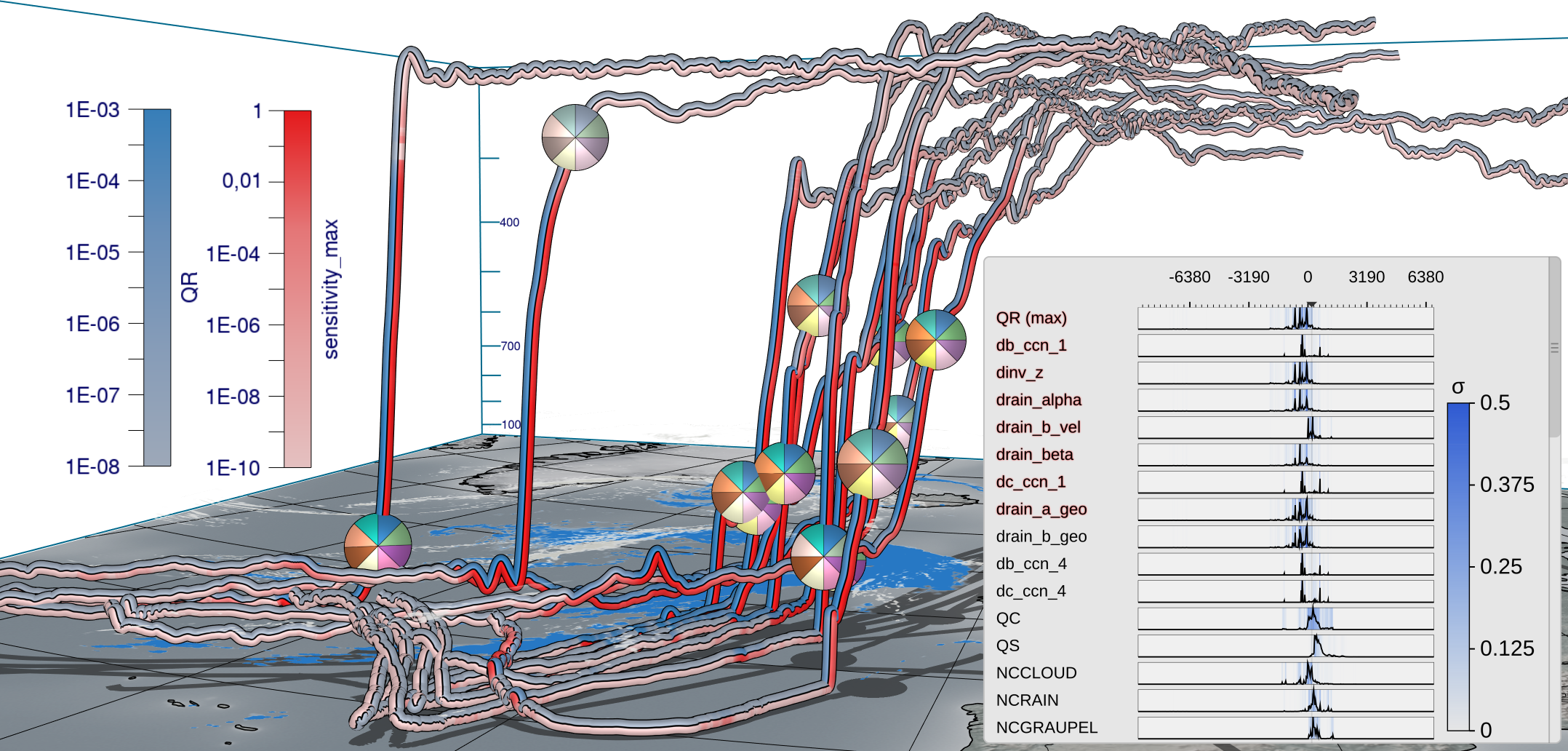}
  \caption{
  Visual analysis of the sensitivity of a prognostic variable to selected model parameters (emphasized in red in curve plot overlay) along warm conveyor belt trajectories in the extratropical cyclone ``Vladiana'', to assess uncertainties 
  %analyze the sensitivity of %\christoph{state variables along} 
  of parameterizations in numerical weather prediction models.
  Prognostic variable (blue) and maximum sensitivity (red) are color coded along the trajectories in bands that are view-aligned, to avoid occlusions that occur when bands are fixed to the trajectory surface. Multiple sensitivities at a selected time step
  %and state variables 
  %(in top to bottom order) 
  are visualized via pie charts that are mapped onto spheres in the 3D view. A consistent view-aligned mapping of sensitivities to pie charts enables an effective comparison across the trajectories. Embedded curve plot shows statistical summaries of prognostic variables, sensitivities, and model parameters to which sensitivities are computed. 
  %show values of  or state variables along trajectories at selected time steps, or (as shown in Fig. 10) aligned to the time of ascent of each trajectory. pie charts enable an effective comparison of sensitivities on multiple trajectories. 
  Surface precipitation is shown on the ground in blue.
%   Visual analysis of the sensitivity of a state variable to selected model parameters (red names in curve-plot overlay) along warm conveyor belt trajectories in the extratropical cyclone ``Vladiana'', to assess uncertainties 
%   %analyze the sensitivity of %\christoph{state variables along} 
%   of parameterizations in numerical weather prediction models.
%   State variable (blue) and maximum sensitivity (red) are color coded in view-aligned bands along trajectories. Embedded curve-plot shows statistical summaries of state variables and sensitivities. Selected sensitivities and state variables (in top to bottom order) are assigned to pie charts in the 3D view in clockwise order. Pie charts show values of multiple sensitivities or state variables along trajectories at selected time steps, or (as shown in Fig. 10) aligned to the time of ascent of each trajectory. View-aligned pie charts enable an effective comparison of sensitivities on multiple trajectories. The blue color on the ground shows surface precipitation.
}
  \label{fig:teaser}
\end{figure*}

\IEEEdisplaynontitleabstractindextext

% \fi
\IEEEpeerreviewmaketitle

\IEEEraisesectionheading{\section{Introduction}\label{sec:introduction}}

\IEEEPARstart{I}{n} meteorology, the warm conveyor belt (WCB) is a well-defined moist airstream, which originates in the lowermost levels of the atmosphere within an extratropical cyclone’s warm sector and generally ascends poleward to the upper troposphere within two days \cite{wernli_lagrangian-based_1997}. WCBs play a critical role in cloud formation and precipitation in the extratropics (e.g., \cite{madonna_warm_2014, bueler_potential_2017, pfahl_warm_2014}).
%Embedded in the WCB, there are trajectories with different ascending rates and different dominant cloud microphysical processes \cite{oertel_convective_2019}.
In numerical simulation output from, e.g., numerical weather prediction (NWP) models, WCBs are typically detected and analyzed by means of path lines (in the atmospheric sciences usually simply referred to as ``trajectories'') computed from the simulated time-dependent 3D wind field. Domain scientists use coherent sets of trajectories to analyze structures which are not directly visible using the time-dependent 3D fields, such as, e.g., the origins of moist airflow or how precipitation patterns emerge from different trajectory ascent.
WCB trajectories are characterized by different ascent rates. The majority of WCB trajectories ascend gradually and slantwise from the lower into the upper troposphere \cite{madonna_warm_2014}, however, faster convective ascent can be embedded.
% These can be categorized as slantwise if they ascend 400 hPa or 600 hPa within 1.5 h to 3.5 h or 6.5 h to 22 h, respectively, or as convective with 400 hPa and 600 hPa ascent times of less than one hour and three hours, respectively \cite{oertel_potential_2020}. 
Convective WCB trajectories play an important role in weather forecasting, e.g., since they have a significant impact on intense surface precipitation rates \cite{oertel_potential_2020, oertel_observations_2021, jeyaratnam_upright_2020}.

%A large source of uncertainty for investigating these trajectories stems from the cloud microphysical schemes implemented in the numerical models. 
The scale of cloud microphysical processes, which are responsible for precipitation formation, is too small to be explicitly resolved in NWP models. Hence, parameterizations are used to calculate the integrated effects on the resolved prognostic variables. These parameterization schemes are still associated with large uncertainties that influence, e.g., rain density formation and trajectory ascent. A thorough analysis of the impact of model parameters on, e.g., rain formation and precipitation, can clarify how, when and where trajectories' properties are sensitive to specific parameterized processes.

Algorithmic Differentiation (AD) is a method to compute derivatives of an implemented model \cite{griewank_evaluating_2008}, which can be used to quantify the impact of multiple model parameters on a prognostic variable at once. AD exploits the fact that any implemented model is a sequence of differentiable, elemental operations on a low level. By repeatedly applying the chain rule, the derivative for any code can be calculated automatically alongside the usual run of the code.
AD has been applied on a warm-rain microphysics scheme for idealized trajectories \cite{baumgartner_algorithmic_2019}, and recently on convective and slantwise WCB trajectories \cite{hieronymus_algorithmic_2022}.
%\maicon{for the two-moment cloud microphysics scheme by Seifert and Beheng~\cite{seifert_two-moment_2006}.
%Hieronymus et al.~\cite{hieronymus_algorithmic_2022} defined}
Sensitivity is defined as the linearly predicted change of a prognostic variable if a model parameter is perturbed by 10\%. The prognostic variable can be any of the simulation output, such as temperature or hydrometeor content. The linear prediction is the gradient computed via AD times 10\% of the model parameter value.

The application of AD to a prognostic variable along WCB trajectories results in one sensitivity value of this variable for each model parameter and on each simulation point along the trajectories. 
%Since the used microphysical schemes can have up to hundreds of model parameters, this results in large sets of sensitivity values that have to be set in relation to the trajectories.
%\maicon{, i.e., Hieronymus et al.~\cite{hieronymus_algorithmic_2022} gathered 4071 sensitivity values per time step and trajectory.}
% Below, a glossary is given for terms used in this work referring to the different types of trajectory and simulation data.
% \begin{itemize}
% \item A \textbf{(state) variable} is a physical quantity like the temperature T or the rain mass density QR.
% %\item A \textbf{(model) parameter} is a simulation setting knob of the used NWP model. \annika{A \textbf{(model) parameter} is a simulation setting used in the parameterizations of the applied NWP model.}
% \item A \textbf{(model) parameter} is a simulation setting used in the parameterizations of the applied NWP model.
% \item The \textbf{target variable} is the (state) variable for which we compute the \textbf{sensitivities} with respect to the different (model) parameters using Algorithmic Differentiation (AD). In our work, we have chosen QR as the target variable.
% \item A \textbf{(user-selected) quantity} is a (state) variable or a sensitivity shown in the visualization.
% \end{itemize}
We build upon the work by 
%Hieronymus et al.~\cite{hieronymus_algorithmic_2022} and the study by Oertel et al. \cite{oertel_potential_2020}. 
Hieronymus et al.~\cite{hieronymus_algorithmic_2022}, who provide a tool and data to analyze microphysical sensitivities in a WCB case-study by means of statistical approaches. Sensitivities along trajectories are analyzed by plotting time-curves and manually comparing the curves associated to different model parameters. The spatial and temporal behavior of sensitivities along individual trajectories or groups of trajectories has not been considered. Oertel et al. \cite{oertel_potential_2020} show that large spatio-temporal variability in terms of trajectories' ascent and characteristics exists - to which extent this also applies to sensitivities remains an open question and motivates the study at hand.
% MH: The old version with my unpublished paper below
% In this study, we build upon the studies by Hieronymus et al. \cite{hieronymus_algorithmic_unpublished_2022} and Oertel et al. \cite{oertel_potential_2020}. Hieronymus et al. \cite{hieronymus_algorithmic_unpublished_2022} analyzed microphysical sensitivities in a WCB case-study by means of statistical approaches. They did not consider the spatial and temporal behaviour of sensitivities along individual trajectories or groups of trajectories. However, Oertel et al. \cite{oertel_potential_2020} showed that large spatio-temporal variability in terms of trajectories' ascent and characteristics exists - to which extent this also applies to sensitivities remains an open question and motivates the study at hand.

In our application, 
%we are concerned with data including 
%several thousands of 
%trajectories with hundreds of potential sensitivities each. 
trajectories are calculated from the resolved 3D wind field using the online trajectory module \cite{miltenberger_online_2013} in COSMO. Every 2\,h, multiple trajectories are started from a predefined starting region at seven vertical levels.
%where the region has been identified using ECMWF offline trajectories \cite{oertel_convective_2019}. 
The sensitivities are then calculated by re-simulating the microphysics along slantwise and convective ascending trajectories from this set.
%\using \cite{hieronymus_algorithmic_2022}.}
%Hence, analysis of such data requires more sophisticated visualization approaches than standard techniques.
Hence, analysis of such data requires more sophisticated visualization approaches than standard techniques.
In particular, through the sustained collaboration between experts from the fields of visualization, computational science and meteorology in the scope of the ``Waves to Weather'' Transregional Collaborative Research Project\footnote{https://www.wavestoweather.de/index.html}), the following questions have turned out to be of considerable relevance to meteorologists:
\begin{itemize}
\item[Q1] Do similar trends regarding selected sensitivities and prognostic variables occur across a group of selected trajectories?\\
%Coherent trajectory clusters are often characterized by specific properties relevant for process understanding in atmospheric sciences. So far, the similarity of certain properties of pre-selected trajectory groups has been mostly plotted manually as individual sub-panels (e.g., Joos and Wernli, 2012, Madonna et al., 2014, Crezee et al., 2017 \cite{MicrophysicalBuildingBlocks}, Raveh-Rubin, 2017 \cite{DryIntrusionsLagrangianClimatology}).
Coherent trajectory clusters are often characterized by specific properties relevant for process understanding in atmospheric sciences. So far, the similarity of certain properties of pre-selected trajectory groups has been mostly plotted manually as individual sub-panels
\cite{madonna_warm_2014, DryIntrusionsLagrangianClimatology}.
%\cite{MicrophysicalBuildingBlocks, JoosWernli2012,madonna_warm_2014, DryIntrusionsLagrangianClimatology}. 
%\textcolor{red}{TODO: Refs.}
\item[Q2] Do different sensitivities and prognostic variables show similar statistical characteristics across a selected trajectory group?\\
Similarities of trajectory properties provide insights in process understanding. So far,  correlations are mostly identified based on individually calculated correlations or based on visual analysis.
\item[Q3] How do sensitivities depend on the time and location along the trajectories, and how are they related to, e.g., surface precipitation and cloud formation?\\
Detailed process understanding requires the current atmospheric state, which is why the combination of trajectories with 3D NWP output is very useful. 
\item[Q4] Do coherent sensitivity patterns emerge if trajectories ascending at different times are considered relative to their time of ascent?\\
In atmospheric sciences, coherent airstreams related to specific weather phenomena can be identified based on their spatio-temporal evolution, such as, for example, strong ascent or descent. Thus, their common characteristics are most pronounced if the trajectories' properties are centered relative to their coherent evolution (e.g., Oertel et al. \cite{oertel_potential_2020}), which in this case-study is the time of the WCB trajectories' ascent start. 
\item[Q5] Do sensitivities differ with respect to different types of trajectories (i.e., convective vs. slantwise)?\\
Although pre-selection of trajectories often results in coherent trajectory clusters, their detailed characteristics still often differ substantially which typically needs to be analyzed based on pre-defined criteria. 
\end{itemize}

%Q1: Do similar trends regarding single sensitivities and state variables occur across a group of selected trajectories;
% Considering single parameters, do similar trends regarding the parameter occur across a group of selected trajectories;
%Q2: Do different sensitivities and state variables show similar statistical characteristics across a selected trajectory group;
%Q3: How do sensitivities depend on the time and location along the trajectories, and how are they related to, e.g., surface precipitation and cloud formation;
% How do sensitivity values depend on the time and location along the trajectories, e.g., with regard to surface precipitation and cloud formation;
%Q4: Do coherent sensitivity patterns emerge if trajectories ascending at different times are considered relative to their time of ascent;
%Q5: Do sensitivities differ with respect to different types of trajectories (i.e., convective vs. slantwise)?

%\christoph{These questions have been developed in an interdisciplinary team in a cooperation of experts from meteorology, simulation and visualization.}
An intuitive analysis of these aspects will enable domain scientists to further investigate the uncertainties in the employed numerical models - which is much needed for further improvement of NWP models.

\textbf{Contributions.}
We propose a visual analytics workflow combining novel application-specific and standard visualization techniques to answer the raised questions. The workflow embeds a curve plot view into a 3D trajectory view and links these two in a bidirectional way. For display in the 3D view, the user selects a prognostic variable, i.e., the target variable, and a set of sensitivities, and can now interactively select time points at which the sensitivities are visualized. 

% Standard visualization techniques, such as curve-plots, are useful for analysis of sensitivities along a given trajectory. The simultaneous visualization of, e.g., the evolution of trajectory height and a sensitivity value reveals occurrences of high sensitivity along the ascent and its relation to trajectory height. Visualizing the evolution of multiple parameters or for several individual trajectories side by side or in one single figure can disclose correlations between different parameters and differences between trajectories regarding a single parameter.

%To resolve Q1, we provide a multi-parameter curve-plot showing for the target variable and each sensitivity the temporal development of the maximum and standard deviation (stdev) to the maximum over all trajectories. 
To address Q1, we provide a multi-parameter curve plot showing the time evolution of the maximum and standard deviation (stdev) to this maximum over all trajectories for any selected sensitivity. Q2 is resolved by automatically sorting sensitivities according to the similarity of their temporal development to a selected sensitivity.
%or prognostic variable. 
In addition, the user can select a short-term trend in the curve plot of a sensitivity and let the system search for similar short-term trends in the temporal developments of other sensitivities. 
%To resolve Q3, the user can select a parameter and the behaviour of this parameter over an arbitrary time interval, and let the system search for similar short-term trends in other parameters (Q3).
The curve plots view enables an interactive comparative visualization of the statistical similarities of local and global trends in the temporal evolution of sensitivities across the set of selected trajectories. Likewise, the trend analysis can be performed for other prognostic variables and model parameters, then showing the mean and stdev to the mean.  
%regarding specific ensemble statistics.

While Q3 can be partially answered via the curve plots,
%which shows how sensitivities depend on time, 
this view cannot reveal the relationships to the locations and shapes of the trajectories and surrounding atmospheric fields. Therefore, the curve plots view is embedded into Met.3D, an open-source 3D visualization system dedicated to meteorological analyses \cite{met3d}. Met.3D visualizes the trajectories in their spatial context (i.e., the 3D trajectory view), including visualizations of additional data sources like textured terrain fields, and in particular 3D atmospheric field data. In principle, Q3 can then be addressed by showing multiple sensitivities along a trajectory via colored bands. This, however, becomes unsuitable as soon as more than 2-3 sensitivities are visualized simultaneously, because the single bands become too thin and can hardly be distinguished. Therefore, we propose a visual mapping using enlarged spheres, which can be moved along the trajectories to specific time points and, due to their extended surface area, can effectively show more bands simultaneously. To also use the trajectory surface for conveying information, the target variable and the maximum sensitivity over all trajectories at each time step are shown via two colored bands along the trajectories. We introduce view-aligned bands, to avoid occlusions that occur when bands are fixed to the trajectory surface.     
By automatically determining for each trajectory its unique time of ascent and interpreting the current time relative to this times, differences and similarities in the sensitivities during the ascend phases can be revealed (to resolve Q4). 
%To simultaneously show multiple quantities at a trajectory point, we propose a mapping using view-aligned color bands and pie charts on enlarged spherical objects. 
%to enable a reduced set of parameters to be perceived effectively. Linking also works in the other direction, by picking a sphere and moving it along the selected trajectory, so that the spheres on all other trajectories as well as the time slider move accordingly. 

The 3D view shows at a glance whether the sensitivities selected for visualization are similar or dissimilar across the set of trajectories. However, since the bands on a sphere are aligned with the trajectory, the bands' orientations vary across the set of trajectories so that a comparative visual analysis becomes difficult. To overcome this limitation (to resolve Q5), we propose to graphically depict the sensitivities via a pie chart that is consistently oriented in view-space and mapped onto the spherical shape. We assess the effectiveness of both visual mappings via a user study. 

%By using functionality in Met.3D, i.e., to select a plane at constant height over ground, Q5 is resolved. At each plane-trajectory intersection point, pie charts are visualized showing selected parameters to indicate the sensitivities at that height and let the user perceive differences between the different trajectories \christoph{@Rüdiger: in \autoref{fig:use_case:overview}, we are not using spheres at the same height, as the slice is near the ground}. By using functionality in Met.3D to visualize atmospheric quantities on the earth surface, patterns of, e.g., intense surface precipitation, can be emphasized and put into relation to the trajectories' locations and shapes, as well as the sensitivities along them. By this, Q6 can be resolved.
% By using functionality in Met.3D, i.e., to select a plane at constant height over ground, Q5 is resolved. By using functionality in Met.3D to visualize atmospheric quantities on the earth surface, patterns of, e.g., intense surface precipitation, can be emphasized and put into relation to the trajectories' locations and shapes, as well as the sensitivities along them (cf.~\autoref{fig:use_case:overview}). By this, Q6 can be resolved.

%The specific contributions visualization 
%The specific visualization-related contributions of this work consist of the following points.
The specific contributions to solve questions Q1 to Q5 are: 
\begin{itemize}
    \item A tube-based trajectory view showing the trajectory data in its geospatial context, using view-aligned color bands on each tube to avoid occlusions and twists.
    \item A sphere-based focus view acting both as a time step marker and magnifying lens. It encodes multiple sensitivities via view-aligned bands or pie charts to enable an effective comparison of the sensitivities on different trajectories.
    \item A curve plots view showing statistical summaries of prognostic variables and sensitivities, which provides options to automatically sort the individual plots according to similarity and occurrence of user-selected subsequences.
    %ly convey similarities and dissimilarities between different trajectories.
    % \item A curve-plot view showing statistical information about the trajectory data in vertically aligned bands, including sorting of bands according to similarity and occurrence of selected subsequences.
    % \item A tube-based trajectory view showing the trajectory data in its geospatial context, using view-aligned color bands on each tube to indicate the target variable and maximum sensitivity.
    % \item A sphere-based focus view acting both as a time step marker and magnifying lens. It encodes multiple quantities via view-aligned bands or pie charts to quickly convey similarities and dissimilarities between different trajectories.
\end{itemize}

The remainder of this paper is structured as follows. After reviewing work that is related to ours in \autoref{sec:relwork}, we provide an overview of the proposed workflow, including a description of the input data and visualization options. In \autoref{sec:vistechniques}, we introduce the design and functionality of the curve plots view and the 3D trajectory view. A description of the technical realization of both views is given in \autoref{sec:implementation}. \autoref{fig:use_case:detail} demonstrates the application of the proposed workflow to analyze convective trajectories within the extratropical cyclone ”Vladiana”. We conclude the paper with a summary of our main contributions and ideas for future work.

\section{Related Work}\label{sec:relwork}

The workflow and techniques we propose are related to multi-parameter and ensemble visualization techniques, as well as feature-based visualization in meteorology. 

\textbf{Multi-Parameter and Ensemble Visualization.} In the curve plots view, we use simple statistical summaries to quantify the spread of variables and sensitivities across a set of trajectories \cite{love2005visualizing,PotterEtAl2010SummaryStat}. 
%Ensemble visualization is related to uncertainty visualization [4], [36], yet it is assumed that the uncertainty is represented by a set of possible data occurrences rather than a stochastic uncertainty model. 
Previous works in ensemble visualization have proposed more advanced graphical depictions to visually convey the ensemble spread of scalar and vector fields (see the survey by Wang et al.~\cite{wang2018} for an overview), and research has been pursued regarding visual abstractions of the major trends in ensembles of line and surface features \cite{pang1997approaches,ContourBoxplotsWhitaker,FerstlEtAl2016EuroVis}. This is different from our application scenario, where the spread of scalar sensitivities across a set of 3D lines is to be analyzed. 
This requirement also distinguishes from approaches that represent the spread in ensembles of multi-dimensional scalar fields, for instance, via histogram encodings \cite{thompson2011analysis}, linearized spatial representation using bar charts \cite{Demir2014MultiCharts}, or probability density functions and distributions \cite{liu2012gaussian,JarDemCVAEns,dutta2015distribution,wang2017statistical,hazarika2017uncertainty}.

Furthermore, since we are dealing with trajectories in 3D space, stacking-based trajectory visualization as, e.g., proposed by Tominski et al.~\cite{StackingBasedTrajectory}, is unsuited for reaching our goals. Such approaches visualize a set of 2D trajectories by stacking them on top of each other as bands along which attributes are encoded via color. However, since the trajectories we consider live in 3D space, stacking is impossible without projecting the trajectories into a common 2D space where shape information is lost. Composite density maps~\cite{CompositeDensityMapsMultivariate} or visitation maps~\cite{Buerger}, on the other hand, aggregate 2D or 3D trajectories depending on specific attributes or shape similarities. Since such techniques do not allow distinguishing individual trajectories, they become inappropriate in our application scenario.

Instead of curve plots as in our workflow, violin plots \cite{ViolinPlots1998} could be used in principle. Violin plots have been proposed for location-based time-series visualization \cite{HoelltViolin2013,HE2020}. Recently, they have been extended to compare multi-parameter data across a set of simulation ensembles \cite{kumpf21}, by showing the histograms of many parameters simultaneously on both sides of a vertical center line. 
Multi-parameter violin plots can be used in our scenario to show side-by-side the temporal evolution of multiple parameters along the trajectories. On the other hand, for the possibly many trajectories we have, the visualization becomes quickly overloaded and requires significantly more space than the statistical summary visualizations used in this work. 

% Another alternative to visualize multi-parameter data are radar charts, pair-wise scatterplots and correlation heatmaps [38], [43]. Most of these techniques, however, even though effective for rather moderate amounts of data points, do not scale well in the number of points. Thus, they are problematic in our scenario. 
Parallel coordinate plots \cite{Inselberg1985, inselberg1991parallel} are another option to directly visualize large sets of multi-parameter data points in a single view. A multitude of methods have been proposed to improve the visibility of single data points in such plots. The survey of Heinrich and Weiskopf~\cite{heinrich2013state} discusses many of these improvements. Parallel coordinate plots, however, are problematic in the current scenario, since it becomes difficult to differentiate the data points belonging to different trajectories and convey the temporal evolution of the sensitivities. Furthermore, information about the location and shape of the trajectories along which sensitivities are analysed is difficult to integrate. 
%spatial or shape information is missing abandoned in difficult to convey in such plots, and scalability issues arise when so many parameters as in our scenario need to be displayed. 
%Johansson and Forsell [35] give an overview of user-centered evaluations of parallel coordinates, Dasgupta et al. [12] focus on the use of PCPs to convey uncertainty in the data. 
%For an overview of techniques using multiple coordinated linked views including PCPs let us refer to the summary report by Roberts at al. [55]. 

Regarding its goals, our proposed approach also shares similarities with techniques that are used to investigate the relationships between multiple input parameters and a single simulation output parameter \cite{BrucknerMoeller10resultDriven,bergner2013paraglide,torsney2011tuner,Sedlmair2014}. These approaches do not visualize sensitivities directly, but they visualize the changes in a simulated phenomenon, like a flow field, due to variations in the input configurations. Thus, they work with an ensemble of simulation runs, while underlying our approach is the existence of a single multi-parameter simulation on an ensemble of paths through the atmosphere. 

Closest to our approach regarding the visualization of quantities along trajectories in 3D space is the technique proposed by Sadlo et al.~\cite{Sadlo06} for visualizing the transport of vorticity along such trajectories. They introduce striped pathlines and slices along the trajectories with extended radius to encode vorticity-specific information. Both the stripe pattern as well as the slices repeat in discrete steps along the trajectory, so that values can be missed. The proposed coloring also has the problem that the stripes become thinner or thicker depending on the bending of a trajectory. Due to the orientation of slices orthogonal to the trajectory tangent, the coloring on these slices is obscured when viewing trajectories from the side.  

\textbf{Feature-based Visualization in Meteorology}
Our driving science domain in this study is meteorology. Meteorological applications have frequently motivated work on visualization in recent years. Overviews were provided by Rautenhaus et al.~\cite{RautenhausEtAl2018VisMet}, Azfal et al.~\cite{AzfalEtAl2019} and Yohizumi et al.~\cite{YoshizumiEtAl2020}. Different aspects of visualization in meteorology have been addressed, including feature-based visualization, flow visualization and 3D interactive visual analysis techniques, all relevant to the techniques discussed in our work. Recent examples in the literature include 3D visual analysis of severe weather (e.g., of supercell thunderstorms \cite{OrfEtAl2017BAMS} and polar lows \cite{MeyerEtAl2021}), and feature-based (mostly 3D) techniques to track cyclone paths \cite{EngelkeEtAl2021} as well as more generic scalar extremal features \cite{KAPPE2022}, and to analyze potential vorticity banners \cite{BaderEtAl2019}, jet stream flow \cite{gmd-15-1079-2022, KernEtAl2017VIS}, atmospheric fronts \cite{KernEtAl2019}, and -as in this study- WCB trajectories \cite{RautenhausEtAl2015GMDb}.
Visual analysis of simulation uncertainty in meteorology has mostly been addressed in the context of visualizing output from simulation ensembles \cite{wang2018,RautenhausEtAl2018VisMet}, e.g., to characterize uncertainty in NWP with respect to uncertainty in initial values. In contrast, in our work, uncertainty information is computed by means of algorithmic differentiation to yield sensitivities with respect to parameters in a numerical simulation.
%Notably, the domain-specific, open-source visual analysis framework Met.3D \cite{met3d} has been used for several studies on meteorological visualization (e.g., \cite{RautenhausEtAl2015GMDb,KernEtAl2017VIS,KernEtAl2019,MeyerEtAl2021}), and poses an ideal framework to implement the techniques discussed in this article.

\begin{figure*}[ht]
 \centering
 \includegraphics[width=\textwidth]{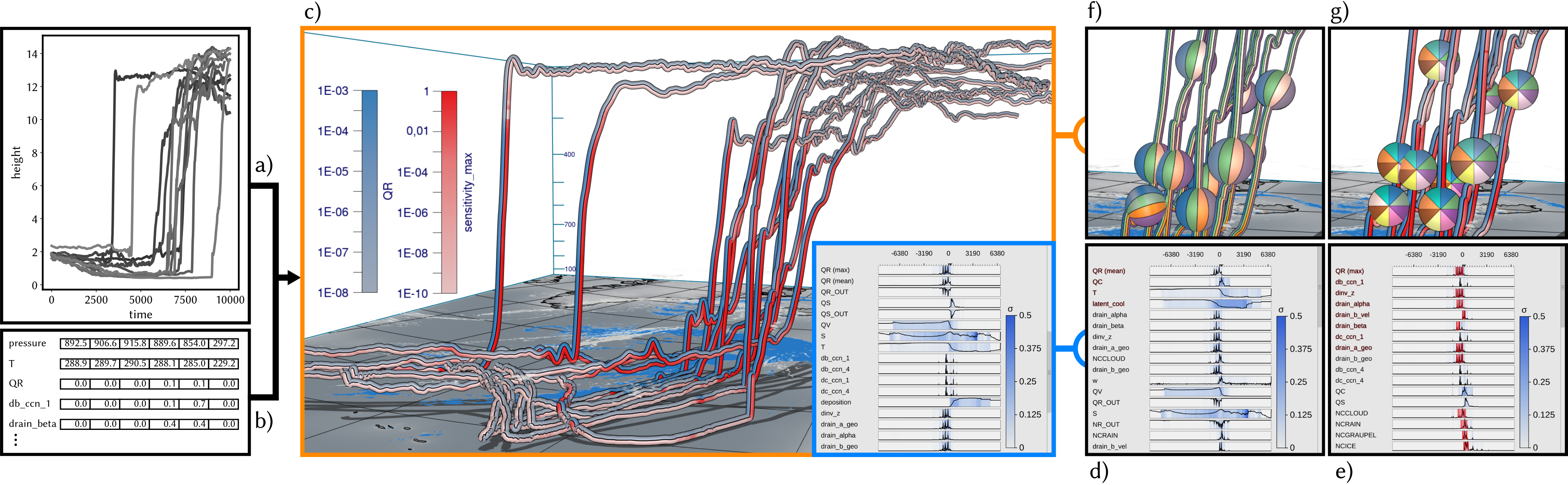}
 \vspace{-0.7cm}
 \caption{Workflow overview: Met.3D reads a) 3D trajectory data and b) tables of model variables and sensitivities along the trajectories. c) The visualization canvas of Met.3D, including the 3D trajectory view that is linked to the curve plots summary view. d) Statistical summaries of the temporal development of variables and sensitivities, which can be ordered automatically regarding the similarity of their temporal development to a selected variable or sensitivity. e) Variables exhibiting a selected sequence of events can be determined automatically and shown first. f) and g) Focus view options using sphere-based multi-parameter visualization via view-aligned bands and pie charts.
 %r visualization great circle spheres, which are only suited for displaying two \christoph{/a limited number of?} parameters, here the target value rain mass density (QR) and the maximum sensitivity, and e) pie chart spheres, which can show more attributes than only the target value and maximum sensitivity. The spheres and the diagram data are aligned using the time of ascent of the WCB trajectories like in \autoref{fig:diagram-view}~bottom. Marked in orange is the trajectory view, and marked in blue is the curve-plot view. The parameters in the curve-plot view can f) also be sorted by similarity to the target variable QR, and g) sub-sequences can be automatically selected using a query sequence in the curve-plot view.
 }
 \label{fig:overview}
 \vspace{-0.3cm}
\end{figure*}

\textbf{Algorithmic Differentiation.} 
Algorithmic Differentiation (AD) is used to compute derivatives of a given computer code in an automatic way \cite{griewank_evaluating_2008}. It has been previously used for sensitivity analysis of the MM5 mesoscale modeling system to evaluate the impact of additional observations on the initial temperature field \cite{bischof_sensitivity_1996}. Baumgartner et al.~\cite{baumgartner_algorithmic_2019} applied AD for a sensitivity analysis along idealized trajectories and warm cloud microphysics. Other use cases include optimization problems, such as optimizing model parameters in machine learning, e.g., in PyTorch \cite{paszke_pytorch_2019}, or applying gradient descent or Newton’s method \cite{margossian_review_2019}. Typical test cases for AD tools include numerical optimization for partial differential equations, i.e., solving Reynolds Averaged Navier-Stokes equations \cite{albring_aerodynamic_2015}.
%\maicon{I added a few use cases of AD in different fields. Please comment if this is okay or too much/too little}

% Reviewer 2:
% "There is a large body of literature on trajectory analysis and visualization in
% the visual analytics community that this work seems to miss. Why this domain
% problem extends from flow visualization, visual analytics for trajectory analysis
% has a long history, with a lot of work coming from the Andrienkos. One highly
% related work would be the Stacking-Based Visualization of Trajectory Attribute
% Data. Overall, there is a whole body of literature missing from this paper related
% to trajectory analysis and visual analytics."

%\textbf{Tajectory Analysis and Visual Analytics.}
%\cite{CompositeDensityMapsMultivariate}
%\cite{StackingBasedTrajectory}

% \textbf{Visual Analytics.}

% TODO: SPRING~\cite{SPRING} based on the dynamic time warping distance (DTW) of a query sequence.

\section{Workflow Overview and Data}

% - Describe the visualization in spatial context using Met.3D and diagram-based technique, selection and linking, horizontal slice to convey variables in surrounding space.

The proposed workflow and methodology enables meteorologists to interactively analyze the effects of simulation model parameters on a selected target variable. We apply it to rain mass density along convective warm conveyor belt trajectories, which are responsible for heavy rainfall on the earth surface. 
The analysis 
%reveals similarities and differences in the distributions of sensitivities over time and space, and 
hints on relationships between the trajectories' spatial locations and shapes, and the occurrence of specific features in the sensitivities of the selected variable to different model parameters.
%The workflow is implemented in Met.3D, a system for the interactive, three-dimensional visualisation of numerical weather predictions and atmospheric model datasets \cite{met3d}. \marc{Met.3D wurde in der Intro schon eingeführt, kann hier denke ich als bekannt behandelt werden.5}
From its existing support to display a single parameter along 3D trajectories \cite{RautenhausEtAl2015GMDb}, Met.3D has been extended according to the specific visualization options required to support a comparative analysis as mentioned.
%The analysis reveals similarities and differences in the distributions of these sensitivities over time, and hints on relationships between the spatial location and shape of the trajectories, and the occurrence of specific features in the multi-parameter distributions. The workflow is implemented in Met.3D, a system for \christoph{the interactive, three-dimensional visualisation of numerical ensemble weather predictions and atmospheric model datasets (taken from the web-page; are we allowed to use this formulation 1:1?)} \cite{met3d}. Met.3D has been extended according to the specific visualization options required to support a comparative analysis as mentioned.

\autoref{fig:overview} shows an overview of the workflow. The input (\autoref{fig:overview}a) is a set of convective WCB trajectories which have been computed over a time interval of interest, and a set of model parameter sensitivities along these trajectories with respect to a selected prognostic variable (\autoref{fig:overview}b).
Sensitivities are named ``d[...]'', which stands for $\partial QR / \partial \text{[...]}$, where rain mass density (QR) is the selected target variable, and ``[...]'' is the model parameter in question.
``sensitivity\_max'' is the per-time maximum of all sensitivities.
We consider WCB trajectories that are computed for the extratropical cyclone ``Vladiana'', which developed from 22-25 Sep 2016 in the North Atlantic during the North Atlantic Waveguide and Downstream Impact Experiment field campaign
\cite{schafler_north_2018}
%\cite{schafler_north_2018, KernEtAl2019,wcd-2022-7}. 
The trajectory data of the case-study shown here is taken from a simulation described in detail by Oertel et al. \cite{oertel_potential_2020} with the NWP model COSMO version 5.1 \cite{baldauf_operational_2011}. In addition, an online trajectory scheme \cite{miltenberger_online_2013} was applied to calculate the positions and properties of the trajectories from the resolved 3D wind field at every model time step, here 20\,s. %This results in an exceptionally high temporal resolution required for AD along WCB trajectories. 

AD has been applied to convective and slantwise trajectories in ``Vladiana'' %\cite{hieronymus_algorithmic_2022}.
%which developed from 22-25 Sep 2016 in the North Atlantic \cite{ schafler_north_2018}. 
with the tool from \cite{hieronymus_algorithmic_2022}, which implements the Seifert and Beheng two moment cloud microphysics model~\cite{seifert_two-moment_2006}
%seifert_parameterization_2008
%together with CCN activation by \cite{hande_parameterizing_2016} 
including routines for the ice phase \cite{karcher_physically_2006,phillips_empirical_2008} and augmented with CoDiPack~\cite{sagebaum_high-performance_2019} to evaluate the Jacobian of the implemented model at every time step in an efficient way. %\ruediger{Maicon: Which tool and a little more details. }\maicon{I added some details. I can expand on that but I figured we can save the space for other things.}
Overall, the sensitivities of rain mass density with respect to 177 model parameters have been computed via AD, of which the 40 most important parameters are used in this work.

The user first inspects the parameter sensitivities by means of curve plots which are shown as overlays in Met.3D (blue outline in \autoref{fig:overview}c). Curve plots show the temporal development of statistical summaries over the set of trajectories (i.e., the maximum and stdev to the maximum) of the target variable and sensitivities. To enable a comparison of different prognostic variables, their means and stdevs to the mean can be shown in the curve plots as well. 
%y is considered a member in an ensemble of trajectories, with the same sensitivity parameters along each member. 
Initially, the target variable is shown on top, and all sensitivities and other variables are shown below it in random order. Both the sensitivities and variables can be sorted with respect to the similarity of their temporal development to a user-selected sensitivity or variable (cf.~\autoref{fig:overview}d), and those whose curve contains a temporal pattern that is similar to a user-selected pattern can be determined automatically and shown together (cf.~\autoref{fig:overview}e). 

%a parameter and time interval, and let ( system search for time intervals where other parameters show a similar distribution than the selected one (to resolve Q3, \autoref{fig:overview}g). 
%This initial inspection is used to quickly obtain an overview of the temporal behavior of parameters. It shows when and for which parameter high sensitivities occur, and whether different parameters show a similar temporal development (cf. sorting and sub-sequence matching in \autoref{fig:diagram-view}). The user can also select a single parameter or single trajectory and let, respectively, all other parameters or the same parameter for all other trajectories being visualized via curve-plot (to resolve Q1).
The initial inspection is used to quickly obtain an overview of the temporal behavior of the variables and sensitivities. It shows when and for which model parameter high sensitivities occur, and which sensitivities or prognostic variables show a similar temporal development. 

In the 3D view of Met.3D, all trajectories are rendered as tubes (orange outline in \autoref{fig:overview}c). Additional visualizations of the earth surface and shadows place trajectories in spatial context. Along each tube, two colored bands are used to visualize the target variable and one other quantity. For close-up views, more bands using a selected set of parameters can be shown (cf.~\autoref{fig:use_case:detail}). Per default, the maximum of all sensitivities is mapped to the second color, to indicate locations of possibly high uncertainty, i.e., high effect of a small change in the model parameter on the selected variable. 
%This color can also be changed to show a selected parameter or its sensitivity. 

%\begin{figure*}[ht]
% \centering
% \includegraphics[width=0.33\textwidth]{technique/diagram/Normal}
% \includegraphics[width=0.33\textwidth]{technique/diagram/Sort}
% \includegraphics[width=0.33\textwidth]{technique/diagram/Align_Ascent_Sort}
% \caption{Left: Diagram view with sub-sequences matched using SPRING \cite{SPRING}. Middle: The parameters are sorted using the absolute %normalized cross-correlation relative to the pressure. Parameters with high positive or negative correlation are shown first. Right: Same as %middle, but the trajectory data is aligned by the time of ascent.}
% \label{fig:diagram-view}
%\end{figure*}
\begin{figure*}[ht]
 \centering
 \includegraphics[width=0.33\textwidth]{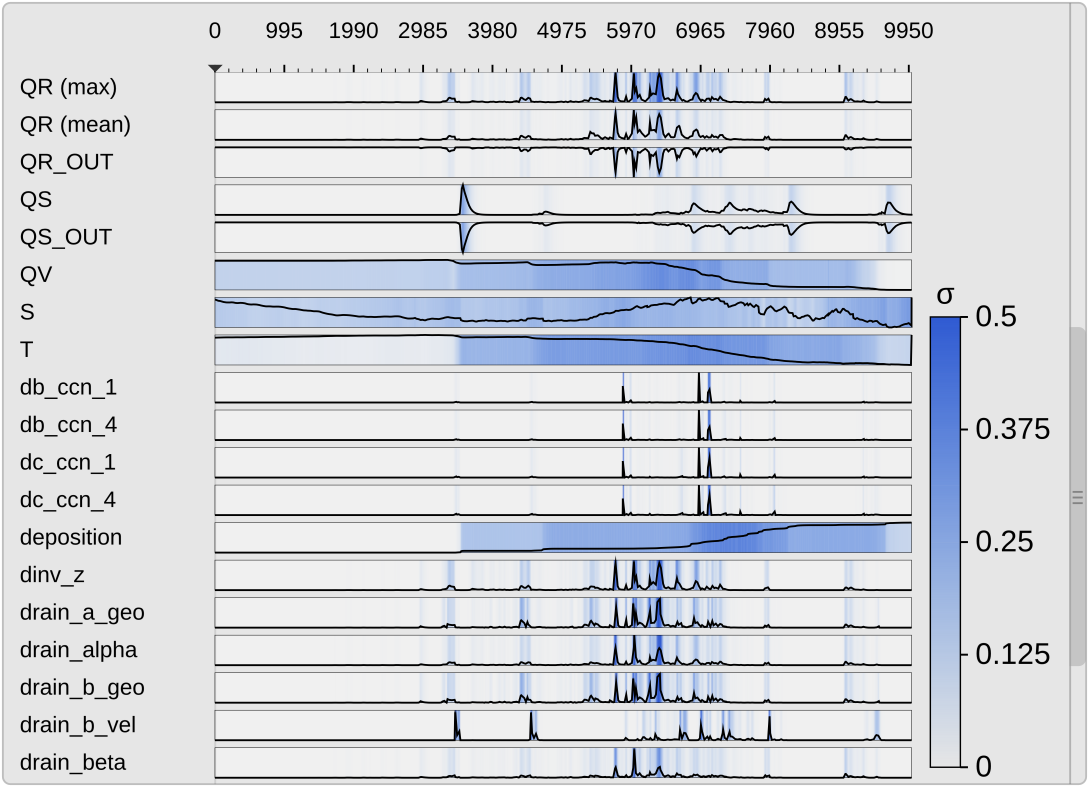}
 \includegraphics[width=0.33\textwidth]{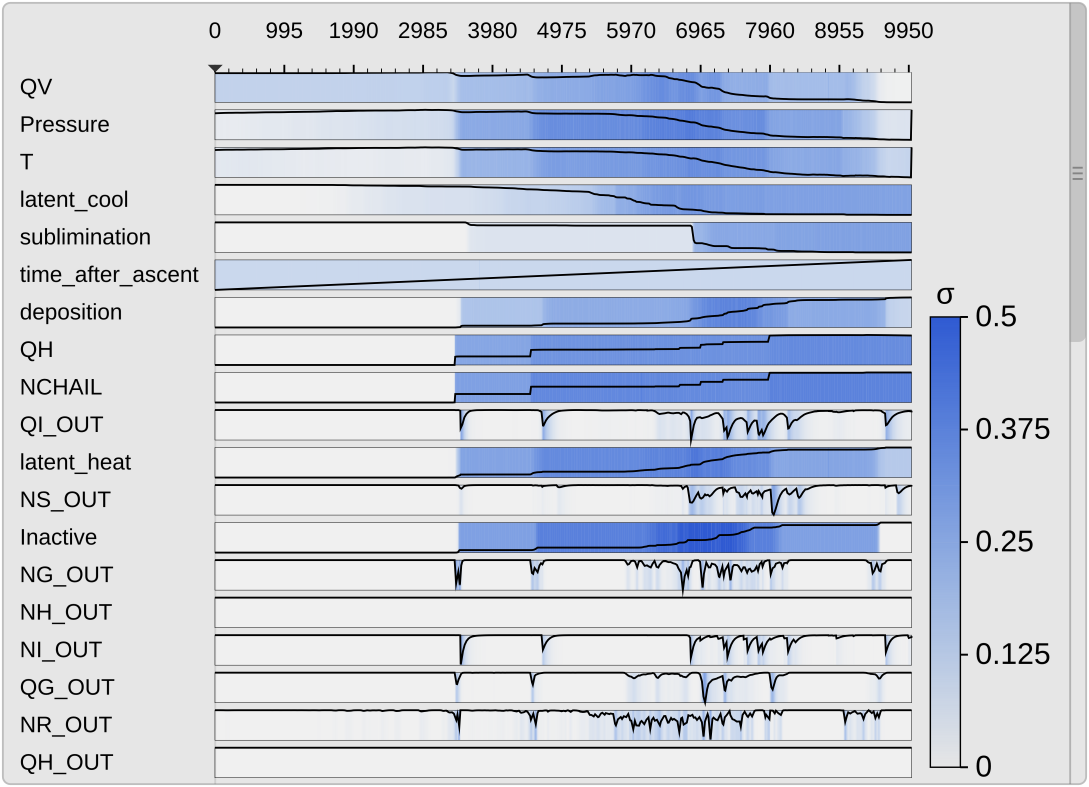}
 \includegraphics[width=0.33\textwidth]{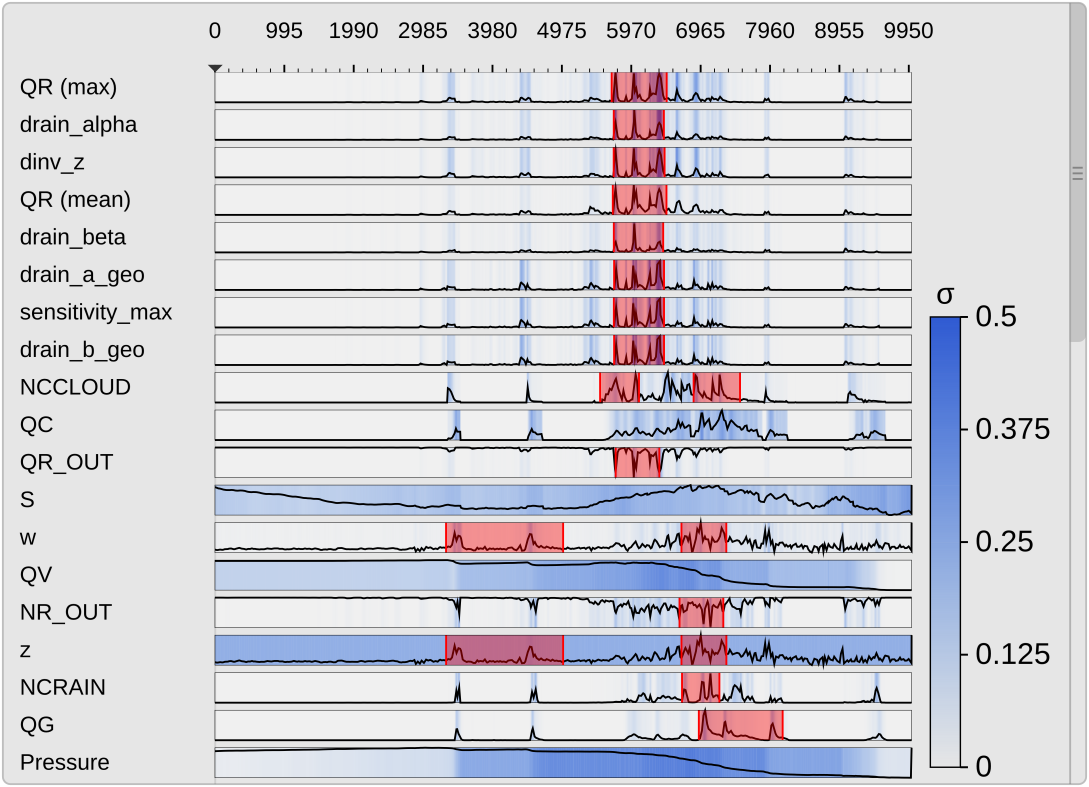}
 \includegraphics[width=0.33\textwidth]{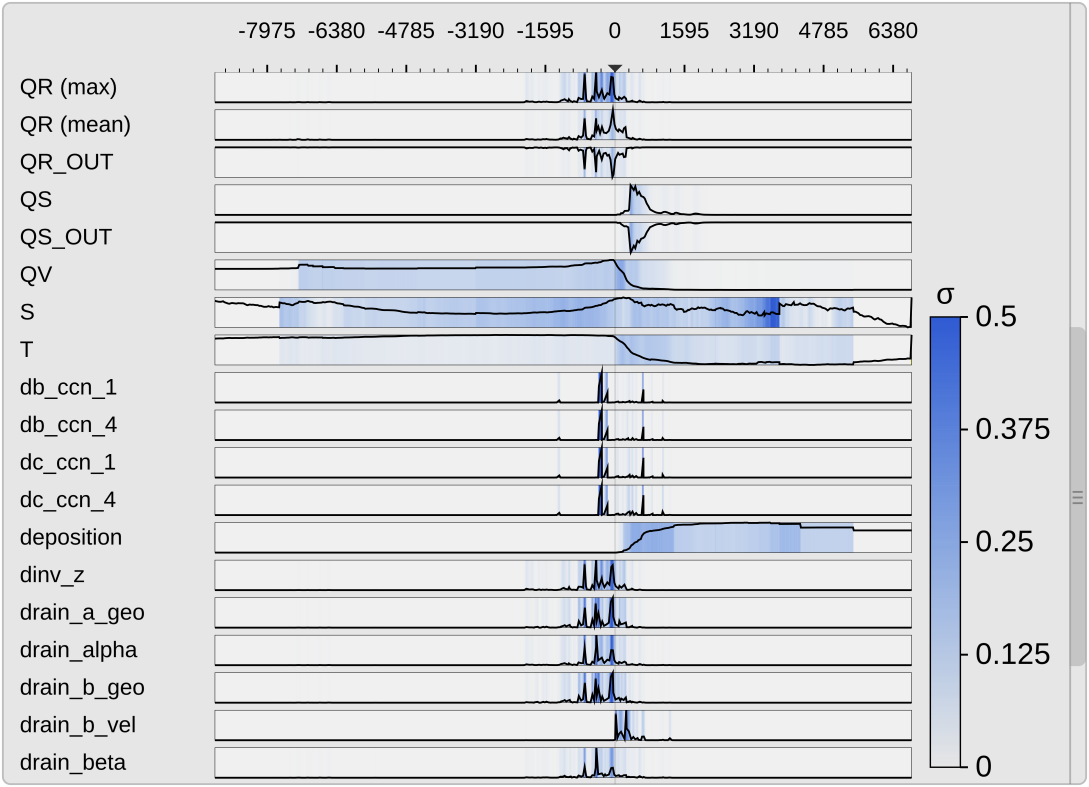}
 \includegraphics[width=0.33\textwidth]{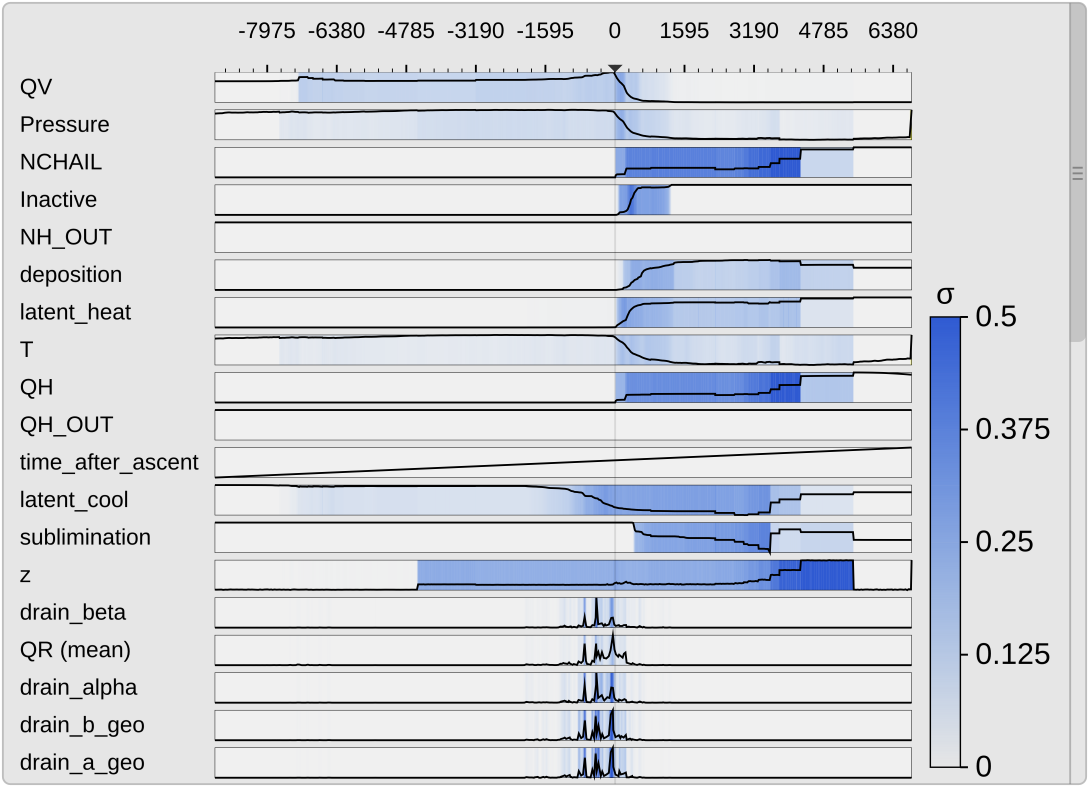}
 \includegraphics[width=0.33\textwidth]{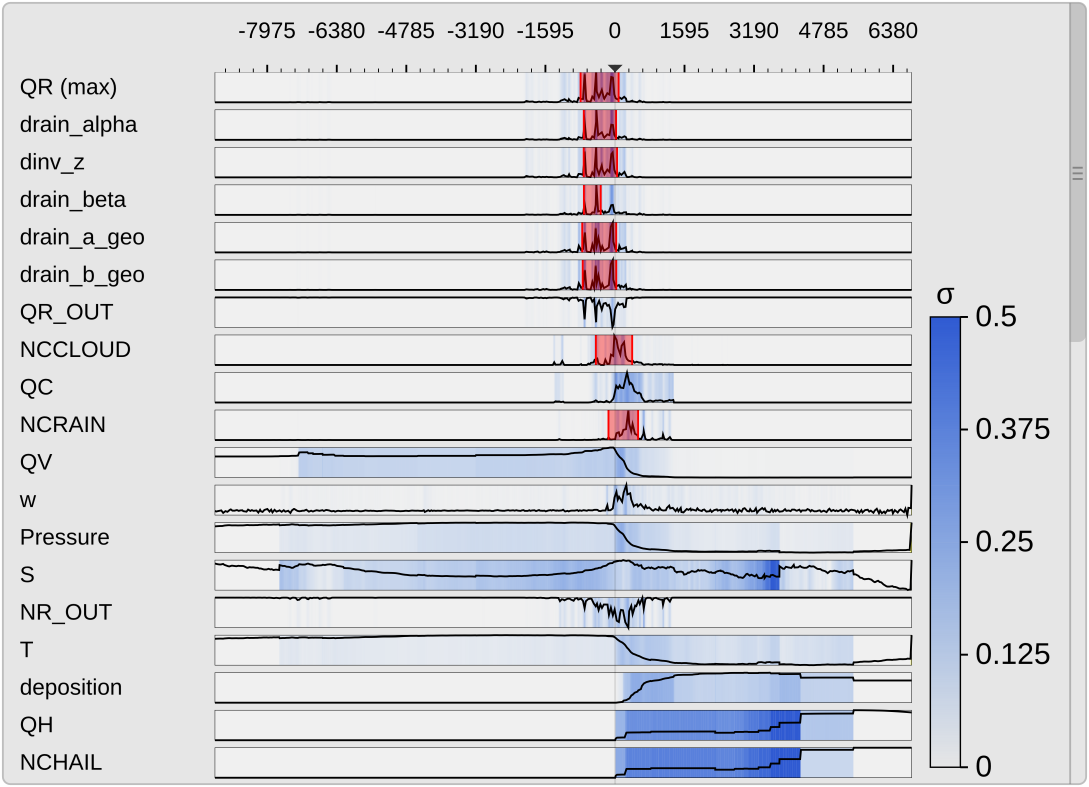}
 \caption{
 %Curve-plots with trajectories aligned by time step (top) and time of ascent (bottom). Curve-plots show mean of physical quantities over all trajectories with color encoded stdev, and maximum of sensitivities and the target variable QR (top band) over all trajectories with color encoded stdev to the maximum. Left: Curve-plots in random order. Middle: Sensitivities are sorted regarding the similarity of their time development relative to rain mass density (QR). Right: A falling edge in pressure is marked, and other sensitivities in which a similar edge has been determined are shown.
 Curve plots with trajectories aligned by time step (top) and time of ascent (bottom). Curve plots show the mean of prognostic variables and the maximum of sensitivities over all trajectories with color encoded stdev to the mean/maximum. Top band shows target variable rain mass density (QR). Left: Curve plots in random order. Middle: Curve plots are sorted regarding the similarity of their time development relative to the target variable QV (cloud mass density). Right: Sorting regarding similarity to max QR. A pattern of consecutive spikes has been selected in QR, and regions in which similar features have been determined are highlighted.
 %a) Curve-plots showing the ensemble mean with color encoded stdev of the target variable (top), and maximum value with color encoded stdev to the maximum of sensitivities in random order. b) Sensitivities are sorted regarding the similarity of their time development using the absolute normalized cross-correlation relative to pressure. c) A falling edge in pressure is marked, and other sensitivities in which a similar edge has been determined using sub-sequence matching and shown right below.
 %They are present, e.g., in the variables pressure, QV and T. 
 %Top row: curve-plot view with trajectories aligned by time step. Bottom row: curve-plot view with trajectories aligned by time of ascent. Left: Standard curve-plot view. Middle: The parameters are sorted using the absolute normalized cross-correlation relative to the pressure. Parameters with high positive or negative correlation are shown first. Right: Falling edges are marked using sub-sequence matching via SPRING \cite{SPRING} using dynamic time warping (DWT).
 %They are present, e.g., in the variables pressure, QV and T.
 }
 \label{fig:diagram-view}
\end{figure*}

The curve plots view is linked to the trajectory view in that the user can move a vertical line along the time axis, and instantly the points on each trajectory corresponding to that time are highlighted by a focus view. The focus view comprises a spherical glyph that is centered at the tube and colored via bands (cf.~\autoref{fig:overview}f) or a pie chart (cf.~\autoref{fig:overview}g), both encoding multiple sensitivities simultaneously . 

The number of bands to be shown and the number of subdivisions of the pie chart is given by the number of sensitivities the user selects. 
Both bands and pie charts are aligned consistently in view-space to allow for an effective visual comparison of the shown values across the trajectories. Within each pie chart, the selected sensitivities in top to bottom order in the curve plots view are mapped to pie chart pieces in clockwise order. In Sec.~\ref{sec:trajectoryview}, we discuss the different visual designs.

Alternatively to moving the time line in the curve plot, the user can pick a sphere glyph and move it along the trajectory. All other glyphs are moved accordingly in time so that via animation the sensitivities on different trajectories can be compared (cf.~supplementary video).%(cf.~\autoref{fig:linking}).

%The user can further select to show the multi-parameter distribution on each trajectory at a selected height over ground (\autoref{fig:use_case:overview}).

%Alternatively to move the time-line in the curve-plot, the user can pick a glyph and move it along the trajectory. All other glyphs are moved accordingly in time so that via animation similar and dissimilar parameter distributions can be seen (\autoref{fig:sphere-types}). The user can further select to show the points on each trajectory with highest sensitivity in a selected variable \christoph{We don't have that option yet, but the user can manually select points with high sensitivity.}, or to show the multi-parameter distribution on each trajectory at a selected height over ground (\autoref{fig:use_case:overview}).

%\christoph{}

\section{Visualization Techniques}\label{sec:vistechniques}

% - Diagram view including mean/stdev coloring, sorting (clustering) and sub-sequence matching.
%  - Trajectory view including coloring and focus-sphere with coloring on it. Explain (show) limitations like many colors on tube and bands vs. pie chart.
% - Slicing and integration in Met.3D, including user interaction.  

The visualization workflow presented in this work builds upon a curve plots view, a 3D trajectory view, and interactive linkage between these two views. Linkage enables to find relationships between locations with high sensitivities along trajectories and the trajectories' locations and shapes.

%\subsection{Multi-Parameter Ensemble View}\label{sec:diagramview}
\subsection{Multi-Parameter Curve Plots View}\label{sec:diagramview}

%In a pre-process, the time interval over which the trajectories have been computed is discretized into a certain number of time points. We initially start with a discretization time step equal to the smallest time step used for trajectory integration, and resample for each trajectory all required parameters and sensitivites at the resulting time points \christoph{I don't think we do that, right?}.
The curve plots view shows the single curve plots of the prognostic variables and sensitivities vertically aligned (cf.~\autoref{fig:diagram-view}).
The time axis is going to the right and the vertical axis represents the value domain.
All values are initially normalized to $[0,1]$.
The trajectories are traced with a time step of $\Delta t = 20s$, which is also the time delta between two data points in the horizontal axis.
When the number of time steps exceeds the number of pixels reserved for showing the curve plots, the algorithm \textit{largest triangle three buckets} (LTTB) \cite{lttb} is used to recursively downsample the data. LTTB takes into account the perceptual importance of points during the downsampling process by assessing the area of triangles formed by points in neighboring buckets.
In this way, the performance penalty of drawing too many points can be avoided, simultaneously ensuring that no features are lost. By generating the curve plots at multiple resolutions, the user can zoom into interesting time intervals and analyze the variables and sensitivities over these intervals in more detail.

%In each band, a curve is shown over a colored background. The height of the curve encodes the mean value (for the target and other physical variables) 
%\christoph{In our use-case, the maximum is also used for the target variable QR, as ...}
%and maximum sensitivities over all trajectories. 
For the target variable and sensitivities, in each band the maximum over all trajectories is shown via a curve. For all other prognostic variables and model parameters the mean over all trajectories is shown. 
%is shown over a colored background. The height of the curve encodes 
%and the mean over all trajectories for all other variables. For the target variable, the mean can be selected alternatively.  
Since the sensitivities are often close to zero, resulting in very small mean values, the maximum values and corresponding stdevs can far more effectively indicate the spread of the distributions and the overall trend regarding their strengths. In particular, regions of potential local instability are emphasized and high sensitivities aren't missed. 
The background is colored according to the stdevs with respect to the values represented by the curves, i.e., stdev is mapped to a color ranging from white (low value) to blue (high value).
By utilizing mouse controls, the user can scroll through the set of parameters and zoom into individual regions in the curve plots view. A moveable vertical line indicates the currently selected time step. 

Since there are many parameters and not all can be shown in one single view, the system proposes an automatic ordering to quickly identify sets of parameters with similar sensitivity development over time. Therefore, the user selects an individual curve plot, and the system sorts all curve plots in descending order regarding the similarity to the curve in this plot. 
%maximum and stdev of the selected sensitivity or mean and stdev of the select physical quantity (Q1). 
As a measure of similarity we use the absolute normalized cross-correlation 
\begin{equation}
NCC(X, Y) = \frac{1}{N} \sum_i \frac{(X_i - \mu_x) (Y_i - \mu_y)}{\sigma_x \sigma_y}.
\end{equation}
Here, $X_i$ and $Y_i$ are two time series, and $\mu_x, \mu_y$ and $\sigma_x, \sigma_y$ the corresponding means and stdevs. Note that due to the division by the stdev, A
%n advantage of the NCC, compared to, e.g., the $\ell_1$ and $\ell_2$ metric, 
NCC becomes independent of the scale of the two time series. 
%due to the division by the stdev. 
%NCC is a similarity metric often used in image processing \cite{NCCImageProcessing}.

\begin{figure}[t]
 \centering
 \includegraphics[width=1.0\columnwidth]{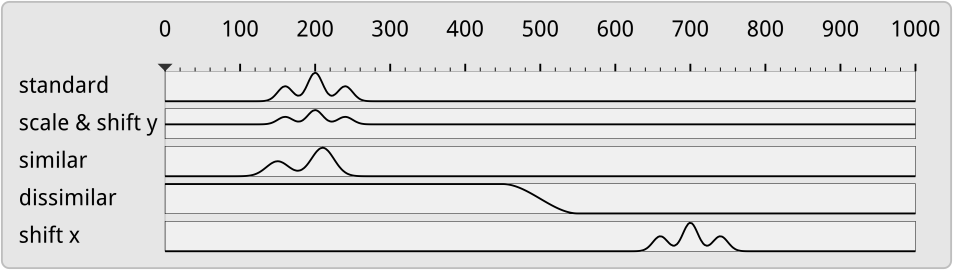}
 \caption{Test sequences sorted by their similarity to ``standard'' using the absolute NCC. The NCC can deal with scaling and shifting in the data axis, but not with shifting in the time axis. We address this limitation by aligning curves relative to the time of ascent of the corresponding WCB trajectories.}
 \label{fig:comparison-curve-sort}
\end{figure}

%Besides the $\ell_1$ and $\ell_2$ metrics, 
%which gave unsatisfactory results due to their dependence on data scale, 
%we also compared NCC to an information theoretic approach based on the mutual information \cite{ShannonMI, FanoMI} of the two sequences. However, in all of our experiments NCC produced almost equal results at considerably lower computation time, enabling real-time interaction with the data. 
We further considered CrossMatch \cite{CrossMatch} and the ``edit distance on real sequence'' (EDR) ~\cite{TrajectorySimilarity} as alternatives for similarity sorting. However, since the former does not support data normalization, and the latter may suppress relevant sensitivities due to built-in noise suppression, both turned out to be less effective in our scenario.  

\autoref{fig:diagram-view} left and middle show, respectively, the initial curve plots using a random ordering of variables, and the ordering with respect to the selected temporal distribution of the variable QV. \autoref{fig:diagram-view} right shows the ordering with respect to QR. As can be seen, a number of sensitivities behave very similarly to QR and, in particular, show a significant change at the point in time where QR changes significantly. Note here that by using the absolute value of the NCC, it is ensured that parameters with high negative correlation are shown before those with low absolute correlation.

A limitation of NCC is that time series which show a similar but time-shifted behavior are found to be dissimilar (cf.~\autoref{fig:comparison-curve-sort}). Even though this can be avoided by computing NCC for successively delayed versions of the original series and finding the peak in the sequence of similarities, we provide a different alternative that takes into account that it is in particular the ascent phase of a trajectory which is of interest. We define the start of the ascent of a trajectory as the start of the most rapid ascent within a 2\,h window. This is calculated by using a sliding window of 2\,h and calculating the total ascent within this time window. Finally, the trajectories are shifted in time so that they all start their ascent at the same time, and the shifted versions are then sorted via NCC. 

\begin{figure}[t]
 \centering
 \includegraphics[width=1.0\columnwidth]{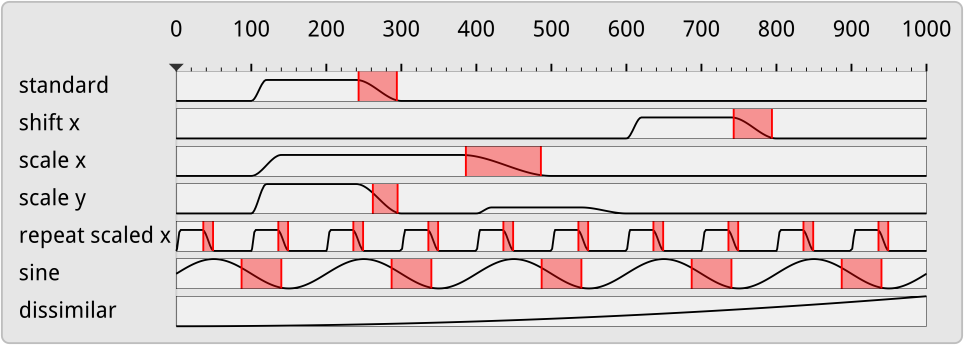}
 \caption{Subsequence matching in the curve plots view using SPRING.
 %top) and NSPRING (bottom). 
 SPRING, due to dynamic time warping, can pick up patterns that are shifted and scaled in the time axis. 
 %As can be seen in the column ``scale y'', applying SPRING on sequences scaled in data direction does not pick up the different normalizations in data.
 %This problem is fixed by NSPRING, and in our approach by using the data scale independent sorting based on the NCC.
 }
 \label{fig:comparison-curve-match}
\end{figure}

To facilitate an improved comparative analysis of the sensitivities along multiple trajectories, 
%n addition to sorting the sensitivities based on the similarity of their developments over time, users are interested in 
it is furthermore important to find similar reoccurring subsequences in this data.
%within the set of physical parameters and sensitivities. 
In particular, since trajectories are seeded at different locations and times, they can first travel close to the surface over different time intervals, before similar upstream paths are observed along which specific sensitivity patterns occur. To determine similar patterns, the user can select a time interval using the mouse, and automatically the subsequence of sensitivity values within this interval is searched in the same and all other curves via the subsequence matching algorithm SPRING \cite{SPRING}. SPRING selects all subsequences with a dynamic time warping (DTW) distance less than a user controlled threshold, by warping one sequence so that it best matches another sequence (see \autoref{fig:comparison-curve-match} for a schematic illustration). The DTW distance is the sum of the per-element distances of two such optimally aligned sequences.
When searching for all subsequences in a sequence of length $n$ with respect to a query sequence of length $m$ with a DTW distance less than a user-specified threshold, a naive algorithm has a time complexity of $O(n^3m)$.
Due to its time complexity of $O(nm)$, SPRING enables interactive use even for long sequences.
 
As SPRING is based on dynamic time warping, the time scale of subsequences may be both stretched or compressed. As can be seen in \autoref{fig:diagram-view}~right, this enables to select, e.g., all falling edges in the temporal developments, independently of their duration. The found subsequences are underlined by red background color.
Compared to
%It is worth noting that we have also tested other techniques for subsequence matching like 
NSPRING \cite{NSPRING}, an extension of SPRING that adds support for data normalization, in all of our experiments SPRING gave most plausible results in line with our perception of similarity (i.e., that the similarity of two sub-sequences is also dependent on their scale).

\subsection{Trajectory View}\label{sec:trajectoryview}

\begin{figure}[t]
 \centering
 \subfloat[\label{subfig:primitives-t2s0}]{\includegraphics[width=0.24\columnwidth]{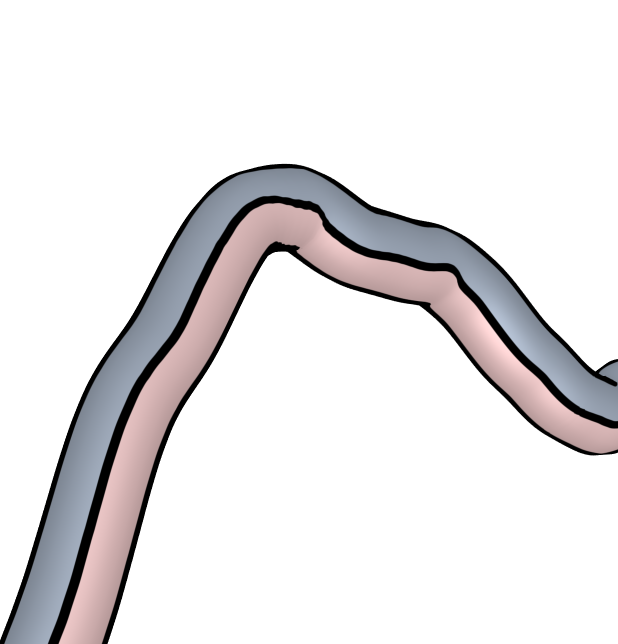}}
 \hfill
 \subfloat[\label{subfig:primitives-t2s2}]{\includegraphics[width=0.24\columnwidth]{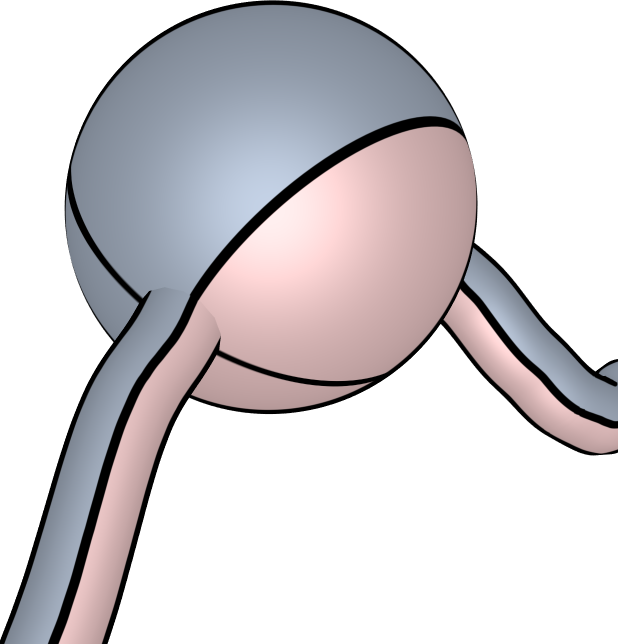}}
 \hfill
 \subfloat[\label{subfig:primitives-t2s8}]{\includegraphics[width=0.24\columnwidth]{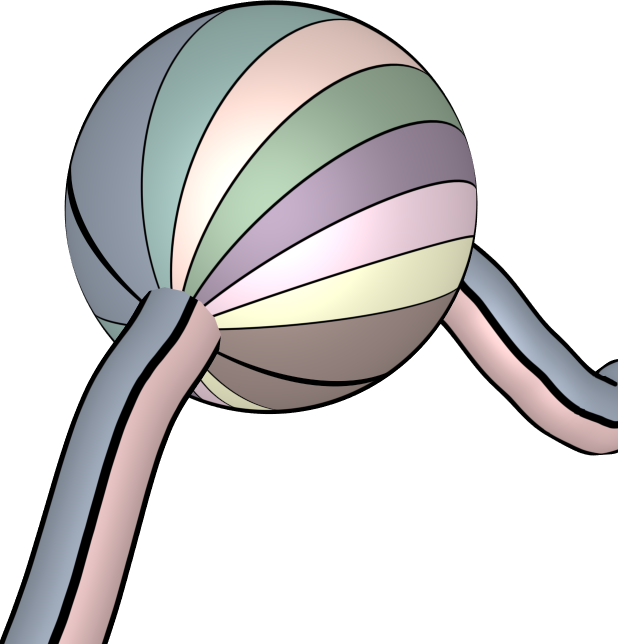}}
 \hfill
 \subfloat[\label{subfig:primitives-t2s8-pie}]{\includegraphics[width=0.24\columnwidth]{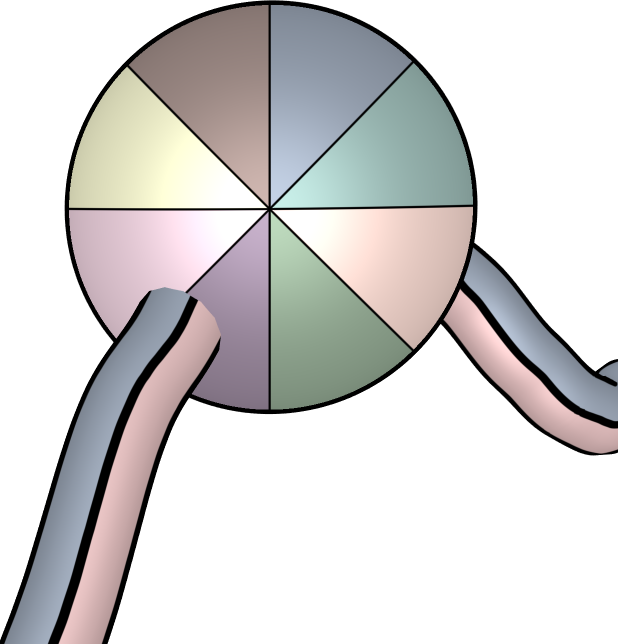}}
 \caption{a) Target variable (bluish colormap) and maximum sensitivity (reddish colormap) are mapped to bands along a tube's surface. b) A focus sphere is used to increase the area on which sensitivities are encoded.
 c) Multiple sensitivities are mapped to bands running along the trajectory and across the sphere's surface. d) Multiple sensitivities are displayed via a pie chart that is mapped onto the sphere surface.
 }
 \label{fig:sphere-types}
\end{figure}

In the trajectory view, all given trajectories are shown in their geospatial context using Met.3D (cf.~\autoref{fig:overview}). Each trajectory is rendered as a colored and illuminated tube with black outlines to let it stand out against the background. Per default, the target variable and the maximum sensitivity are encoded by two different colors, and they are shown on the tube via two bands running into the direction of the trajectory's tangent (see \autoref{fig:sphere-types}a for an illustration).  

%However, when defining these bands in object space, i.e., the assignment of points on the tube surface to either band is fixed, parts of a band can disappear and become visible on the opposite surface part when rotating about the trajectory or the tube twists. This makes it difficult to match a band with its corresponding quantity, and it is especially critical when multiple trajectories are shown and need to be compared regarding the data that is shown in the bands. To avoid this problem, we have developed a rendering technique that renders the bands so that each band covers always one half of the visible tube surface regardless of the current view and the tube's orientation (cf.~\autoref{fig:implementation-geometry}). This rendering is used in all trajectory views throughout this work.%~\autoref{fig:sphere-types}.  
However, when defining these bands in object space, i.e., the assignment of points on the tube surface to either band is fixed, parts of a band can disappear and become visible on the opposite surface part when rotating about the trajectory or when the tube twists. This makes it difficult to match a band with its corresponding quantity, and it is especially critical when multiple trajectories are shown and need to be compared regarding the data that is shown in the bands. To avoid this problem, we have developed a rendering technique that renders the bands so that each band covers always one half of the visible tube surface regardless of the current view and the tube's orientation (cf.~\autoref{fig:sphere-types}b). This rendering is used in all trajectory views throughout this work.%~\autoref{fig:sphere-types}.  

%By selecting individual parameters in the diagram view, the user can choose which parameters to also display in the trajectory view. We propose to show the individual parameters 
% on the tube surface as view-oriented bands, i.e. 
% (cf.~\autoref{fig:sphere-types} and \autoref{fig:linking}). This way, the information displayed on the tube surface maintains its relative position independently from the viewing direction. We outline the exact used rendering technique for this in \autoref{sec:appendix-tube}.

While in principle it is possible to show more than two bands on each trajectory, quickly with increasing view-distance the bands cannot be distinguished anymore. 
%(cf.~\autoref{fig:sphere-types}b).
%the maximum number of parameters that can be simultaneously displayed in this way is limited. As can be seen in \autoref{fig:linking}, the further we are from a trajectory, the harder it gets to distinguish individual parameters on the trajectory surface.
To circumvent this restriction, we propose a focus view that utilizes a locally enlarged surface to obtain more space for the shown sensitivities. 
%On the surface of the trajectories, only two context bands are shown, i.e., one band for a target value and one band for the maximum sensitivity of all computed gradients w.r.t. this target value. 
On each trajectory, a sphere with adjustable radius is rendered at the currently selected time. The sphere acts both as a time marker and a magnifying glass enabling the display of more sensitivities at once. By showing the focus sphere on each trajectory only at the selected time, occlusions that are introduced when increasing the radii of the trajectories everywhere can be avoided. Initially, only the two bands along the trajectory tubes are continued over the sphere (cf.~\autoref{fig:sphere-types}b). 

To show more than two sensitivities on a focus sphere, we introduce two different visual mappings. The first mapping
%called \textit{great circles}, 
acts like a magnifying glass when multiple sensitivities or variables were encoded via bands on the tube and continued across the sphere (cf.~\autoref{fig:sphere-types}c). When crossing over the sphere, the bands become wider so that the different colors can be better perceived and distinguished. As for the bands on the tube, also the bands crossing over the sphere are view-aligned, i.e., while they orient according to the trajectory tangent they cover the same parts of the visible sphere surface. In \autoref{sec:implementation}, the rendering approach that is used to generate this view is explained in detail. The advantage of this mapping is that the alignment of the bands with the trajectories gives a visually smooth and fairly uncluttered appearance. On the other hand, due to the illumination of the sphere surface, towards the silhouettes the bands become darker so that the relationships between colors and sensitivities are disturbed. We counteract this by reducing light-dependent shading of the bands, i.e., the coefficients of the Blinn-Phong shading model are reduced for diffuse and specular lighting, while being increased for ambient lighting.
Nevertheless, since at some time steps spheres can become positioned at trajectory points with vastly different tangents, a visual comparison of the seen band patterns --- which are then differently oriented --- becomes difficult. 
%Nevertheless, since at one time all spheres are usually positioned at trajectory points with vastly different tangents, a visual comparison of the seen band patterns --- which are then differently oriented --- becomes difficult. 
%(cf.~\autoref{fig:many_traj_band}).

%sdisplays the same of view-oriented bands as shown on the tube (\autoref{fig:sphere-types}~left). 
%In \autoref{sec:appendix-sphere-great-circle}, we explain how we maximize the smoothness of the transition between the trajectory and sphere view by utilizing great circles. 

The second mapping intends to avoid the aforementioned drawbacks by using a coloring that neither mimics the use of bands nor is aligned with the trajectory. Our proposed solution is the use of a pie chart-based coloring of the sphere, with each piece given equal area and colored according to a certain sensitivity (cf.~\autoref{fig:sphere-types}d). 
%the analogously to the view on the trajectory (\autoref{fig:sphere-types}~right). 
The values are taken at the selected time step from the trajectory and used to color the pie pieces. The user selects the sensitivities to be shown on the pie chart, and the pie chart is automatically subdivided into an equal number of pieces. Also the pie charts are view-aligned, i.e., they are aligned with the view up-axis (cf.~\autoref{sec:implementation}). As for sphere coloring using bands, $N$ best distinguishable colors are chosen from the Brewer colormap~\cite{brewer}. Per default, we offer users the 8-class Set1 qualitative color map from ColorBrewer plus turquoise. 
%\christoph{(one color is reserved for the target variable, the other 8 can be used for sensitivities)}, 
Sensitivities from low to high are mapped from 20\% saturated to fully saturated colors. This avoids that adjacent pieces with low sensitivities fade out to almost indistinguishable colors. Since each piece of a pie chart is equally affected by shading, the use of shading is less problematic than for bands. Furthermore, each view-aligned chart has a consistent orientation, which makes it easier to compare charts on multiple trajectories.
%, and slightly more sensitivities can be encoded before the differences become indistinguishable. 
A disadvantage is that pie charts might seem to stand apart from the trajectories, since they are not aligned with their tangents. %\autoref{fig:many_traj_band} provides a comparison of the two different mappings when shown along the 3D trajectories in Met3D.
%\begin{figure}[t]
% \centering
% \includegraphics[width=0.3\columnwidth]{implementation/ImgGreatCircles}
% \hspace{1.0cm}
% \includegraphics[width=0.3\columnwidth]{implementation/ImgPieCharts}
% 
% %\vspace{0.2cm}
% 
% %\includegraphics[height=3.2cm]{implementation/SketchTube}
% %\includegraphics[height=3.2cm]{implementation/SketchGreatCircles}
% %\includegraphics[height=3.2cm]{implementation/SketchPieCharts}
% %\caption{View-aligned bands on the tube, and transition from tube to sphere when rendering view-aligned bands and pie charts on the sphere. 
% \caption{View-aligned bands on the tube, and transition from tube to sphere when using view-aligned bands and pie charts on the sphere. 
% %Bottom: Screen-aligned assignment of parameters to visible surface parts.
% }
% \label{fig:implementation-geometry}
%\end{figure}

% , and they are assigned to the pies so that more similar colors are opposite to each other
\begin{figure*}[t]
    \centering
    \includegraphics[width=\columnwidth]{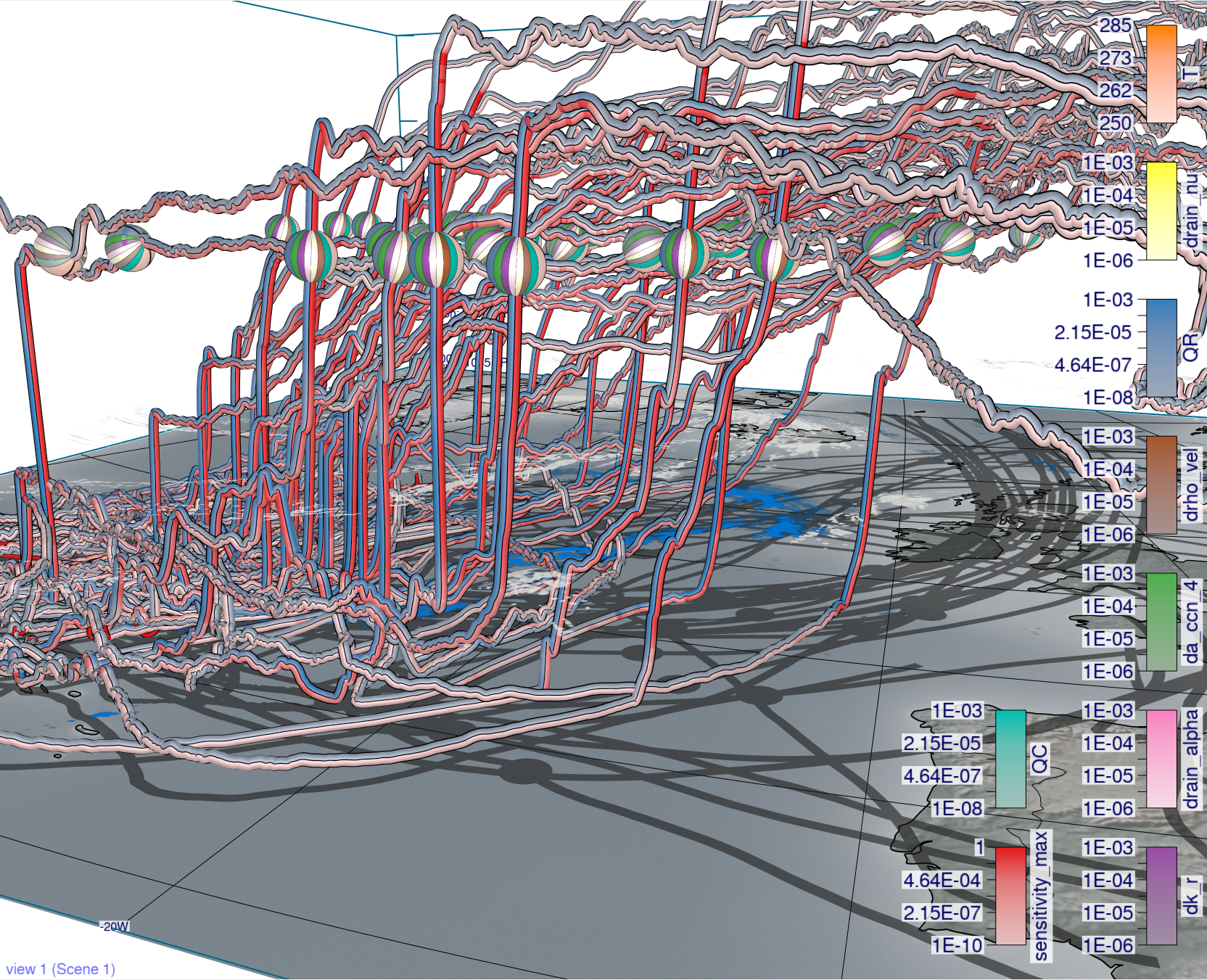}
    \includegraphics[width=\columnwidth]{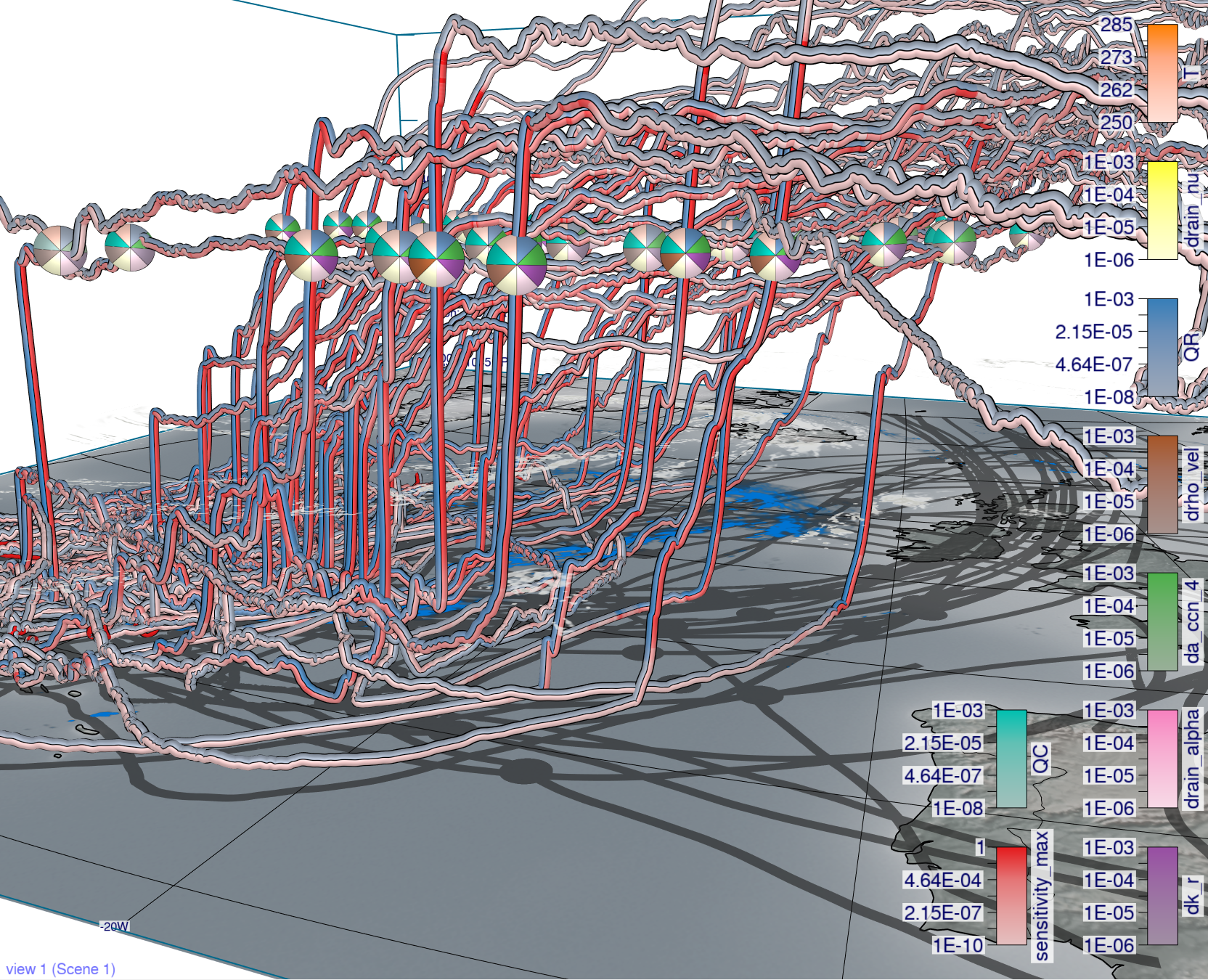}
    \caption{50 trajectories from the use-case ``Vladiana'' using trajectory-oriented bands on spheres and view-aligned pie charts. The different orientations of the bands make it difficult to compare the multi-parameter distributions on different trajectories. With pie charts it is unambiguous which variable is shown, independently of the trajectory orientation.     
    }
    \label{fig:many_traj_band}
\end{figure*}

\section{User Feedback and Study}

According to our collaborators from computational science and meteorology, questions Q1, Q2 and Q4 can be effectively resolved by the curve plots view, which has been designed specifically to answer these questions. Due to the possibility to show statistical summaries side-by-side for many sensitivities, in combination with similarity sorting and subsequence matching, large numbers of sensitivities can be handled in a fairly efficient and effective way.
The number of sensitivities that can be read by the system is not limited, yet beyond a certain number the corresponding curve plots cannot be shown simultaneously and the user needs to scroll through them. However, especially in this case the functionality of the system allows quickly identifying the sensitivities of interest through similarity sorting and subsequence matching.
Our collaborators were especially intrigued about how quickly similar trends can be resolved with only little user intervention. On the other hand, they clearly stated that the curve plots view alone, without any linkage to a 3D view in which relationships between sensitivities, the shape and location of trajectories, and surrounding atmospheric fields can be monitored, is too limited to enable detailed process understanding.    
%through the curve plot view. The user can easily see trends in the data through similarity sorting and subsequence matching. The number of sensitivities 
%quantities that can be displayed on the surface of the tubes and the 

There was common agreement that the linked 3D trajectory view effectively overcomes the aforementioned limitations. In particular, the embedding of multi-parameter visualizations in the form of bands and pie charts were perceived effective for answering Q3 and Q5, since these options enable performing certain analysis tasks exclusively in the 3D view and can, thus, avoid frequent attention shifts between the curve plots and 3D views. 

% without any convenient Our collaborators clearly While the curve plots view has been designed that can be displayed on the spherical focus glyphs, however, is limited. When the pie chart is split into more than 6-8 sectors, the sectors become too small and cannot be well distinguished. Increasing the size of the glyphs can in principle weaken this problem, yet this increases significantly the amount of occlusion.  

%Another problem is that the number of colors we can use on the pie charts is limited. In \autoref{fig:many_traj_band}, we have chosen colors from the 8-class Set1 qualitative color map from ColorBrewer plus turquoise. 
% However, there is not that many more well discernible colors that can be used for color mapping, and some color combinations might also be inappropriate for viewers with some form of color blindness. The saturation of the colors indicates the value of the displayed quantity and strongly desaturated colors can look quite similar. This might not be a severe problem when visualizing sensitivites, where a viewer is usually interested in the high sensitivities, however, we have implemented an optional mouse hover-over that shows in a tooltip the value of the quantity below the mouse cursor.

%To study the relative effectiveness of band- and pie chart-based multiple parameter visualization, we have further investigated the impacts of these design choices on identifying certain multi-parameter patterns and comparing them across trajectories. 
To study the relative effectiveness of band- and pie chart-based multi-parameter visualization, we have further investigated the impacts of these design choices on identifying certain multi-parameter patterns and comparing them across trajectories. 
%These design choices were in our case the bands-based and pie chart-based focus spheres. 
We conducted a task-oriented user study to measure the performance and effectiveness of each mapping,
%for assessing single sensitivities and comparing sensitivities across multiple trajectories, 
by varying the sensitivity values, the surrounding spatial context and the viewing parameters. 

\subsection{Tasks}
We choose the following three representative tasks for the user study, to measure and compare the validity of the different visual designs:

\begin{itemize}
    \item[T1] Finding on which trajectories sensitivities behave against the trend at the current time point. 
    \item[T2] Comparing the distributions of sensitivities to a given target distribution.
    \item[T3] Finding groups of trajectories with similar multi-parameter distributions.  
    %for which trajectories the value of a specific quantity lies within a certain range.
\end{itemize}

%We have conducted a user study with $19$ \textcolor{red}{TODO} participants who were presented images using either the oriented bands or pie charts and needed to answer different questions regarding the visualized data. In total, there were 12 visualization-related questions, 4 general questions and one free text feedback from. The general questions were regarding the experience of the participant in the fields of meteorology and visualization, whether they had any form of color blindness, and which of the two techniques they subjectively liked the most.

The study was browser-based to ensure remote participation under the same settings, i.e., participants did not need to install and run Met.3D. 
Participants were presented images showing either the oriented bands or pie charts, similar to the images shown in \autoref{fig:many_traj_band} c) and d). In total, there were 12 design-specific questions, 4 general questions and one free text feedback form. The general questions were regarding the experience of the participant in the fields of meteorology and visualization, 
%whether they had any form of color blindness, 
and their subjective preference of the two designs. 

%\subsection{Participants}

%We recruited 19 participants. 13 ($68.4\%$) stated that they had (experienced) knowledge in 3D visualization, but less knowledge in multi-parameter visualization and meteorology. 10 ($52.6\%$) came from the field of meteorology, but were rather un-experienced in 3D and multi-parameter visualization. 
%We recruited 21 participants. \textcolor{red}{TODO!} 6 ($31.6\%$) stated that they had (experienced) knowledge in 3D visualization, but less or no knowledge in multi-parameter visualization and meteorology. 3 ($15.8\%$) came from the field of meteorology, but were rather un-experienced in 3D and multi-parameter visualization. 7 ($36.8\%$) considered themselves as knowledgeable both in the fields of visualization and meteorology.
We recruited 27 participants. 5 ($18.5\%$) stated that they had (experienced) knowledge in 3D visualization, but less or no knowledge in multi-parameter visualization and meteorology. 3 ($11.1\%$) came from the field of meteorology, but were rather un-experienced in 3D and multi-parameter visualization. 5 ($18.5\%$) considered themselves as knowledgeable both in the fields of visualization and meteorology.
%We recruited 21 participants. 5 ($23.8\%$) stated that they had (experienced) knowledge in 3D visualization, but less or no knowledge in multi-parameter visualization and meteorology. 1 ($4.8\%$) came from the field of meteorology, but were rather un-experienced in 3D and multi-parameter visualization. 3 ($14.3\%$) considered themselves as knowledgeable both in the fields of visualization and meteorology.
Participants were first briefed about the study and provided information about the problem, the data, and the visual designs.
%Then, they were given few  warm-up tasks for exercise, were we gave hints on how to complete the tasks. 

\subsection{Procedure}

In the final study, the participants were asked to perform the tasks T1, T2, and T2 in the following ways: 
%tasks to analyse how well the two different visual designs :
\begin{itemize}
    %\item[M1] Participants were asked to select those trajectories where the shown multi-parameter distributions seem to be outliers (no more than $30\%$ outliers were shown). 
    \item[M1] Participants were asked to select those trajectories where the shown multi-parameter distributions seem to be outliers (at most $25\%$ outliers were shown). 
    \item[M2] Participants were asked to select those trajectories where the distribution of all shown quantities is similar to a selected trajectory or a reference distribution. 
    \item[M3] Participants were asked to group trajectories based on the shown multi-parameter distributions. 
\end{itemize} 
% Out of the visualization-related questions, $4$ questions were aimed at T1, $4$ at T2 and $4$ at T3. T1 and T2 are related, but T2 is more difficult to answer, since values that are not peaks not getting mapped to eye-catching, high-saturation colors. T3 refers to questions whether users can, e.g., see correlations between different quantities or find quantities fulfilling certain criteria. The system then measured the task completion times and calculated the scores according to their operations.

\begin{figure}[t]
 \centering
 \includegraphics[width=1.0\columnwidth]{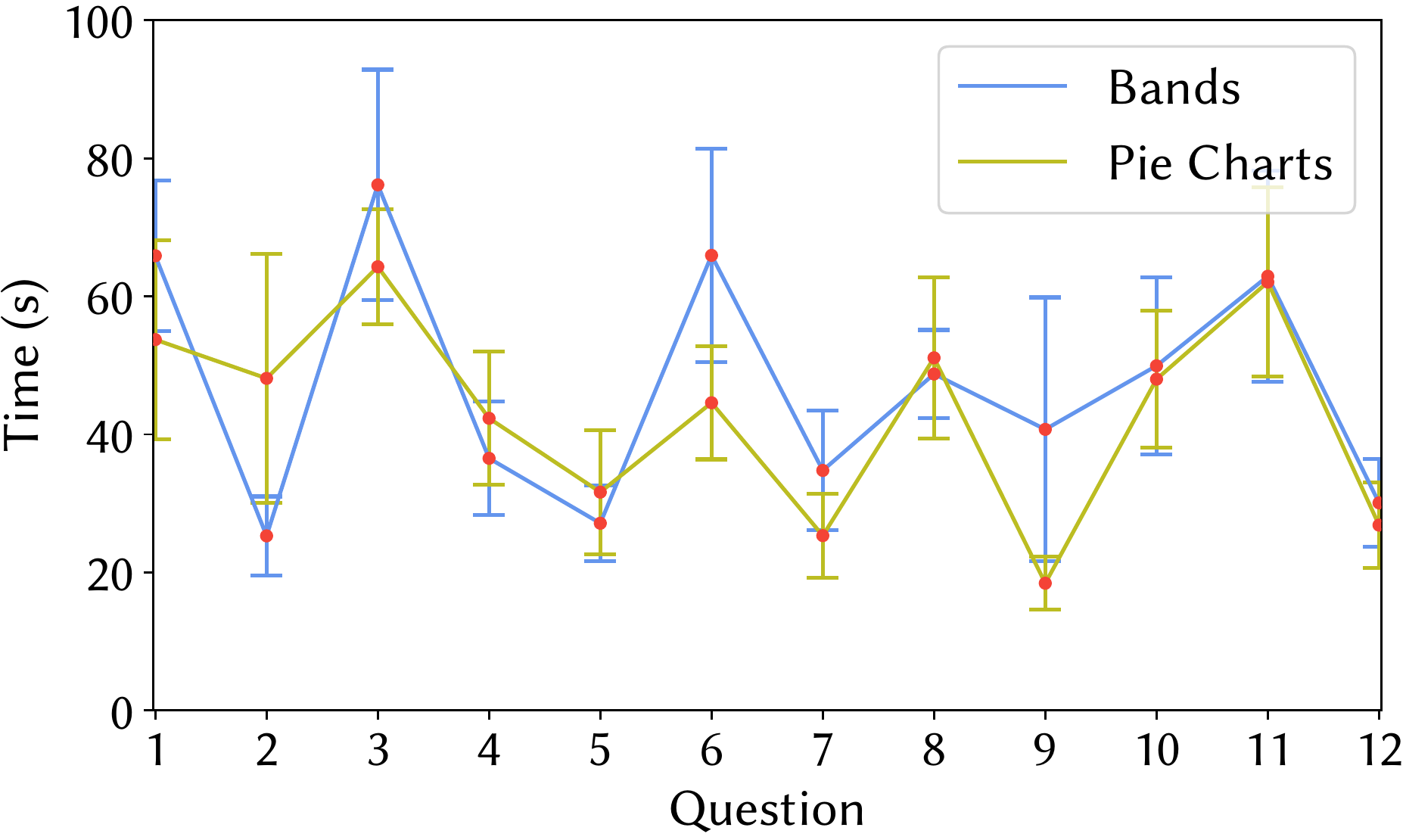}
 \vspace{-0.7cm}
 \caption{Average and confidence interval of time participants needed to answer questions in the user study using bands and pie chart-based spheres.}
 \label{fig:user-stufy}
 \vspace{-0.4cm}
\end{figure}

\subsection{Analysis}

For all tasks, we measured the time it took the users to perform the task, and how accurately the task could be performed.
%Therefore, we defined a similarity score between two different multi-parameter distributions based on the sum of pairwise differences, and consider distributions similar if ... and a peak if...
%Fig.~\ref{fig:user-stufy} shows the accumulated results for the time, including in the caption the means and stdevs of the measured times over all participants. 
Fig.~\ref{fig:user-stufy} shows the accumulated results over all participants. 
Participants were able to answer the questions in $48.1s$ (bands) and $42.8s$ (pie charts). A Student's t-test yields a t-value of $1.83$ and a p-value of $6.84\%$, hinting at statistical significance of the deviation between the two distributions. Participants were able to answer the questions correctly with an accuracy of $79.0\%$ (bands) and $79.6\%$ (pie charts). Applying a Student's t-test to this data yields a t-value of $-0.138$ and a p-value of $89.1\%$. Thus, we can conclude that there was no statistically significant difference in the accuracy of the participants to solve the tasks between pie charts and bands, however, it is statistically significant that participants could solve the tasks more quickly with the pie charts.
%This shows that both visual mappings perform almost equally well in analysis tasks. Still, about two thirds of the participants stated in the general questions that they preferred the pie charts over the bands-based spheres, while users in the remaining third commended the lens-like character of the bands on the spheres in the free text feedback. Consequently, we believe it should be left up to the users to choose the visual mapping of their choice, with the default being the pie charts.
About two thirds of the participants stated in the general questions that they preferred the pie charts over the bands-based spheres, while users in the remaining third commended the lens-like character of the bands on the spheres in the free text feedback. Consequently, we believe it should be left up to the users to choose the visual mapping of their choice, with the default being the pie charts.

Since many time steps are shown along the bands, while only one single time step is shown on a pie chart, it was conjectured that peaks could be overlooked in an animation when using pie charts while they remain visible for a longer time when using bands.

Multiple users asked for a mouse hover-over to inspect the values of the quantities below the mouse cursor. This feature is already supported, but could not be provided during the image-based study. Two users asked for discrete, quantized color maps instead of the continuous color maps we used in the study. This feature is also already available in Met.3D, and we show an image using quantized 8-class single hue color maps in Appendix G. %Still, the users got a respectable accuracy of $85.5\%$ for this type of question even without any of these additional assistances.

Another issue that was raised is the number of different sensitivities that can be well distinguished. In particular, multiple user said that no more than 4-6 different sensitivities should be shown simultaneously, in particular because bands and pie pieces become too small when viewed from a distance. Furthermore, due to decreasing sensitivity with decreasing value, low values were perceived similarly. However, this was not seen as a limitation, because attention is put on the high sensitivity values. What was perceived as problematic by many participants is the number of trajectories that are simultaneously displayed, as comparing the corresponding multi-parameter distributions then becomes tedious. Thus, we support deselecting individual trajectories with the mouse. These trajectories are then desaturated in the 3D view.%, and no sphere glyph is shown for them. 

\section{Implementation}\label{sec:implementation}

All techniques presented in this paper have been integrated into Met.3D, which uses the OpenGL API for GPU-based rendering. 
For drawing the curve plots view, the vector graphics library NanoVG\footnote{\url{https://github.com/memononen/nanovg}} is embedded. It provides hardware-accelerated rendering of vector graphics elements like anti-aliased lines and polygons, and the specification of scissor geometry to restrict rendering to a rectangular screen region. This is necessary for providing a scroll bar for the content of the curve plots view. 
%Compared to dedicated UI libraries, NanoVG enables more flexibility for designing a custom UI. 
%All of the UI logic has been implemented anew for this work.

Met.3D offers functionality to render three-dimensional trajectories via illuminated polygonal tubes, including the base map showing the earth surface and shadows cast by the trajectories. However, the specific rendering options required by our approach, i.e., showing view-aligned bands on trajectories and spheres, as well as view-aligned pie charts on spheres, are not available. Notably, these options cannot be realized using object-space texture mapping or standard pixel shaders due to the requirement to keep the color patterns fixed in screen space. 

A detailed description of our implementation is given in the Appendix. In the following, we outline the basic concepts underlying the implementation, including additional rendering options.
%Sketches of the resulting visual representations can be found in \autoref{fig:implementation-geometry}.

\begin{figure}[t]
 \centering
 \includegraphics[width=0.98\columnwidth]{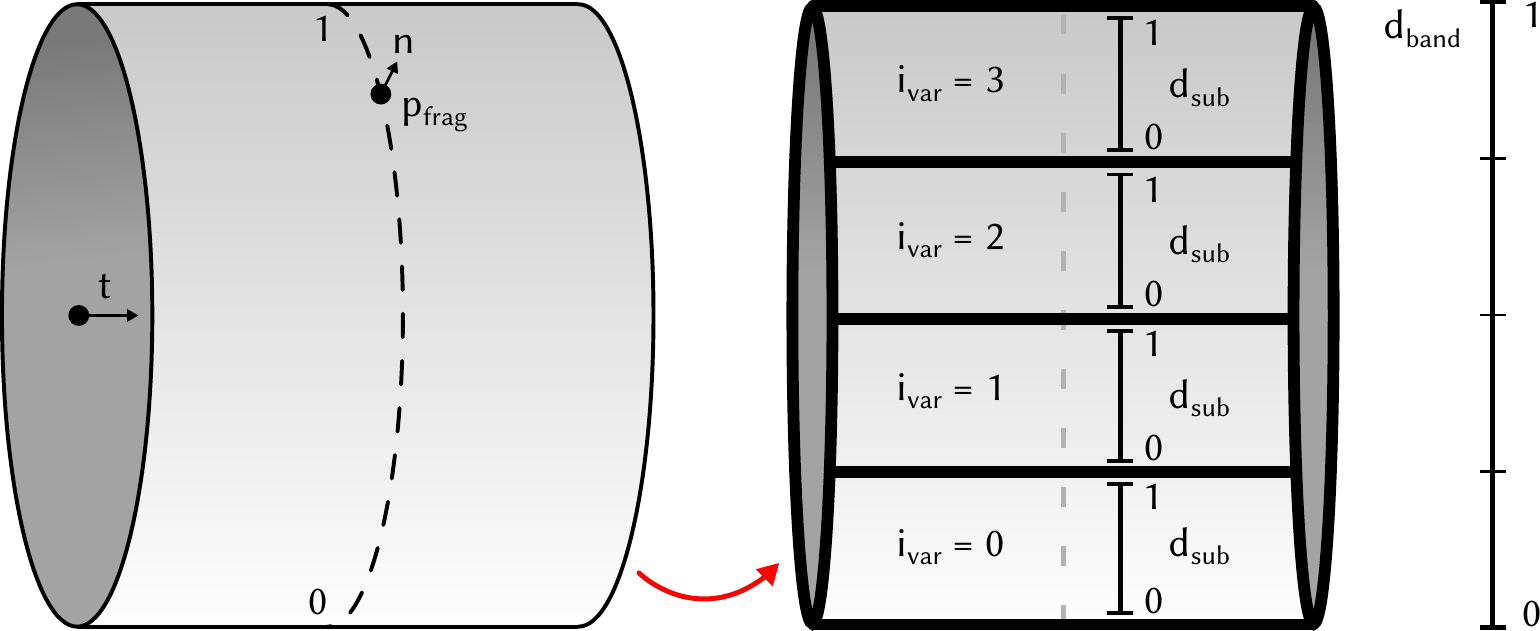}
 
 \vspace*{0.1cm}
 
 \includegraphics[width=0.98\columnwidth]{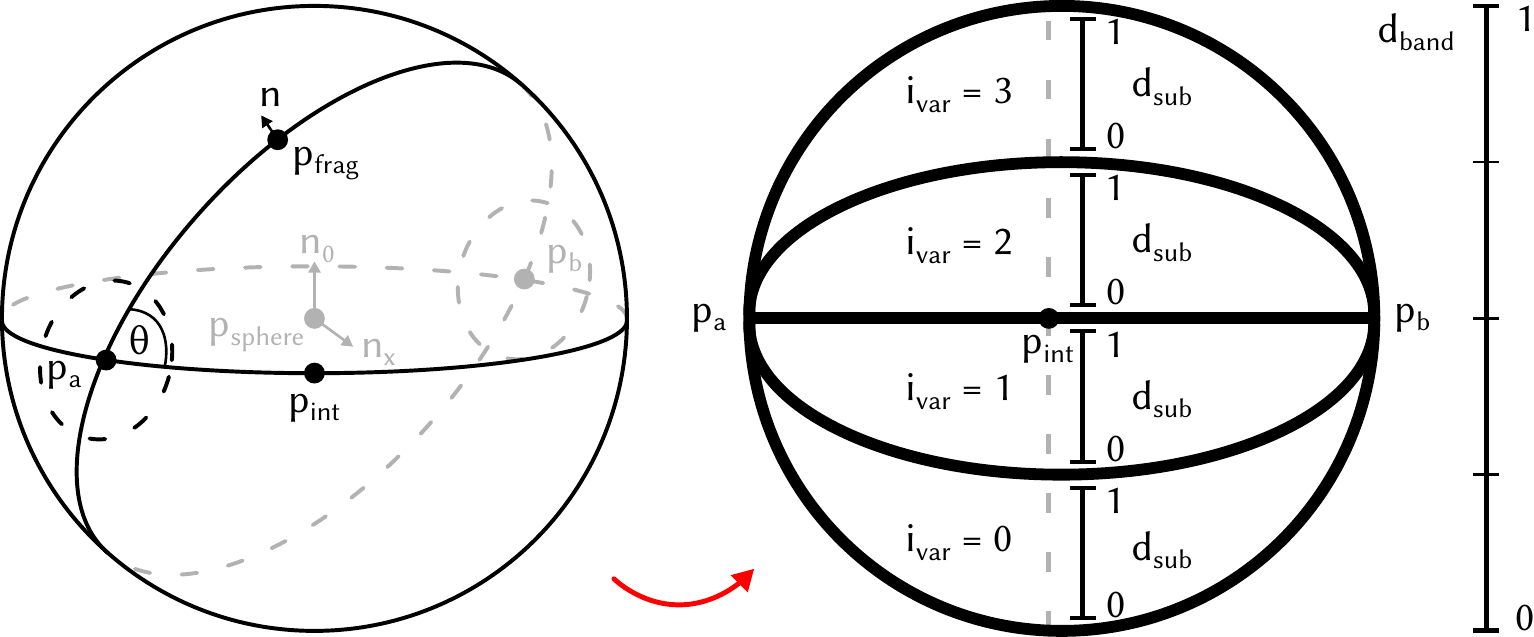}
 \caption{Top: Illustration of local surface properties and subdivision of the visible part of a trajectory to determine a fragments band position $d_{band}$, the sub-band position $d_{sub}$ and its corresponding variable ID $i_{var}$. Bottom: Same as top, but now to compute the band information on a sphere. 
 %Illustration of how the input vectors and points on the sphere are used to compute the band position $d_{band}$, the sub-band position $d_{sub}$ and its corresponding variable ID $i_{var}$.
 }
 \label{fig:tube-band}
 \vspace{-0.5cm}
\end{figure}

\textbf{View-aligned bands} For rendering the trajectories, 
%parameters are mapped onto the tube surface as view-aligned bands, as outlined in \autoref{sec:trajectoryview}. 
%Therefore, 
it needs to be determined for each fragment that is rendered for the tube surface to which of the $n$ bands in screen space it belongs. 
Each fragment lies on a circular arc orthogonal to the trajectory tangent (cf.~\autoref{fig:tube-band}~top). 
%\maicon{I do not see a label "a" in Fig. 8.} 
The bands run perpendicular to this arc along the trajectory's tangent direction. In order for the bands to have equal thickness, the angle along the arc to the fragment position is projected onto a line perpendicular to the tangent, which removes the curvature of the arc from the individual bands. The projected arc is then subdivided into $n$ sectors which all have the same height in screen space, and the fragment is classified according to the sectors by computing its relative position $d_{band}$ in the projection and assigning the corresponding variable index $i_{var}$ to it.
%(cf.~\autoref{fig:tube-band}b \maicon{Label "b" in Fig. 8?}). 
All required parameters can be derived solely from local properties of the rendered surface, i.e., the surface normal vector $n$, the trajectory tangent vector $t$ and the camera view vector $v$. In particular, by projecting the camera view direction into the plane orthogonal to the trajectory's tangent direction, the problem of computing the circular arc and the angle it subtends can be reduced to a two-dimensional problem
(cf.~Fig. 2 in Appendix A). 
%(cf.~\autoref{fig:tube-projection} in \autoref{sec:appendix-tube}). 

% \begin{figure}[tb]
%  \centering
%  \includegraphics[width=\columnwidth]{sphere/GreatCircle}
%  \caption{Illustration of how the input vectors and points on the sphere are used to compute the band position $d_{band}$, the sub-band position $d_{sub}$ and its corresponding variable ID $i_{var}$.}
%  \label{fig:sphere-band}
% \end{figure}

% Since For this, the band position $d_{band}$ and variable index $i_{var}$ need to be computed (cf.~\autoref{fig:implementation-geometry} bottom left and \autoref{sec:appendix-tube}). These values can be derived solely from local properties of the rendered surface, i.e., the surface normal vector $n$, the trajectory tangent vector $t$ and the camera view vector $v$. The bands run perpendicular to the tangent direction of the trajectory along a circular arc. In order for the bands to have equal thickness, the angle of the arc is projected onto a line perpendicular to the tangent, which removes the curvature of the circular arc from the individual bands. \christoph{@RW: Is it clear that how to do this is explained in the appendix?}

%Bands on a sphere are generated by first rendering each sphere as a polygon model on the GPU. In a pixel shader
Bands on a sphere are generated by first rendering each sphere as a polygon model on the GPU. In a preprocess
%, each generated fragment has access to the camera position $p_{cam}$ and 
the first ($p_a$) and last ($p_b$) intersection points between the trajectories and the spheres are computed, including the times $t_a$ and $t_b$ at which these intersections occur.
%along the trajectory are given. $p_{cam}$ is used to compute the view vector, 
$p_a$ and $p_b$ are then used in the pixel shader to determine the trajectory direction, and $t_a$ and $t_b$ are used to interpolate the parameters continuously along the bands travelling across the surface.
The mapping is obtained by first computing the projection of the sphere center $p_{sphere}$ onto the sphere under the used camera transformation, i.e. $p_{int}$ (cf.~\autoref{fig:tube-band}~bottom). The circle going through this point and the points $p_a$ and $p_b$ serves as a reference circle relative to which the bands are arranged.  
%For this, a ray-sphere intersection algorithm \cite{RaySphereIntersection} is used with a ray starting at the camera position with direction $p_{sphere} - p_{cam} $, which yields the point $p_{int}$.
Therefore, the normals of the disks $P_0$ spanned by $p_a, p_b, p_{int}$ and $P_{frag}$ spanned by $p_{a}, p_{b}, p_{frag}$ are computed. These two disks meet under a certain angle $\theta$.
By using this angle and the disks' normals, a measure of the distance to the border of the sphere in screen space can be computed, from which the band position and derived sub-band positions are obtained.  
To continuously map a parameter onto a band, it is assumed that $t_a$ and $t_b$ are linearly interpolated along the line from $p_a$ to $p_b$. The time at the point on this line closest to $p_{frag}$ is used as the time at $p_{frag}$, and the parameter value at this time is read from the trajectory data.

% on the line segment $(p_{a}, p_{b})$ connecting the entrance and exit point is computed. The distance of this point to $p_{a}$ and $p_{b}$ is then used to linearly interpolate between the time steps $t_a$ and $t_b$ at the entrance and exit point to get the time step $t_{frag}$ used for the fragment on the sphere surface. Then, $t_{frag}$ is used to look up the parameter values of the trajectory at time step $t_{frag}$ from a shader storage buffer stored on the GPU.

\textbf{View-aligned pie charts} To color a sphere with a pie chart that encodes the values of multiple parameters into its pieces, the screen space projection of the sphere is subdivided into a predefined number of individual pieces. To achieve a consistent assignment of parameters to pieces for all spheres, first the angle $\alpha_{band}$ representing the angular distance of a fragment $p_{frag}$ to the up-axis of the camera is computed. The global band position $d_{band}$ is then given by 
\begin{equation}
d_{band} =\frac{\alpha_{band} \mathrm{\ mod\ } 2\pi}{2\pi}.
\end{equation}
When mapping $N$ parameters onto the sphere, the band position $d_{band} \in [0, 1)$ is subdivided into multiple sub-band positions $d_{sub}$.

\section{Results: Case-Study ``Vladiana''}\label{sec:case-study}

\begin{figure*}[!htb]
    \centering
    \begin{tabular}{cc}
        \begin{minipage}{0.716\textwidth} \includegraphics[width=\textwidth]{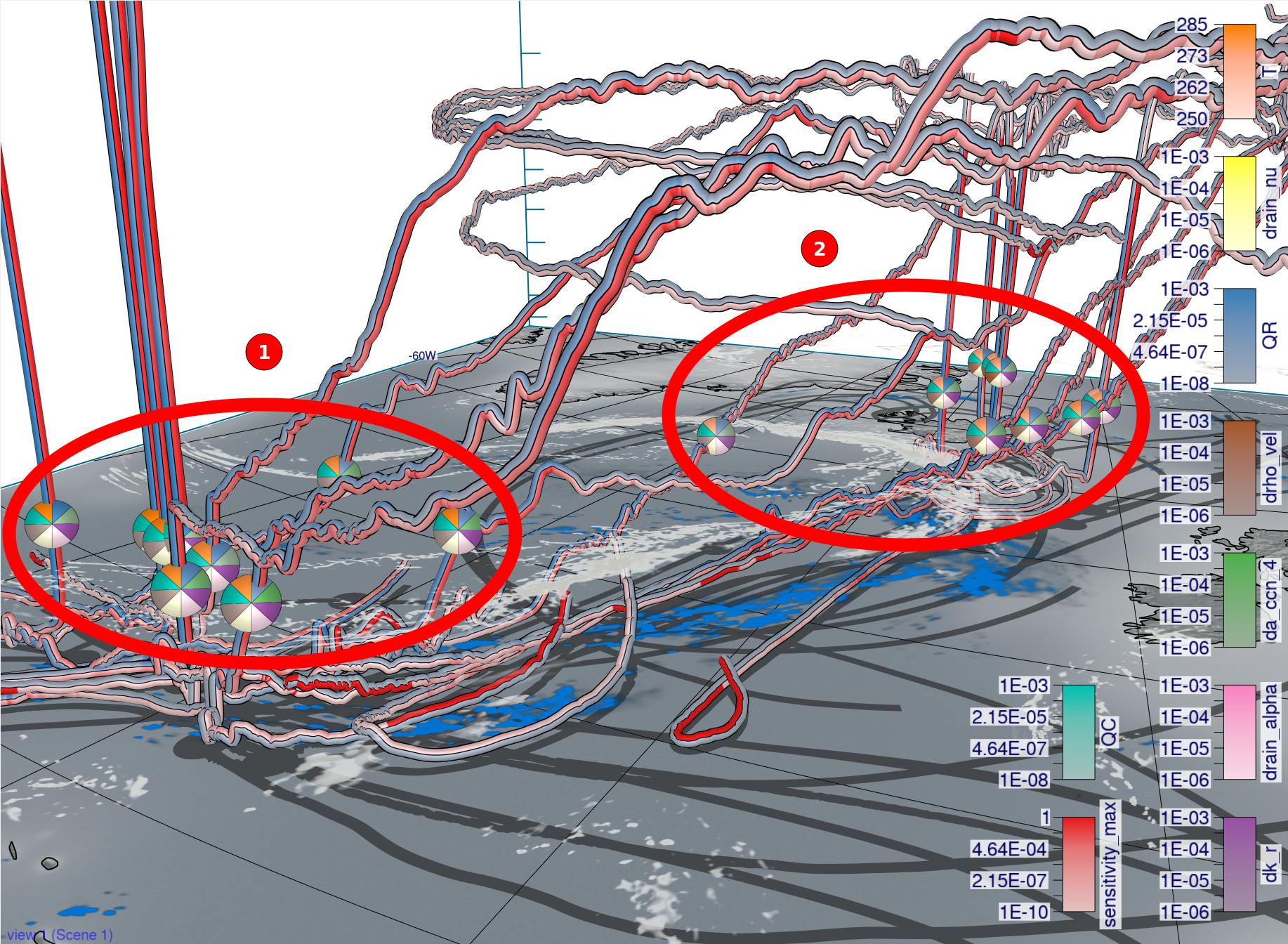} \\ \footnotesize{a)} \end{minipage}& \begin{minipage}{0.24\textwidth} \includegraphics[width=\textwidth]{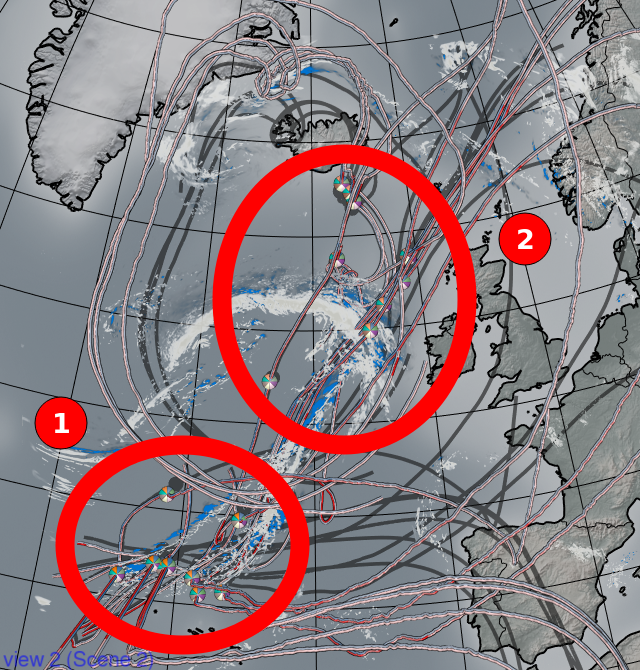} \\ \footnotesize{b)} \\ \includegraphics[width=\textwidth]{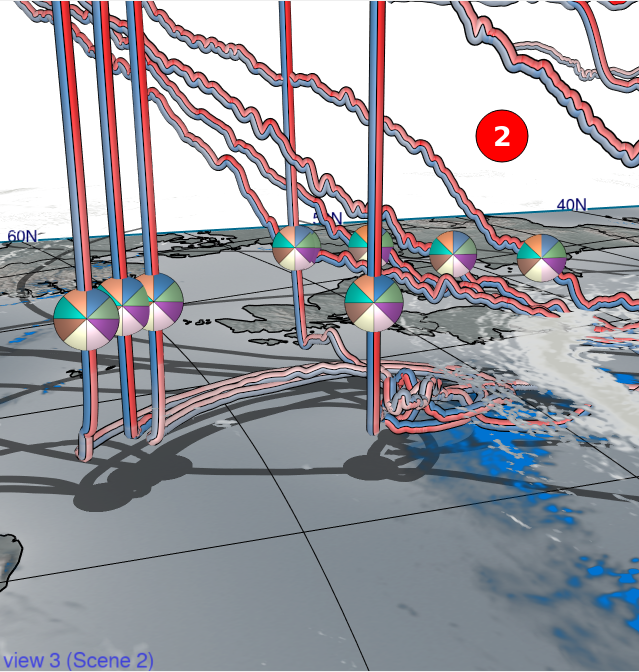} \\ \footnotesize{c)} \end{minipage}
    \end{tabular}
    \caption{Overview of selected trajectories and first insights with spheres at the same height. Low-level clouds at approximately 1500\,m altitude (gray) and surface precipitation (blue) are shown at 07 UTC 23 September 2016 when multiple trajectories start their ascent. a) Trajectories ascending in the south (group 1) and in the north (group 2) with spheres showing eight variables each. b) View from the top with the northern group 2 near clouds and precipitation and the southern group 1 with less clouds and precipitation. c) A close-up view of group 2. \label{fig:use_case:overview}}
\end{figure*}
%We demonstrate the value of our method by discussing first investigations of the sensitivity of the rain mass density (QR) simulated by the NWP model to microphysical parameters along WCB trajectories within \annika{cyclone} “Vladiana” concerning the amount of rain mass density simulated by the numerical model. 
We demonstrate the value of our method by discussing first investigations of the sensitivity of the rain mass density (QR) simulated by the NWP model to microphysical parameters along WCB trajectories within cyclone ``Vladiana''.
We are particularly interested in whether there are differences in the sensitivities along the ascent based on the location and the ascent rate.
For this analysis, we first employed the exploratory visual analysis capabilities of the proposed method to identify the regions and times of interest. Appendix E demonstrates how the domain scientists selected the trajectories of interest.
%We demonstrate in Appendix E how the domain scientists were able to use the exploratory visual analysis capabilities to discover these regions and times of interest. 
In this section, we will focus on these specific selected trajectories to analyze the joint development of multiple sensitivity parameters.
%For this analysis, we first employ the exploratory visual analysis capabilities of the proposed method to discover regions and times of interest, before focusing on specific trajectories to analyze the joint development of multiple sensitivity parameters.
% All data shown in the figures presented in this section are produced via the AD-based sensitivity analysis tool by Hieronymus \cite{hieronymus_algorithmic_2022} using the dataset described in detail by Oertel et al.~\cite{oertel_potential_2020}.
WCB trajectories ascend in a wide region near the extratropical cyclones' fronts %during the period 
between 23 September 2016 and 26 September 2016. For the example presented here, we compare sensitivities related to QR along trajectories in different regions of the cyclone.
% For the example presented here, we are interested in comparing sensitivities of WCB trajectories in different regions of the cyclone. 
From a total of 8744 available WCB trajectories %(cf.~\cite{oertel_potential_2020}), 
we select two groups of trajectories, one in the north and one in the south of the region of interest (cf.~\autoref{fig:use_case:overview}). From both groups, we select ten trajectories each, those five with the slowest ascent and those five with the fastest ascent. This results in 20 trajectories with a large variance in their ascent rate and location.

\begin{figure}[ht]
    \centering 
    \subfloat[\label{figure:use_case:bands:south}]{\includegraphics[width=0.5\columnwidth]{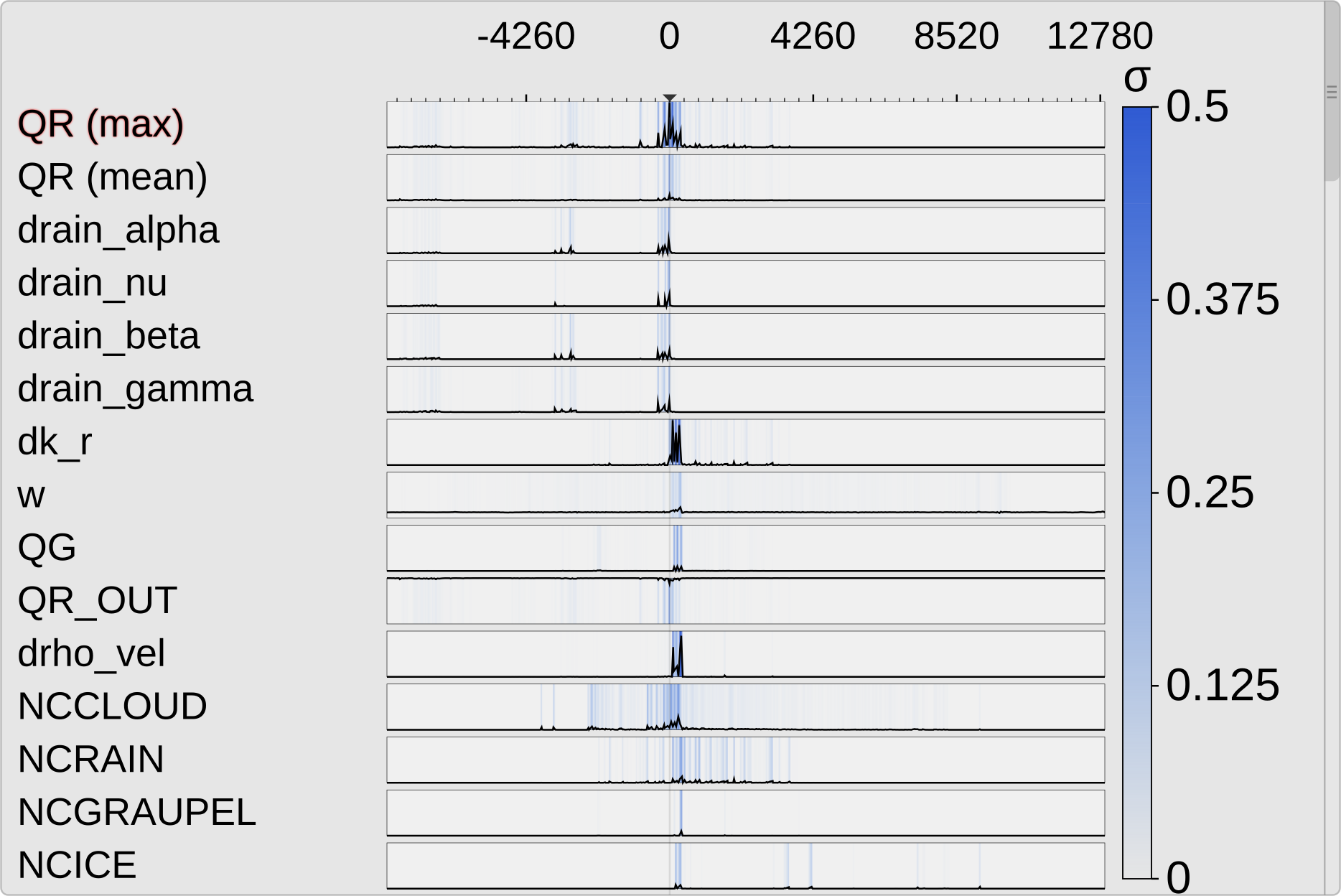}}
    % \hfill
    \subfloat[\label{figure:use_case:bands:north}]{\includegraphics[width=0.5\columnwidth]{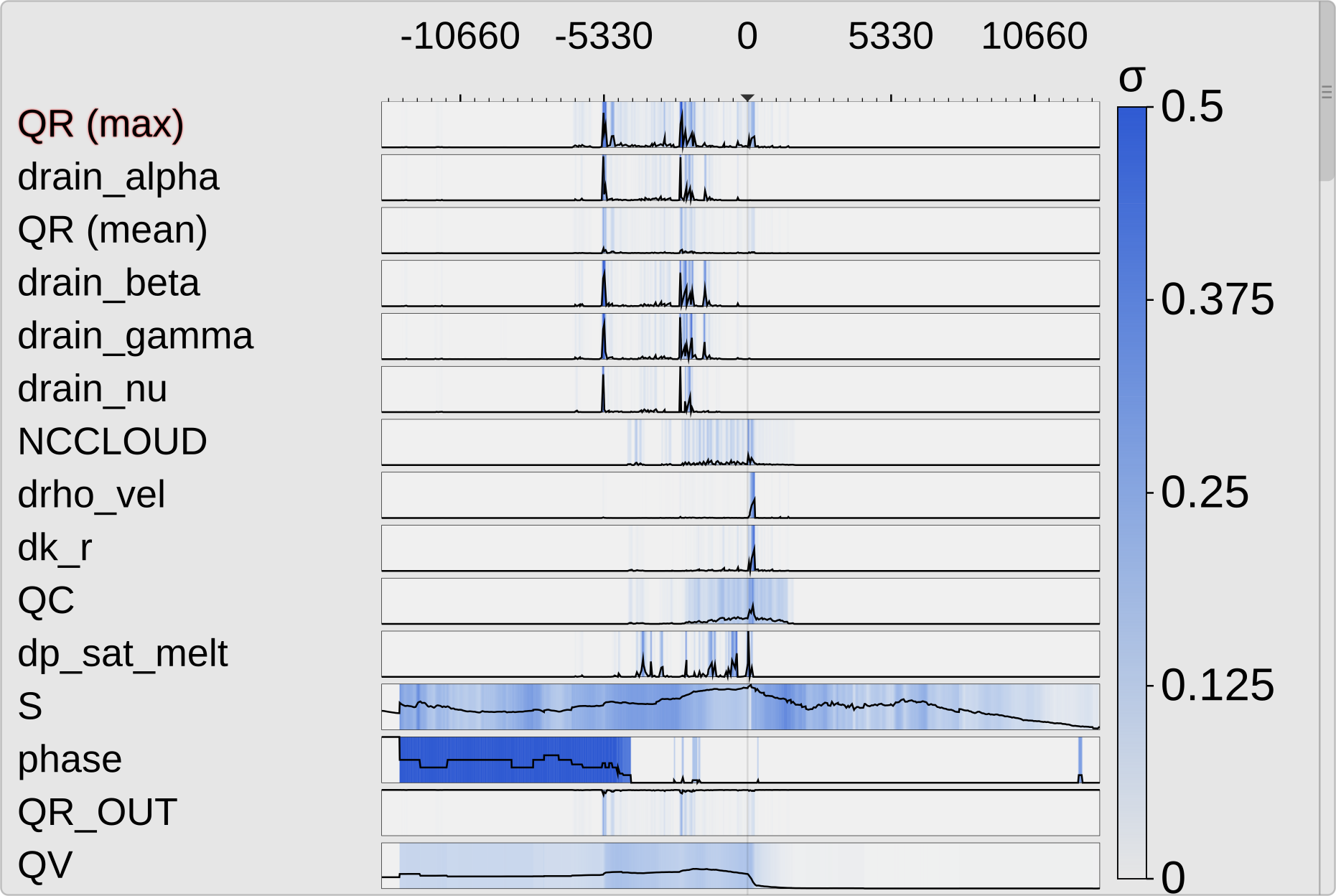}}
    \caption{The curve plots are centered at the start of the ascent of the trajectories. The labels for the x-axis show the simulation time step, where each simulation step stands for 20\,s. a) Only trajectories from the southern group have been selected. There are large peaks for rain mass density (QR) around the start of the ascent, indicating precipitation from above. 
    % The variance for water vapor mass density (QV) is large before the ascent starts such that we expect different precipitation intensities for the trajectories. The sharp rise in variance is evidence for different lead times of the trajectories before their ascent starts. 
    b) Trajectories from the northern group with a peak in QR several hours before their ascent starts. 
    %Those have a peak in rain mass density hours before the ascent starts.
    }
    \label{figure:use_case:bands}
\end{figure}

% MH: I commented the old start of this section
% Our dataset consists of a sample of 291 trajectories associated with the extratropical cyclone ``Vladiana'' over the North Atlantic.
%\maicon{I can, in principle, increase the amount to a couple thousand, but I do not see the benefit if we show the usage and do not do a thorough analysis; or I could start this part with the clustered dataset already; I am currently running the simulations for all trajectories and I am going to cluster those. The intermediate result with 5000 trajectories does not look that different. Edit: I created a set from 8744 trajectories with $k=9$.}
% As a starting point, we cluster the trajectories by the location of their rapid ascent with k-means and choose a cluster northwest of Britain and one west of Spain. From these clusters we choose the five trajectories with the slowest and the five with the most rapid ascent each and load this dataset in Met.3D.

\autoref{figure:use_case:bands} shows curve plots with trajectories selected either from the southern (\autoref{figure:use_case:bands}a) or northern (\autoref{figure:use_case:bands}b) group, to analyze trends of parameters across one or more groups of trajectories (Q1, Q2, and Q3). 
We select 
%the maximum of 
QR as target variable, and center the x-axis by the time of rapid ascent of each trajectory.
%Our target variable is the rain mass density (QR), and the x-axis is sorted by the time of the rapid ascent of each trajectory. 
The different locations of peaks in QR for both groups stand out. The southern group has QR maxima at the start of the ascent, while the northern group has larger QR maxima a few hours before the ascent starts. Such high QR along trajectories can arise from either (i) sedimentation of rain from above (influenced by parameters alpha, beta, and gamma in the numerical model's parameterization) or (ii) local production from collision of available cloud droplets (influenced by the cloud condensation nuclei (CCN), the mass density of cloud droplets (QC), and a cloud collision parameter (k\_r); for detailed description of these parameters see \cite{seifert_two-moment_2006,hieronymus_algorithmic_2022}). % We can add hieronymus_algorithmic_unpublished_2022 once it is published or even remove seifert_two-moment_2006 then.
Hence, we are interested in which process dominates in which region.
The automated ordering (\autoref{sec:diagramview}) of the parameters provides further insight. The parameters are sorted by similarity in each time step to the maximum of QR. The sensitivities to the parameters rain\_alpha, rain\_beta, rain\_gamma (used for sedimentation velocity), and rain\_nu (used in the particle-mass distribution of rain droplets) are the variables with the highest similarity to QR in both cases. 
%%This illustrates that there is enough rain in both groups for precipitation to occur. However, 
% 

\begin{figure}[t]
    \centering 
    \includegraphics[width=\columnwidth]{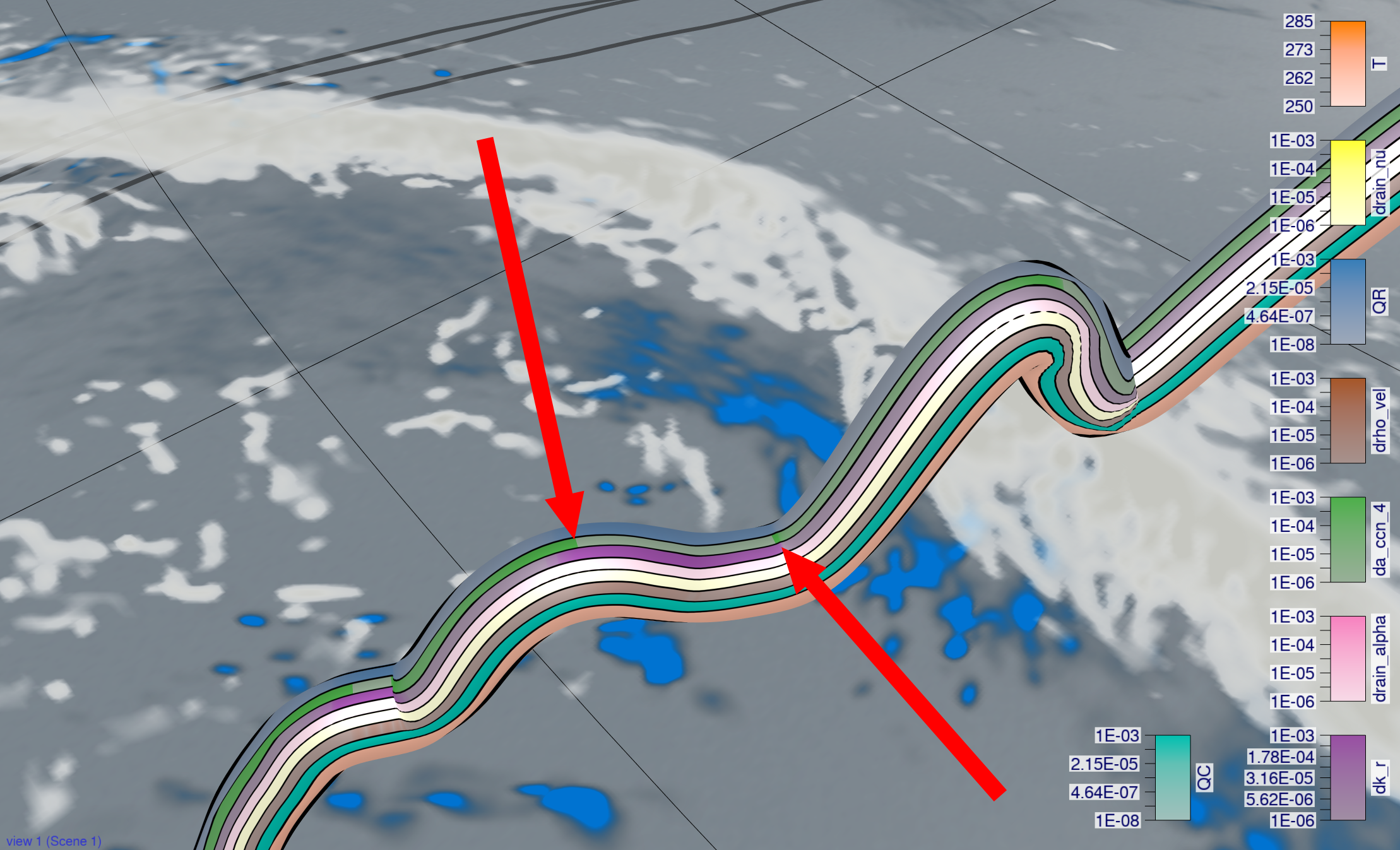}
    \caption{Zoomed in at a slantwise trajectory. The green band (da\_ccn\_4 associated with cloud droplet formation) alternates with the purple band (dk\_r associated with cloud droplet collision to form raindrops).}
    \label{fig:use_case:detail}
\end{figure}

Sensitivities to CCN parameters and to k\_r are ranked higher in the southern group, indicating that rain droplets' formation due to collisions of cloud droplets are closely related to local QR formation.
These correlations are not present in the northern group, which indicates that local QR maxima result from sedimentation of precipitation from above. We conclude that rain formation, specifically local maxima of QR, in the southern group is more closely related to the formation of cloud droplets and, therefore, the updraft $w$ of the trajectories than in the northern group (Q3). 

To elaborate on Q3, we compare the maximum sensitivity of QR to any parameter in \autoref{fig:use_case:overview}. The blue color along trajectories shows QR, whereas red indicates the maximum sensitivity to any parameter. Low sensitivity values (i.e., unsaturated bands) appear mostly when the trajectories descend and after they have reached their maximum height (\autoref{fig:use_case:overview}a,b). 
%We can conclude that processes occur mainly in lower heights or where there is an updraft. 
This emphasizes that processes influencing QR dominate at lower altitude and during updrafts.

The curve plots in \autoref{figure:use_case:bands} are centered at the start of the ascent to address Q4. The variance of the sensitivities (blue shades) is similarly distributed for both groups, but peaks appear at different times.  
% We can infer that there is more variance between trajectories with different locations of ascent than there is between trajectories with a similar location.
We can infer that the variance between trajectories with different locations of ascent is higher than between trajectories with a similar location.
% Furthermore, we see similar variances within each cluster for all parameters (compare blue shades in Figure~\ref{figure:use_case:bands} left and right). We can infer there is more variance between trajectories with different locations of rapid ascent than there is between trajectories with a similar location (Q2 and Q4).

% In Figure~\ref{fig:use_case:overview}, we see an overview of the two clusters. On the left, we see at (1) the southern cluster and at (2) the northern one with the total amount of rainfall on the ground in blue and a cloud band starting in the south going north and then to the west. The blue color on the trajectories indicates the amount of rain mass density, whereas red indicates the maximum sensitivity to any parameter. We can see gray bands mostly when the trajectory goes downwards and once trajectories reach their maximum height (Figure~\ref{fig:use_case:overview} left and top right). 
At last, we investigate differences in sensitivities between convective and slantwise trajectories (Q5). 
We synchronize the height of the spheres along trajectories and select three variables (QC, QR, and temperature) and model parameters with a high ranking from \autoref{figure:use_case:bands}. 
%We synchronize the height of the spheres along trajectories to compare the sensitivities of different parameters and select three model state parameters (the amount of cloud and rain mass density and temperature) and parameters with a high ranking from Figure~\ref{figure:use_case:bands}. 
The color intensities of da\_ccn\_4 (green) in slantwise ascending trajectories (e.g., \autoref{fig:use_case:overview}a) is lower than for convective ones. This indicates that processes associated with a\_ccn\_4 have a smaller effect on QR during slantwise ascent. 
%The color intensities of da\_ccn\_4 (green; sensitivity to CCN parameter a\_ccn\_4) in slantwise ascending trajectories (e.g., the trajectory ascending from south to north on the left panel of \autoref{fig:use_case:overview}) is lower than for convective ones. This indicates that processes associated with a\_ccn\_4 have a smaller effect on the rain mass density. 
% The lower color intensities of green bands in slantwise ascending trajectories (i.e., trajectories ascending from south to north in the left panel of \autoref{fig:use_case:overview}) indicates that processes have a smaller effect on the rain mass density than convective ones (Q5). 
Furthermore, the collision of cloud droplets (dk\_r; purple color) is more important during convective ascent.
% Furthermore, dk\_r (purple) has higher values for convective ascent, which is related to the collection of cloud droplets that become rain droplets.
This agrees with our previous assessment, and shows that the formation of cloud droplets and their subsequent conversion to QR are more important for convective ascent than for slantwise ascent.
% This agrees with our previous assessment that the formation of cloud droplets and their accumulation and conversion to rain droplets affects rain mass density more in convective ascents than in slantwise ascents.

% For the spheres in \autoref{fig:use_case:overview} left and bottom right, we selected three model state parameters (the amount of cloud and rain mass density and temperature) and parameters with a high ranking from \autoref{figure:use_case:bands}. We compare the spheres at the same height during the ascent (Q5). 
% It stands out that the temperature is higher in the south than in the north (yellow compared to orange). Furthermore, the purple segment of the sphere ($k_r$) is more saturated in the south, which is related to the collection of cloud droplets that become rain droplets. 
% We can see in the southern cluster a sphere with no green color (da\_ccn\_4) along a trajectory with a different shape than the others. We will investigate this shape on the example of a slantwise ascending trajectory near the precipitation pattern that stretches from the south to the north.

% \begin{figure}[ht]
%     \centering 
%     \includegraphics[width=\columnwidth]{figures/use-case/trajectory_detail_10.png}
%     \caption{Zoomed in at a slantwise trajectory. The green band (da\_ccn\_4 associated with cloud droplet formation) alternates with the purple band (dk\_r associated with cloud droplet collision to form raindrops).}
%     \label{fig:use_case:detail}
% \end{figure}

\begin{figure}[ht]
    \centering 
    \includegraphics[width=\columnwidth]{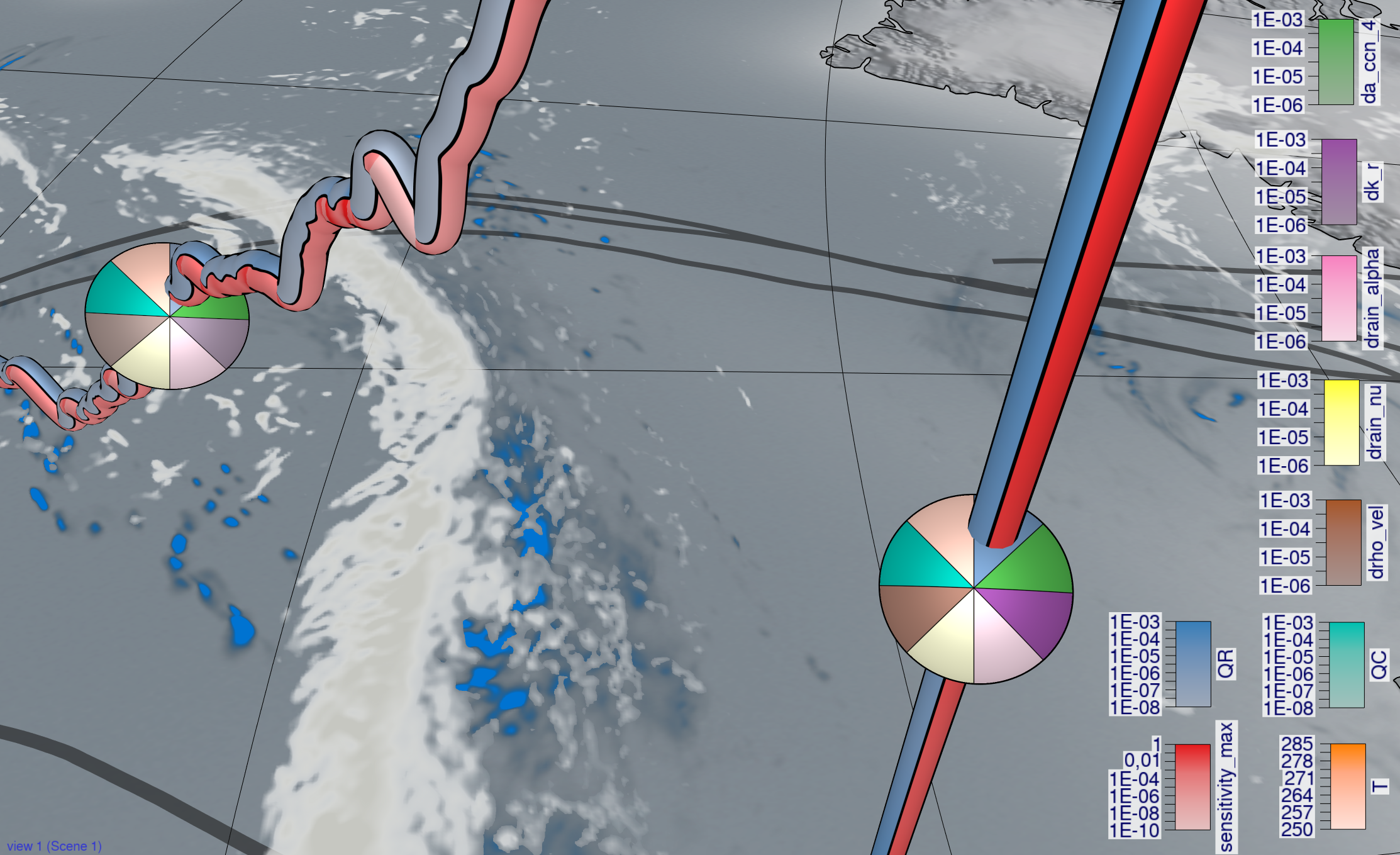}
    %\caption{Zoomed in at a convective trajectory with the slantwise trajectory from \autoref{fig:use_case:detail} on the left. The green (da\_ccn\_4 associated with cloud droplet formation) and purple band (dk\_r associated with cloud droplet collision to form raindrops) are continuously saturated during the convective ascent.}
    \caption{Zoomed in at a convective trajectory with the slantwise trajectory from \autoref{fig:use_case:detail} on the left. The green (da\_ccn\_4 associated with cloud droplet formation) and purple sensitivity (dk\_r associated with cloud droplet collision to form raindrops) have simultaneously high values during the convective ascent.}
    \label{fig:use_case:detail_conv}
\end{figure}

For a more detailed analysis, we zoom in to one slantwise ascending trajectory, and use multiple bands to show each parameter (cf.~\autoref{fig:use_case:detail}).
%To uncover further details regarding Q5, we zoom in to one slantwise ascending trajectory, and use multiple bands to show each parameter (cf.~\autoref{fig:use_case:detail}). Alternatively, one can use the same settings as in \autoref{fig:use_case:overview} and interactively move the sphere along the trajectory. 
\autoref{fig:use_case:detail} reveals an alternating pattern between dk\_r (purple) and da\_ccn\_4 (green). The overall slantwise ascent of the trajectory shows short periods of sharp ascent with strong cloud droplet formation. These periods are interrupted by periods of slower ascent, where the collision of cloud droplets dominates. These processes do not alternate in convectively ascending trajectories (cf.~\autoref{fig:use_case:detail_conv}) and rather occur simultaneously. This can produce and accumulate large amounts of QR quickly, cf.~\autoref{fig:use_case:overview}c with slantwise trajectories in the background and convective trajectories in the foreground,
%\annika{@Maicon: can we refer here to Fig. 10 c, to show that QR for convective trajs is generally larger than for slantwise trajs?} 
%\maicon{Good idea. ", cf. \autoref{fig:use_case:overview}c with slantwise trajectories in the background and convective trajectories in the foreground"}, 
leading to more intense surface precipitation in a limited area. In contrast, slantwise ascent spreads these processes over a larger area, explaining the impact of such trajectories on large-scale precipitation patterns ~\cite{oertel_convective_2019, oertel_potential_2020, oertel_observations_2021, jeyaratnam_upright_2020}. 

%\maicon{I can expand on the curve-plots by using a larger sample and checking for different peaks in different time steps etc. if we need more material.}

\section{Conclusion and Future Work}

We propose a novel visual analysis workflow to investigate the relationships between the sensitivity of a target variable to changes in model parameters and the location and shape of WCB trajectories. This information is required to analyze the validity of physical assumptions on which microphysical parameterizations in the numerical model codes are based. Making the sensitivities accessible along important features such as WCB trajectories offers new insights into the correlation structures between different parameters and differences between trajectories. To perform these analyses in an effective way, we link a curve plot-based summary view with a novel sphere-based focus view that enables comparison of multi-parameter distributions on different trajectories. The curve plots view provides statistical overviews and enables to quickly find parameters with similar temporal evolution.
% \maicon{We mapped parameters to bands to analyze the interaction of multiple parameters on a single trajectory in detail.}
%We develop the workflow in a team of scientists from visualization, high-performance computing, and meteorology, and integrate it into the open-source meteorological visualization software Met.3D. The workflows usability and benefits is demonstrated with a real-world case-study.
We develop the workflow in a team of scientists from visualization, high-performance computing and meteorology, and integrate it into the open-source meteorological visualization software Met.3D. The usability and benefits of the workflow is demonstrated with a real-world case-study.

In the future, we intend to extend the workflow in multiple ways. Firstly, we will investigate how to effectively show additional 3D atmospheric fields, or features in these fields, in the surrounding of trajectories, to reveal specific regional multi-field patterns causing high sensitivities. Secondly, we intend to make the workflow usable with ensembles of trajectories, where multiple sets of trajectories from different simulation runs are considered. In this way, relationships between sensitivities and the ensemble spread can be examined.   
%\maicon{I'm just throwing in an idea. I can generate datasets with an additional dimension for the target variable:}
Thirdly, we are going to support multiple target variables that can be switched interactively.
%\annika{Maicon/Christoph: would it be worth to mention the AD analysis for the case we are currently/or in the future working on? This could be a planned application using this tool to map spatio-temporal variability of MCRPH sensitivities in a new case with the ICON model}
%\annika{@Maicon, please double-check/adjust the following sentence}
Furthermore, the atmospheric scientists are currently analyzing parameter sensitivities in another WCB case-study simulated with the state-of-the-art NWP model ICON \cite{Zaengl2015}. The visualization and analysis framework presented here will also be applied for the systematic analysis and mapping of sensitivities in this follow-up case-study.
\ifCLASSOPTIONcompsoc
  \section*{Acknowledgments}
\else
  \section*{Acknowledgment}
\fi

The authors acknowledge support by the Deutsche Forschungsgemeinschaft (DFG) within the Transregional Collaborative Research Centre TRR165 Waves to Weather, (www.wavestoweather.de), Projects A7, Z2 and B8 as well as funding from JGU Mainz.

\ifCLASSOPTIONcaptionsoff
  \newpage
\fi



\bibliographystyle{IEEEtran}
\bibliography{IEEEabrv,main}
%\begin{thebibliography}{1}


%\end{thebibliography}


%\newpage

\begin{IEEEbiography}[{\includegraphics[width=1in,height=1.25in,clip,keepaspectratio]{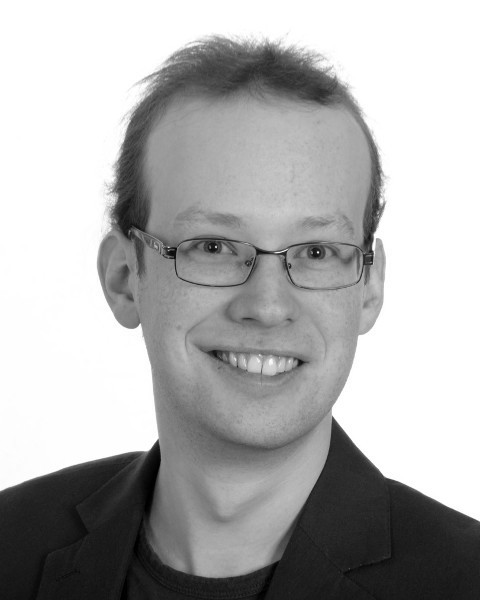}}]{Christoph Neuhauser}
is a PhD candidate at the Computer Graphics and Visualization Group at the Technical University of Munich (TUM). He received his Bachelor's and Master's degrees in computer science from TUM in 2019 and 2020. Major interests in research comprise scientific visualization and real-time rendering.
\end{IEEEbiography}

% if you will not have a photo at all: IEEEbiographynophoto

%\newpage

\begin{IEEEbiography}[{\includegraphics[width=1in,height=1.25in,clip,keepaspectratio]{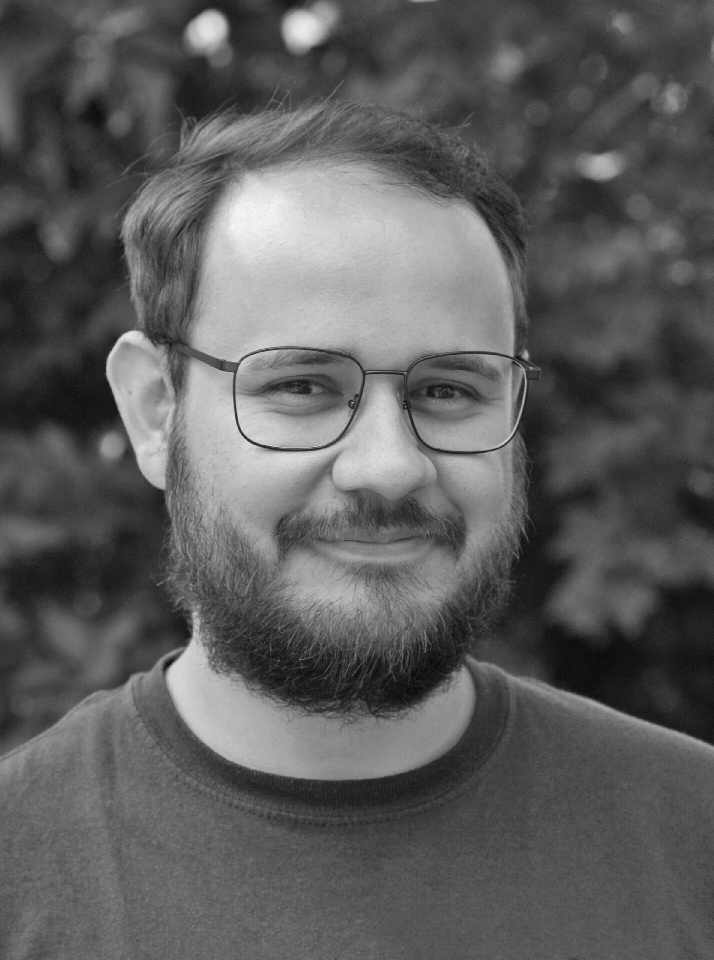}}]{Maicon Hieronymus}
is a PhD candidate at the Efficient Computing and Storage Group at the Johannes Gutenberg University Mainz (JGU) since 2019. He studied computer science (B.Sc.) and computer science with minor in physics (M.Sc.) at JGU. Research interests involve model analysis of cloud microphysical models and algorithmic differentiation in atmospheric physics.
\end{IEEEbiography}

\begin{IEEEbiography}[{\includegraphics[width=1in,height=1.25in,clip,keepaspectratio]{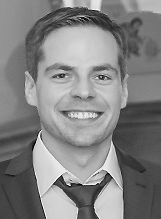}}]{Michael Kern}
is a research scientist and software engineer at Advanced Micro Devices (AMD) with major focus on real-time computer graphics and ray tracing. He studied computer science and received his Ph.D. from TUM in 2020. His thesis was concerned about feature detection and uncertainty visualization in meteorological data. Besides work, his main interests are scientific visualization and high-performance GPU computing. 
\end{IEEEbiography}

\begin{IEEEbiography}[{\includegraphics[width=1in,height=1.25in,clip,keepaspectratio]{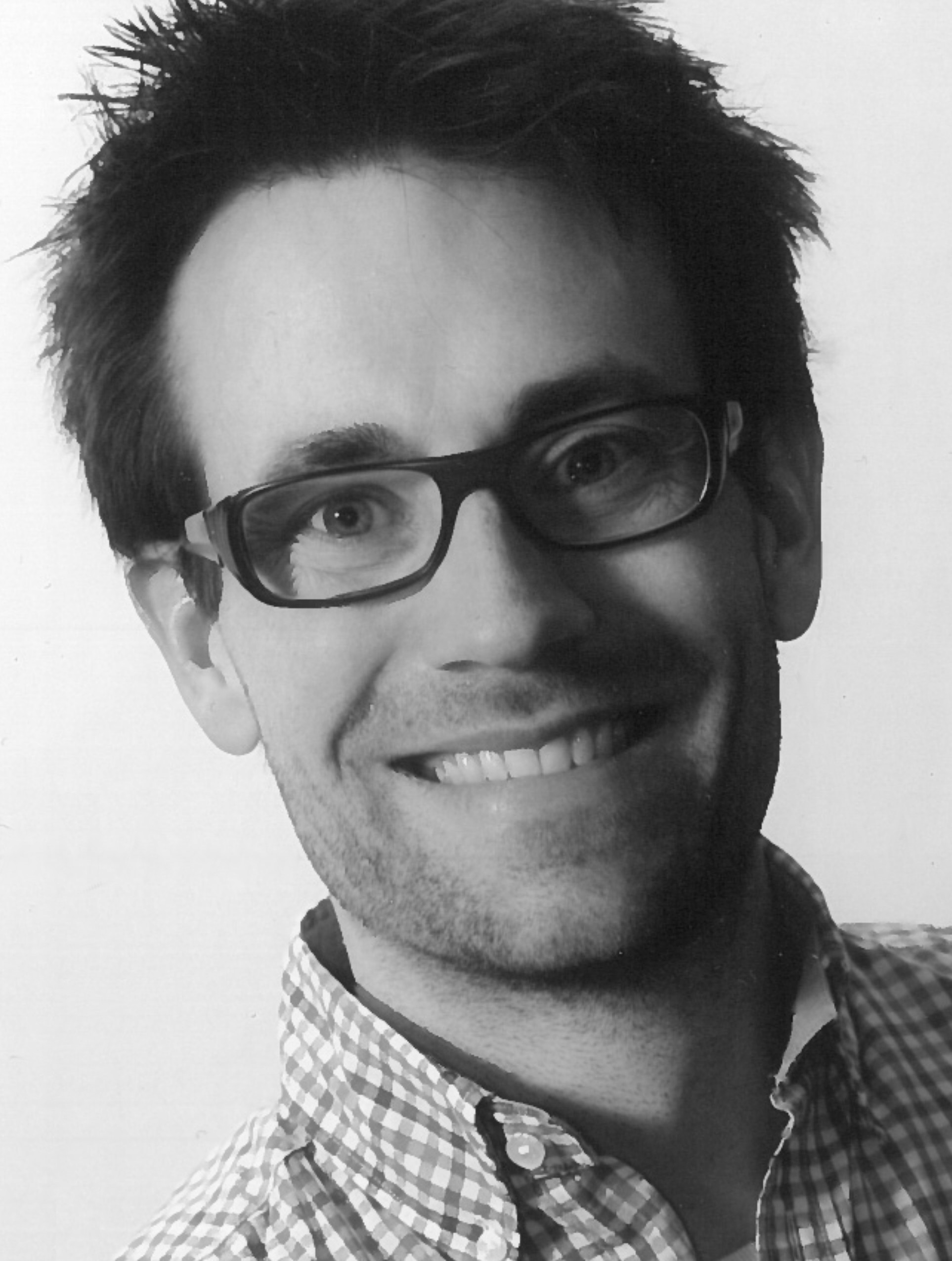}}]{Marc Rautenhaus}
received the M.Sc. degree in atmospheric science from the University of
British Columbia, Vancouver, in 2007, and the Ph.D. degree in computer science from TUM, Munich, in 2015. He currently leads the Visual Data Analysis Group at the Regional Computing Centre of Universit\"at Hamburg. Marc's research interests focus on the intersection of visualization, data analysis, and meteorology.
\end{IEEEbiography}

\begin{IEEEbiography}[{\includegraphics[width=1in,height=1.25in,clip,keepaspectratio]{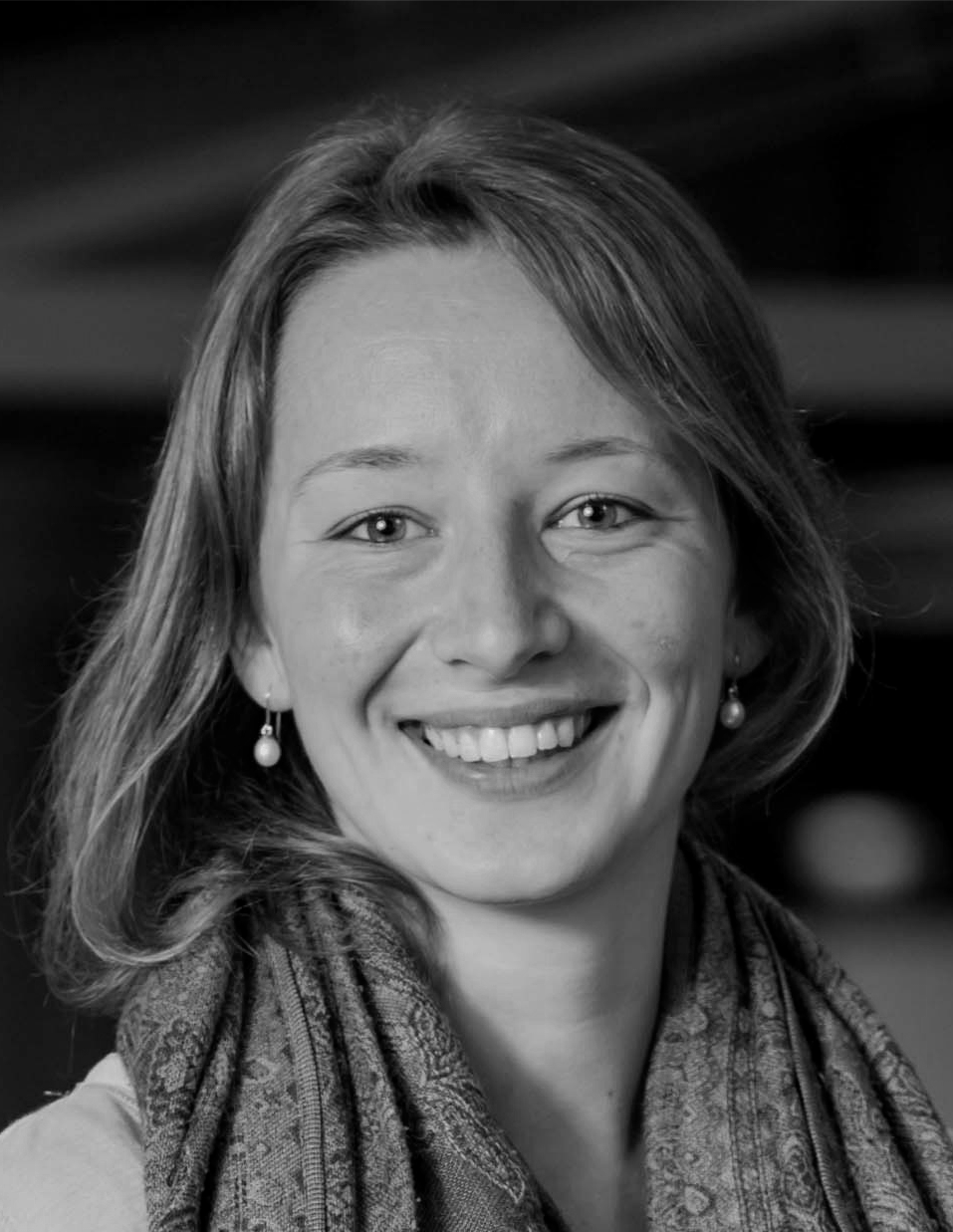}}]{Annika Oertel}
Annika Oertel studied atmospheric science and received a Ph.D. in atmospheric dynamics from ETH Zurich. Since 2020 she works as postdoctoral researcher in the Institute of Meteorology and Climate Research at Karlsruhe Institute of Technology. Her research focuses on the model representation of microphysical processes in warm conveyor belts and their interactions with the larger-scale extratropical circulation.
\end{IEEEbiography}

%\vfill
%\newpage

\begin{IEEEbiography}[{\includegraphics[width=1in,height=1.25in,clip,keepaspectratio]{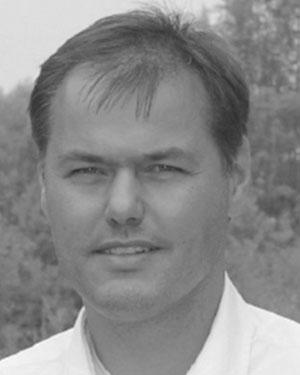}}]{R\"udiger Westermann}
studied computer science at the Technical University Darmstadt and received his Ph.D. in computer science from the University of Dortmund, both in Germany. In 2002, he was appointed the chair of Computer Graphics and Visualization at TUM. His research interests include scalable data visualization and simulation algorithms, GPU computing, real-time rendering of large data, and uncertainty visualization.
\end{IEEEbiography}

\vfill


%\vfill


\end{document}

%\title{Appendix: Visualization of Model Parameter \\ Sensitivity along Trajectories in Numerical \\ Weather Predictions}
\title{Appendix: Visual Analysis of Multiple \\ Dynamic Sensitivities along Ascending \\ Trajectories in the Atmosphere}

\author{Christoph~Neuhauser,
        Maicon~Hieronymus,
        Michael~Kern,
        Marc~Rautenhaus,
        Annika~Oertel,
        and~Rüdiger~Westermann
% <-this % stops a space
\IEEEcompsocitemizethanks{\IEEEcompsocthanksitem Christoph Neuhauser and Rüdiger Westermann are with Technical University of Munich (TUM).\protect\\
E-mail: \{christoph.neuhauser\,$|$\,westermann\}@tum.de.\protect\\
\IEEEcompsocthanksitem Maicon Hieronymus is with Johannes Gutenberg University Mainz.\protect\\
E-mail: mhieronymus@uni-mainz.de. \protect\\
\IEEEcompsocthanksitem Michael Kern is with Advanced Micro Devices, Inc.\protect\\
E-mail: Michael.Kern@amd.com. \protect\\
\IEEEcompsocthanksitem Marc Rautenhaus is with Universität Hamburg.\protect\\
E-mail: marc.rautenhaus@uni-hamburg.de. \protect\\
\IEEEcompsocthanksitem Annika Oertel is with Karlsruhe Institute of Technology.\protect\\
E-mail: annika.oertel@kit.edu.}}
\IEEEtitleabstractindextext{%
\begin{abstract}
Numerical weather prediction models rely on parameterizations for subgrid-scale processes, e.g., for cloud microphysics. These parameterizations are a well-known source of uncertainty in weather forecasts that can be quantified via algorithmic differentiation, which computes the sensitivities of prognostic variables to changes in model parameters. It is particularly interesting to use sensitivities to analyze the validity of physical assumptions on which microphysical parameterizations in the numerical model source code are based. In this article, we consider the use case of strongly ascending trajectories, so-called warm conveyor belt trajectories, known to have a significant impact on intense surface precipitation rates in extratropical cyclones. We present visual analytics solutions to analyze interactively the sensitivities of a selected prognostic variable, i.e. rain mass density, to multiple model parameters along such trajectories. We propose a visual interface that enables to a) compare the values of multiple sensitivities at a single time step on multiple trajectories, b) assess the spatio-temporal relationships between sensitivities and the shape and location of trajectories, and c) a comparative analysis of the temporal development of sensitivities along multiple trajectories. We demonstrate how our approach enables atmospheric scientists to interactively analyze the uncertainty in the microphysical parameterizations, and along the trajectories, with respect to a selected prognostic variable. We apply our approach to the analysis of convective trajectories within the extratropical cyclone ``Vladiana'', which occurred between 22-25 September 2016 over the North Atlantic.
\end{abstract}

\begin{IEEEkeywords}
Meteorology, trajectories, temporal data, multi-parameter data, diagrams, linking, focus+context, sensitivity analysis.
\end{IEEEkeywords}}

\maketitle

\IEEEdisplaynontitleabstractindextext

% \fi
\IEEEpeerreviewmaketitle

\appendices
\IEEEraisesectionheading{\section{Tube Rendering}\label{sec:appendix-tube}}

\begin{figure}[t]
 \centering
 \includegraphics[width=\columnwidth]{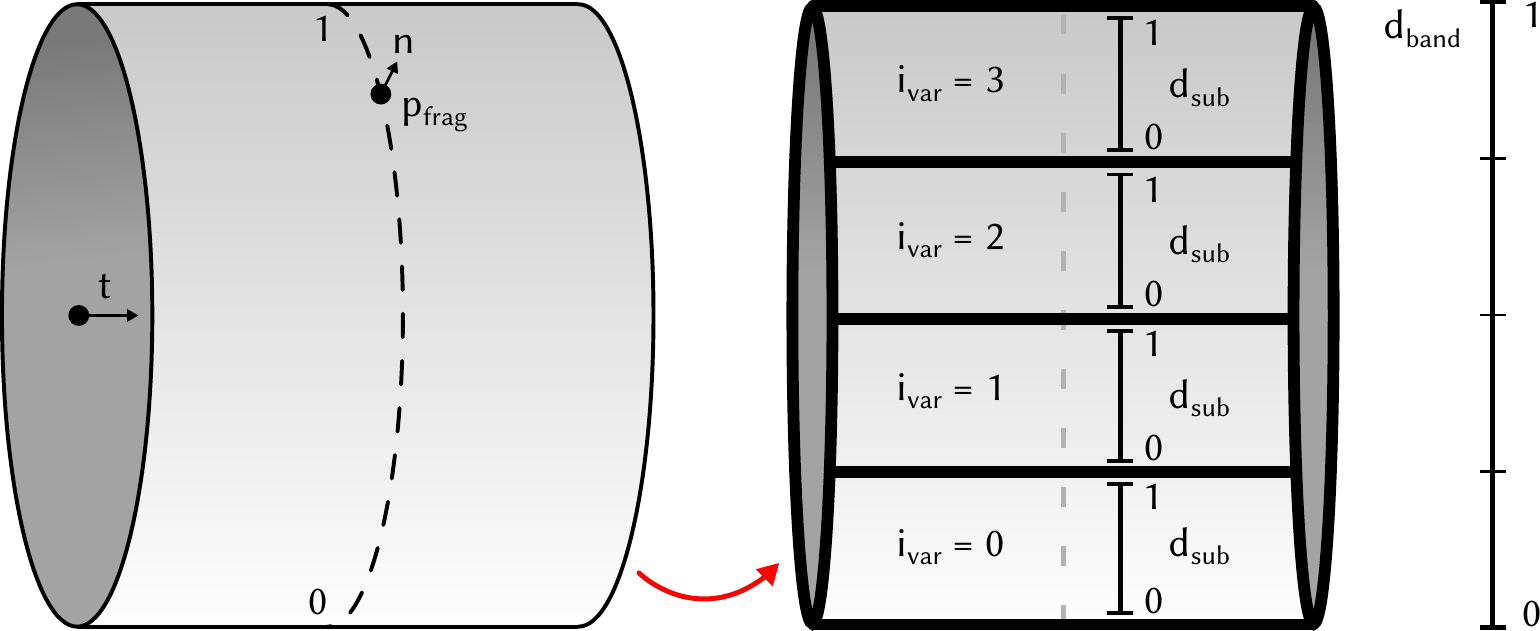}
 \caption{Illustration of local surface properties and subdivision of the visible part of a trajectory to determine a fragments band position $d_{band}$, the sub-band position $d_{sub}$ and its corresponding variable ID $i_{var}$.}
 \label{fig:tube-band-appendix}
\end{figure}

\IEEEPARstart{T}{o} obtain a renderable trajectory representation, the trajectory (i.e. 3D pathlines) are polygonized by extruding them into tubes in a GPU geometry shader. The parameters are mapped onto the surface of the tube as a set of bands running in the direction of the trajectory tangent (cf.~\autoref{fig:tube-band-appendix}). When mapping the bands onto the tube in object space, occlusion effects can occur, as not all parameters may lie in the front, visible part of the tube. Also, due to twist and rotations around the tube, the order in which the bands appear on screen can change and make a comparison between different tubes and the association of parameters to bands more difficult. To avoid this, our rendering technique aligns the bands in view space and keeps their relative order on the screen fix, independent of the viewing direction. For this, a screen space band position $d_{band}$ is computed in the pixel shader on the GPU using only the tangent vector $t$ of the pathline associated with the tube surface fragment, the surface normal $n$ and the view vector $v = \frac{p_{cam} - p_{frag}}{\lVert p_{cam} - p_{frag} \rVert_2}$ pointing from the fragment towards the camera position $p_{cam}$ as inputs.

\begin{figure}[t]
 \centering
 \includegraphics[width=0.45\columnwidth]{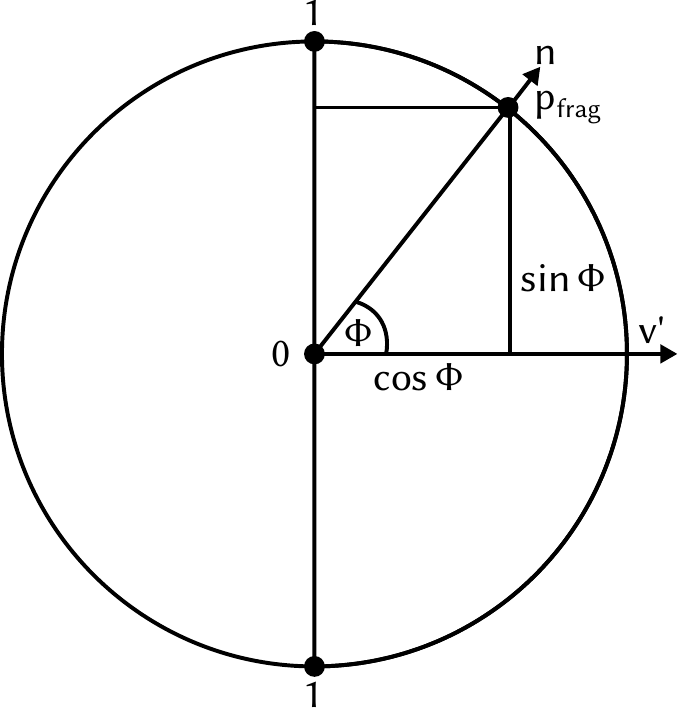}
 \caption{Cross section of the tube with the plane perpendicular to the tangent vector $t$ of the pathline.}
 \label{fig:tube-projection}
\end{figure}

By projecting the camera view direction into the plane orthogonal to the tangent direction of the trajectory, the problem of computing the band position can be reduced to a two-dimensional problem. The projected camera direction $v'$ can be computed by using $v_{aux} = \frac{t \times v}{\lVert t \times v \rVert_2}$ as $v' = \frac{v_{aux} \times t}{\lVert v_{aux} \times t \rVert_2}$. The resulting setting is shown in \autoref{fig:tube-projection}.

Using the angle $\phi = \angle(v', n)$ between the projected view vector $v'$ and the normal vector $n$ would unfortunately not be sufficient as a measure, because it does not change linearly in screen space, thus producing bands of differing width. In order to derive the desired screen space measure, the fragment position needs to be projected onto an imaginary band, illustrated as the vertical line in \autoref{fig:tube-projection}. As can be seen in the figure, the normalized distance of the projected point to the center of the band amounts to the sine of the angle $\phi$. In order to compute the sine, one of the two equalities below can be used:
\begin{equation}
\lvert \sin(\phi) \rvert = \lVert v' \times n \rVert_2 = \sqrt{1 - \langle v', n \rangle ^2}
\end{equation}

These statements hold due to the following mathematical properties of the sine, cosine, cross product and scalar product.

%\begin{equation}
%\begin{gathered}
%\lVert v' \rVert_2 = \lVert n \rVert_2 = 1 \\
%\langle v', n \rangle = \lVert v' \rVert_2 \lVert n \rVert_2 \cos(\phi) \\
%\lVert v' \times n \rVert_2 = \lVert v' \rVert_2 \lVert n \rVert_2 \lvert \sin(\phi) \rvert \\
%\sin^2(\phi) + \cos^2(\phi) = 1
%\end{gathered}
%\end{equation}

\begin{equation}
\begin{gathered}
\lVert v' \rVert_2 = \lVert n \rVert_2 = 1 \\
\langle v', n \rangle = \lVert v' \rVert_2 \lVert n \rVert_2 \cos(\phi) \\
\lVert v' \times n \rVert_2 = \lVert v' \rVert_2 \lVert n \rVert_2 \lvert \sin(\phi) \rvert \\
\sin^2(\phi) + \cos^2(\phi) = 1 \\
\Rightarrow
\lVert v' \times n \rVert_2 = \lvert \sin(\phi) \rvert = \sqrt{1 - \cos^2(\phi)} = \sqrt{1 - \langle v', n \rangle ^2}
\end{gathered}
\end{equation}

As a final step, the resulting distance $\lvert \sin(\phi) \rvert$ needs to be corrected, as the absolute value of the sine doesn't go from 0 to 1 from one end of the imaginary band to the other, but from 1 to 0 in the middle and back to 1 at the other side. In order to correct this problem, we need to compute the sign of the sine by using the winding direction of the angle $\phi$. The sign of the sine can be computed as the sign of the volume of the parallelepiped spanned by t, v' and n.
\begin{equation}
\label{eq:determinant}
vol(t, v', n) = det(t, v', n) = \langle t, v' \times n \rangle
\end{equation}

The equality of the determinant and the combination of the scalar product and cross product can be proven by simple expansion of the respective formulas using the three input vector coordinates as variables. Finally, we can compute the screen space band measure we are looking for as
\begin{equation}
d_{band} = \frac{1}{2} \lvert \sin(\phi) \rvert \cdot sgn(det(t, v', n)) + \frac{1}{2}.
\end{equation}

When mapping $N$ parameters onto the tube, we subdivide the band position $d_{band} \in (0, 1)$ into multiple sub-band positions $d_{sub}$. For this, we compute the variable ID $i_{var} = \lceil d_{band} \cdot N \rceil$ and then finally $d_{sub} = d_{band} \cdot N - i_{var}$ (cf.~\autoref{fig:tube-band-appendix}).

%\newpage
\section{Great Circle-based Sphere Rendering}\label{sec:appendix-sphere-great-circle}

%Additionally to the tubes, focus spheres are rendered at a user selected time step in the geometry view. They serve both as a marker for the current time step and as a magnifying glass for trajectories the viewer is not close to spatially.

The spheres are drawn on the GPU using instanced rendering, where the sphere geometry is specified only once, and instantiated at all points where a sphere should be rendered. The pixel shader is then equipped with the following input variables.

\begin{itemize}
    \item The fragment position $p_{frag}$.
    \item The surface normal $n$.
    \item The camera position $p_{cam}$.
    \item The trajectory entrance and exit points $p_a$ and $p_b$ computed as the first and last intersection of the trajectory with the sphere and their corresponding time steps $t_a$ and $t_b$.
\end{itemize}

With this data, the view vector $v = \frac{p_{cam} - p_{frag}}{\lVert p_{cam} - p_{frag} \rVert_2}$ and the trajectory direction vector $l = \frac{p_{b} - p_{a}}{\lVert p_{b} - p_{a} \rVert_2}$ pointing from the entry point $p_a$ into the direction of the exit point $p_b$ can also be computed in the shader.

\subsection{Band Mapping on the Sphere}

\begin{figure}[t]
 \centering
 \includegraphics[width=0.8\columnwidth]{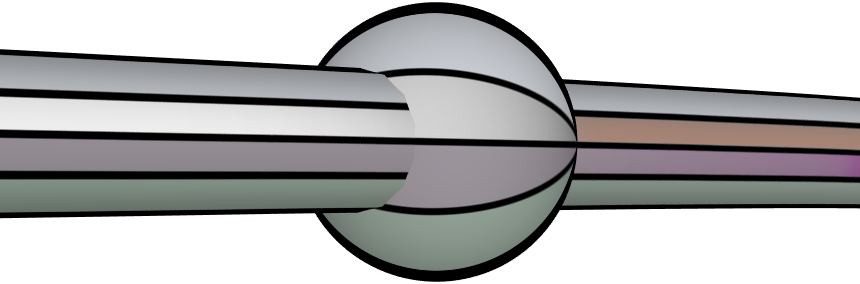}
 \includegraphics[width=0.8\columnwidth]{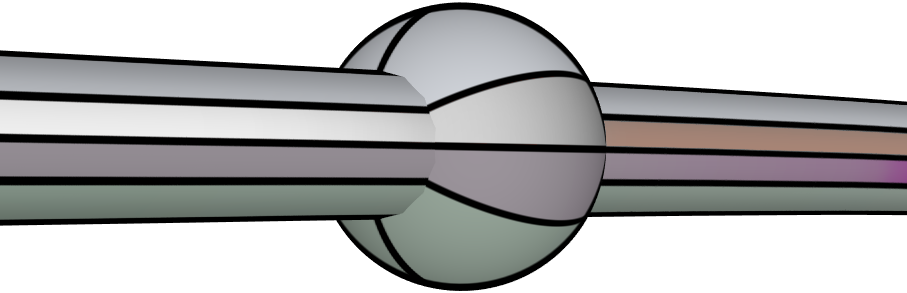}
 \includegraphics[width=0.8\columnwidth]{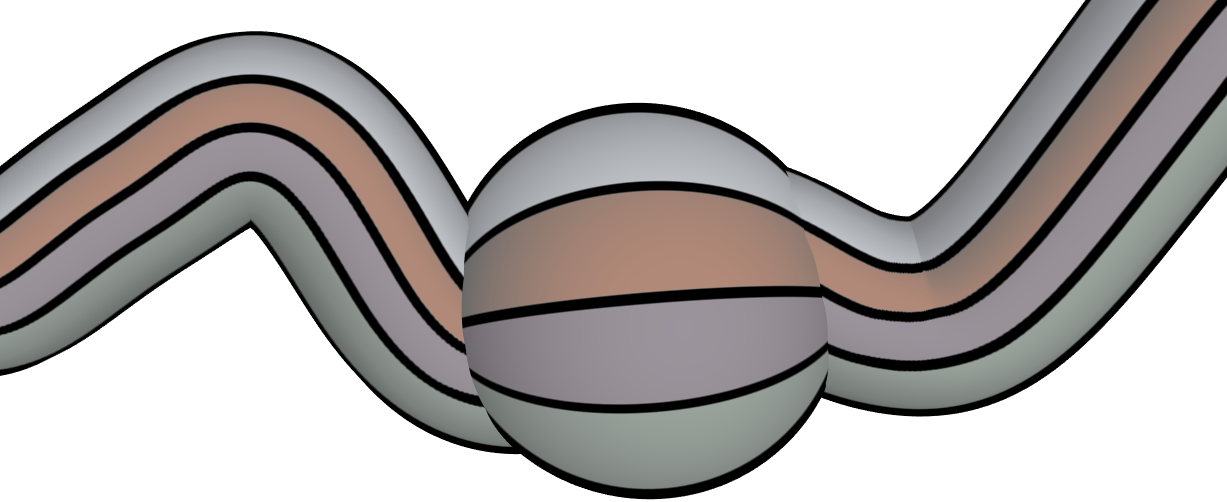}
 \caption{Top: The rejected method using the pure screen space location on the circular projection of the sphere. Middle: Illustration of the smooth transition between tube bands and sphere bands when $p_a$, $p_b$ and $p_{sphere}$ are collinear. Bottom: Illustration of the hard transition between tube bands and sphere bands when $p_a$, $p_b$ and $p_{sphere}$ are not collinear. This effect can be observed at turns in the trajectory data.}
 \label{fig:sphere-transition}
\end{figure}

In order to compute which parameter band a fragment on the sphere belongs to, analogously to the tube, we propose to use the technique presented below. It solves different shortcomings compared to simply using the screen space location on the circular projection of the sphere (cf.~\autoref{fig:sphere-transition}~top compared to middle). If the entry and exit points are collinear with the center of the sphere, then a smooth transition between the color bands on the tube and the color bands on the sphere will be guaranteed independently of the viewing angle.

First, the projection of the sphere center $p_{sphere}$ onto the surface of the sphere under the used camera transformation needs to be computed. For this, a ray-sphere intersection algorithm \cite{RaySphereIntersection} is used with a ray starting at the camera position with direction $p_{sphere} - p_{cam} $, which yields the point $p_{int}$.

Next, the normal vectors of the two disks $P_0$ spanned by the points $(p_{a}, p_{b}, p_{int})$ and $P_{frag}$ spanned by the points $(p_{a}, p_{b}, p_{frag})$ are computed. The intersections of the disks with the sphere can be interpreted as two great circles meeting under a certain angle $\theta$.

Similarly to the tube surface, the value $\lvert sin(\theta) \rvert = \lVert n_0 \times n_{frag} \rVert_2$ computed using the normal vectors of the two disks can be used as a measure in screen space for the distance to the border of the sphere. Finally, the band position is analogously given as
\begin{equation}
\label{eq:dband-great-circle}
d_{band} = \frac{1}{2} \lvert \sin(\theta) \rvert \cdot sgn(det(l, n_0, n_{frag})) + \frac{1}{2}.
\end{equation}

\begin{figure}[tb]
 \centering
 \includegraphics[width=\columnwidth]{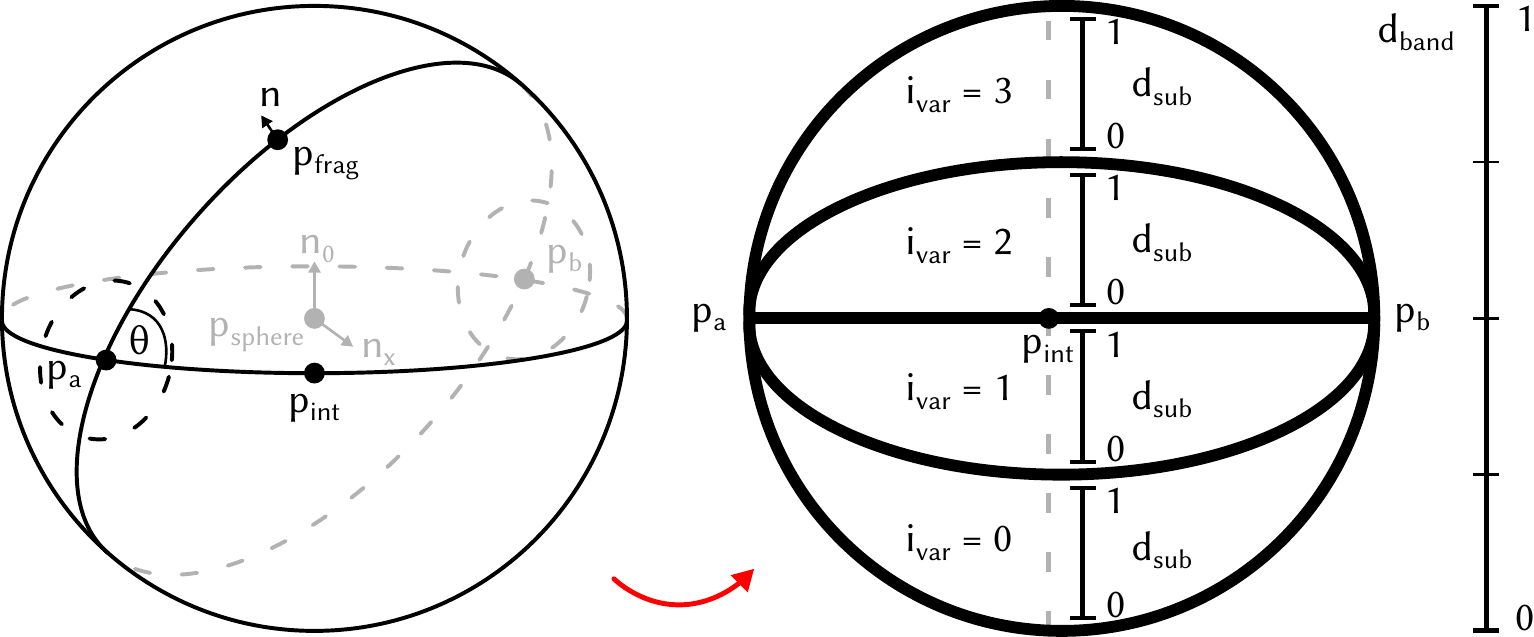}
 \caption{Illustration of how the input vectors and points on the sphere are used to compute the band position $d_{band}$, the sub-band position $d_{sub}$ and its corresponding variable ID $i_{var}$.}
 \label{fig:greatcircles}
\end{figure}

When mapping $N$ parameters onto the sphere, we again subdivide the band position $d_{band} \in (0, 1)$ into multiple sub-band positions $d_{sub}$ (cf.~\autoref{fig:greatcircles}).

A black separator line is drawn between two neighboring sub-bands. A problem that arises is that changes in the sub-band position and the direction orthogonal to it are no longer linear in screen space, especially when the sphere is viewed under an oblique angle, and thus distortions can occur. In the three subsections below, this problem and its solution are outlined. Using the correction factors $f_i$ introduced below, the final separator thickness can be computed as
\begin{equation}
w_{sep}' = \frac{w_{sep}}{f_1 f_2 f_3}.
\end{equation}

\subsubsection{Lateral Band Thickness}\label{sec:lateral-thickness}

In \autoref{eq:dband-great-circle}, the sine of the angle $\theta$ is used. However, if we look at the sphere from a direction that is not perpendicular to the vector connecting the entry and exit points, the problem arises that the sine of the angle of the two great circles is no longer a linear measure in screen space. The closer we get to the edge of the sphere, the higher the distortion. If we were to use the angle $\theta$ directly instead of its sine, then the view dependent distortion disappears. Thus, the thickness is corrected using the derivative of the arcsine $\frac{\partial \arcsin{x}}{\partial x} = \frac{1}{\sqrt{1 - x^2}}$ with x = $\lVert n_0 \times n_{frag} \rVert_2$, i.e.,
\begin{equation}
f_1 = \frac{1}{\sqrt{1 - \lVert n_0 \times n_{frag} \rVert_2^2}}.
\end{equation}

\subsubsection{Longitudinal Band Thickness}\label{sec:longitudinal-thickness}

Another problem is that the separator line gets thinner the closer it gets to the entrance and exit points (cf.~\autoref{fig:correction-factors} middle left). For the derivation of the formula below, it is assumed that the entrance and exit points are poles of the sphere, which is equivalent to assuming they are collinear with the sphere center. In all other cases, the distortion on the sphere surface will increase the sharper the angle $\angle(p_{a}, p_{sphere}, p_{b})$ between center and entry and exit point is. However, as sharp $180^{\circ}$ turns in trajectory lines are very rare, this distortion will usually only be minor and won't impede the visualization strongly under normal circumstances.

\begin{figure}[tb]
 \centering
 \includegraphics[width=\columnwidth]{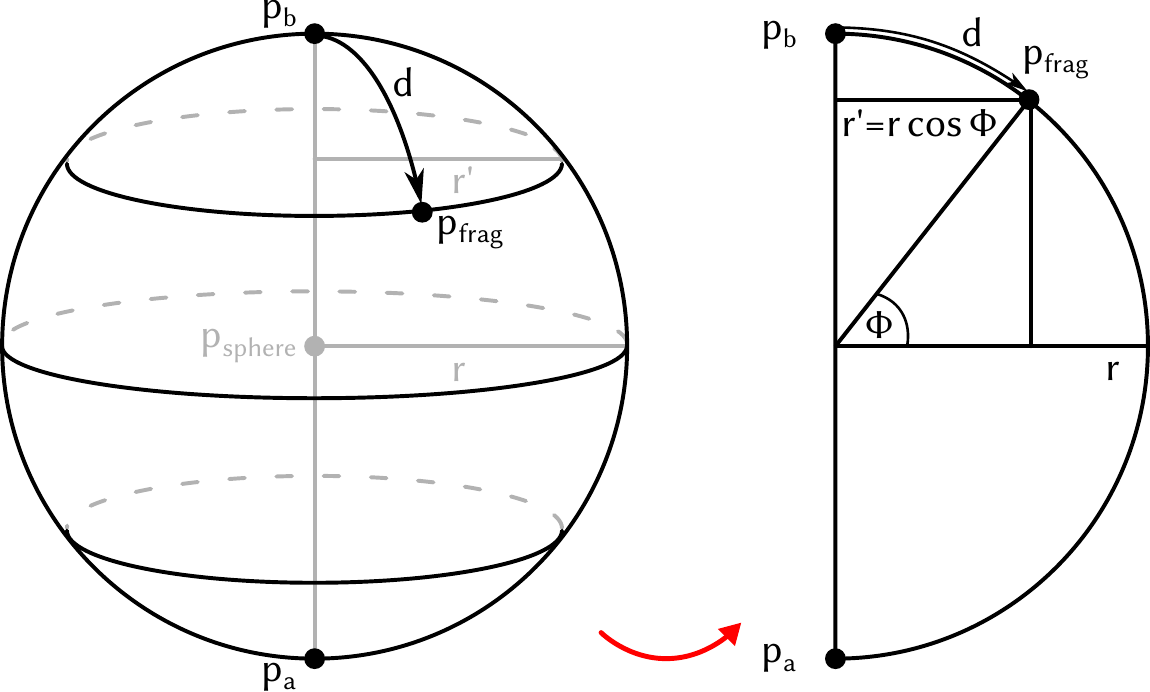}
 \caption{Illustration of how the input points and vectors are used to compute the band position $d_{band}$, the sub-band position $d_{sub}$ and its corresponding variable ID $i_{var}$.}
 \label{fig:longitudinal-correction}
\end{figure}

For computing the correction factor $f_2$, the angular distance of the point $p_{frag}$ from the point $p_x$ in radians $d_x = arccos(\langle p_x - p_{sphere}, p_{frag} - p_{sphere} \rangle)$ becomes necessary (cf.~\autoref{fig:longitudinal-correction}). We compute the angular distance $d = \min(d_a, d_b)$ from one of the poles, and correspondingly the angular distance in radians $\phi = \frac{\pi}{2} - d$ from the equator. The lateral radius of the sphere at $p_{frag}$ is given as $r \cdot \cos{\phi}$, and consequentially, our final longitudinal band thickness correction factor amounts to
\begin{equation}
f_2 = \frac{r \cdot \cos{\phi}}{r} = \cos{\phi}.
\end{equation}

\subsubsection{Sphere Band Thickness}\label{sec:sphere-thickness}

\begin{figure}[t]
 \centering
 \includegraphics[angle=90,origin=c,width=0.24\columnwidth]{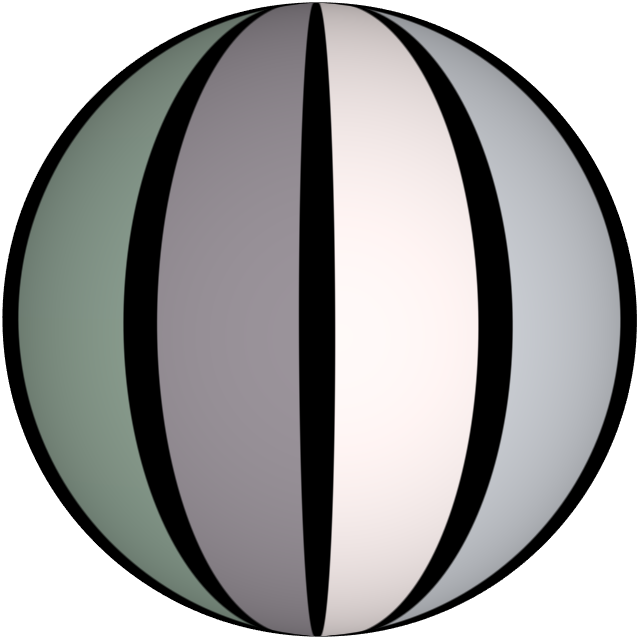}
 \includegraphics[angle=90,origin=c,width=0.24\columnwidth]{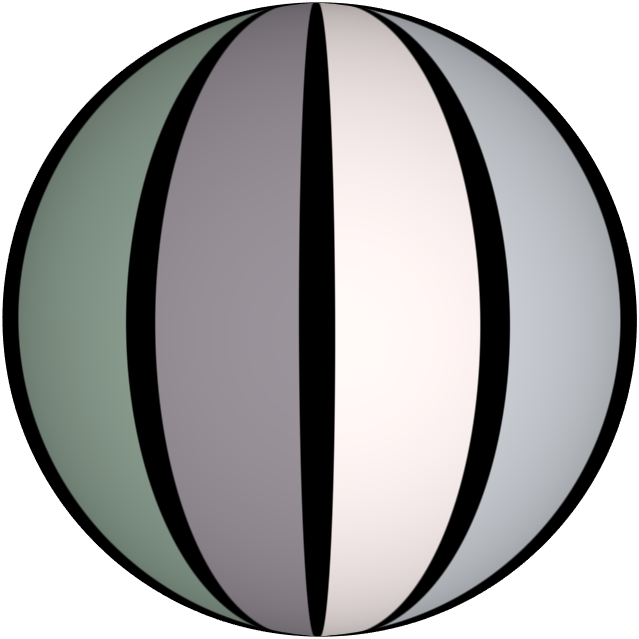}
 \includegraphics[angle=90,origin=c,width=0.24\columnwidth]{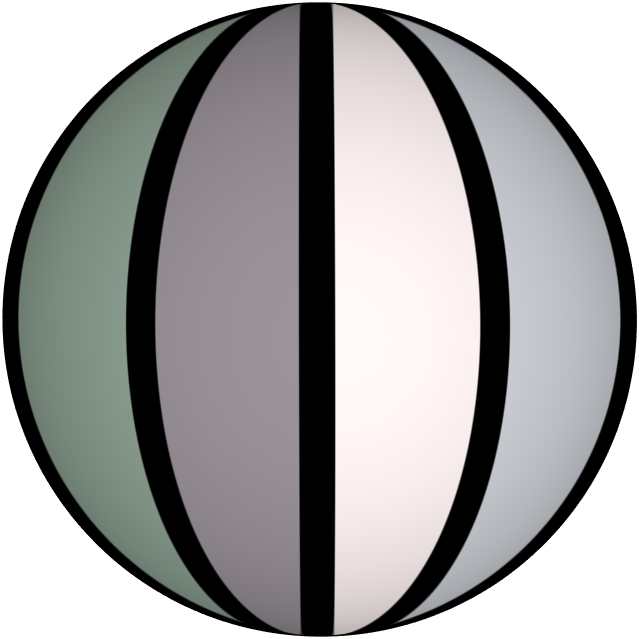}
 \includegraphics[angle=90,origin=c,width=0.24\columnwidth]{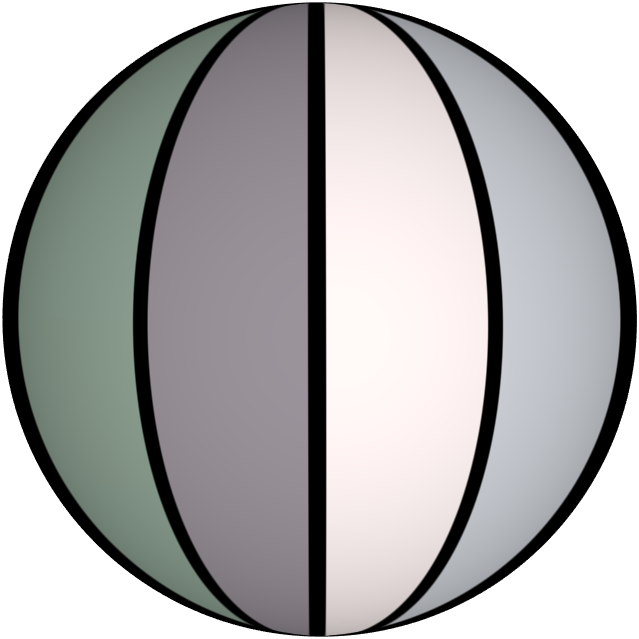}
 \caption{From left to right: The effect of no correction factor and the three factors presented in \autoref{sec:lateral-thickness}, \autoref{sec:longitudinal-thickness} and \autoref{sec:sphere-thickness}. Please note that the difference of the two leftmost spheres is hardly visible, as this type of distortion only appears when the sphere is viewed from an oblique angle not perpendicular to $l$.}
 \label{fig:correction-factors}
\end{figure}

In order to achieve equal sub-band separator thickness of a sphere with its corresponding tube, a final correction factor is used. For example, if the diameter of the sphere $2r$ is twice the width of the tube $w_{tube}$, then the separator width in sub-band space will be halved using the factor below (cf.~\autoref{fig:correction-factors} right).
\begin{equation}
f_3 = \frac{2r}{w_{tube}}
\end{equation}

\subsection{Time Dimension Mapping on the Sphere}

In order to map different time steps onto the sphere, the point closest to $p_{frag}$ on the line segment $(p_{a}, p_{b})$ connecting the entry and exit point is computed. The distance of this point to $p_{a}$ and $p_{b}$ is then used to linearly interpolate between the time steps $t_a$ and $t_b$ at the entry and exit point to get the time step $t_{frag}$ used for the fragment on the sphere surface. Then, $t_{frag}$ is used to look up the parameter values of the trajectory at time step $t_{frag}$ from a shader storage buffer stored on the GPU.

%\begin{figure}[tb]
% \centering
% \includegraphics[width=\columnwidth]{sphere/TangentTransition}
% \includegraphics[width=\columnwidth]{sphere/SmoothTransition}
% \includegraphics[width=\columnwidth]{sphere/HardTransition}
% \caption{Top: The rejected method using the pure screen space location on the circular projection of the sphere. Middle: Illustration of the smooth %transition between tube bands and sphere bands when $p_a$, $p_b$ and $p_{sphere}$ are collinear. Bottom: Illustration of the hard transition between %tube bands and sphere bands when $p_a$, $p_b$ and $p_{sphere}$ are not collinear. This effect can be observed at turns in the trajectory data.}
% \label{fig:sphere-transition}
%\end{figure}

\section{Pie Chart-based Sphere Rendering}\label{sec:appendix-sphere-pie chart}

\begin{figure}[t]
 \centering
 \includegraphics[width=0.5\columnwidth]{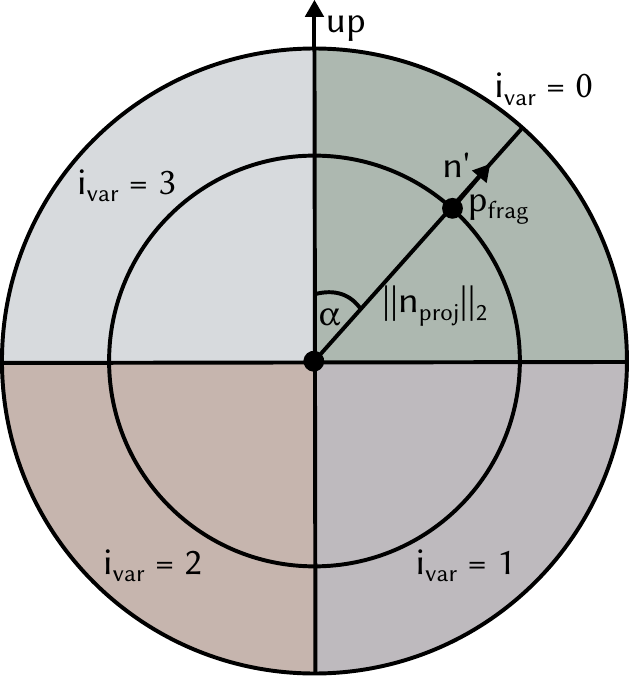}
 \caption{Illustration of how the input vectors and points on the sphere are used to compute the band position $d_{band}$, the sub-band position $d_{sub}$ and its corresponding variable ID $i_{var}$.}
 \label{fig:piecharts}
\end{figure}

For the rendering of a sphere colored via a pie charts, we want to subdivide the screen projection of the sphere in angular bands, i.e., individual pie slices (cf.~\autoref{fig:piecharts}). For this, we want to compute the angle $\alpha_{band}$, which represents the angular distance of the fragment $p_{frag}$ to the up-axis of the camera. As input, we need the surface normal vector $n$, the camera view direction $v$ and the camera up-vector $up$. As a first step, the normal $n$ is projected into the view plane to obtain
\begin{equation}
n_{proj} = n - \langle n, v \rangle \cdot n.
\end{equation}

Then, we set $n' = \frac{n_{proj}}{\lVert n_{proj} \rVert_2}$. The length $\lVert n_{proj} \rVert_2$ is the normalized screen space distance to the center of the sphere. This can be easily checked for the special case $v = (0, 0, 1)^T$, where $\lVert n_{proj} \rVert_2$ becomes $\sqrt{n_x^2 + n_y^2} \in [0, 1)$. We will use this fact later in \autoref{eq:f5}. In the next step, we compute the angle $\alpha_{band}$ as follows:

%\begin{equation}
%\alpha_{band} = \frac{(atan2(det(n', up, v), \langle n, up \rangle) + \frac{\pi}{2}) \mathrm{\ mod\ } 2\pi}{2\pi}
%\end{equation}

\begin{equation}
\label{eq:pie-chart}
\alpha_{band} = atan2(det(n', up, v), \langle n', up \rangle) + \frac{\pi}{2}.
\end{equation}

$atan2(y, x)$ computes the angle between the positive x axis and the line connecting the origin and the point $(x, y)^T$. atan2 returns the angle in mathematically positive direction, i.e., a counterclockwise angle. However, in our case, we do not want the counterclockwise angle to the positive x axis, but the clockwise angle from the positive y axis (the positive y axis being the up vector of the camera). This can be most easily achieved by transposing (i.e., interchanging) the x and y coordinates we feed to atan2. To get the y coordinate of the point we use for calculating the angle, the term $\langle n', up \rangle$ is used in \autoref{eq:pie-chart}. This way, we project the view plane normal onto the up axis vector. For the x coordinate, $det(n', up, v)$ is used. We can again use \autoref{eq:determinant} to get the equality $det(n', up, v) = \langle n', up \times v \rangle$. Here, $up \times v$ can be interpreted as the right axis vector of the view plane. When we project the view plane normal onto this new right axis vector, we get the x coordinate for \autoref{eq:pie-chart}. The pie chart in the view plane can be seen in \autoref{fig:piecharts}.

Finally, we can compute the global band position $d_{band}$ as
\begin{equation}
d_{band} =\frac{\alpha_{band} \mathrm{\ mod\ } 2\pi}{2\pi}.
\end{equation}

When mapping $N$ parameters onto the sphere, we again subdivide the band position $d_{band} \in [0, 1)$ into multiple sub-band positions $d_{sub}$ (cf.~\autoref{fig:piecharts}).

A black separator line is drawn between two neighboring sub-bands. A problem that also arises for the pie chart-based spheres is that changes in the sub-band position are not linear in screen space and dependent on the distance to the screen space center of the sphere. Consequently, two more correction factors are introduced below, and the final separator thickness is computed as
\begin{equation}
w_{sep}' = \frac{w_{sep}}{f_4 f_5}.
\end{equation}

The factor $f_4$ is equal to $\lVert n_{proj} \rVert_2$, which itself, as was shown earlier in this section, is equal to the normalized distance to the screen space center of the sphere. This way, it is guaranteed that the separator thickness doesn't get thinner the closer we get to the center of the pie chart.
\begin{equation}
\label{eq:f5}
f_4 = \lVert n_{proj} \rVert_2
\end{equation}

Finally, the factor $f_5$ is used to make sure that the separator thickness of the pie chart sphere and the trajectory tube is equal. This time, we use the factor $2r\pi$ in the numerator as opposed to $2r$ for the sphere bands, as our measure should no longer be relative to the diameter of the screen space circle/sphere, but its circumference.
\begin{equation}
f_5 = \frac{2r\pi}{w_{tube}}
\end{equation}

%\begin{equation}
%w'_{sep} = \frac{w_{sep} w_{tube}}{2r \pi \lVert n_{proj} \rVert_2}
%\end{equation}

\clearpage

\section{Variable and Parameter Names}\label{sec:appendix-variable-names}

\begin{table}[ht]
  \caption{Variable Names in the Dataset}
  \label{tab:variable_names}
  \centering
  \begin{tabu}{@{}l|X@{}}
  \toprule
  \textbf{Variable} & \textbf{Description}  \\ 
  \midrule
    pressure & Pressure in hPa \\ \hline 
    T & Temperature in Kelvin \\ \hline 
    w & Vertical velocity in $\text{m}\,\text{s}^{-1}$ \\ \hline 
    S & Saturation \\ \hline 
    QV & Water vapor mass density in $\text{kg}\,\text{m}^{-3}$ \\ \hline 
    QC & Cloud mass density in $\text{kg}\,\text{m}^{-3}$ \\ \hline 
    QR & Rain mass density in $\text{kg}\,\text{m}^{-3}$ \\ \hline 
    QS & Snow mass density in $\text{kg}\,\text{m}^{-3}$ \\ \hline 
    QI & Ice mass density in $\text{kg}\,\text{m}^{-3}$ \\ \hline 
    QG & Graupel mass density in $\text{kg}\,\text{m}^{-3}$ \\ \hline 
    QH & Hail mass density in $\text{kg}\,\text{m}^{-3}$ \\ \hline 
    NCCLOUD & Cloud number density in $\text{m}^{-3}$ \\ \hline 
    NCRAIN & Rain number density in $\text{m}^{-3}$ \\ \hline 
    NCSNOW & Snow number density in $\text{m}^{-3}$ \\ \hline 
    NCICE & Ice number density in $\text{m}^{-3}$ \\ \hline 
    NCGRAUPEL & Graupel number density in $\text{m}^{-3}$ \\ \hline 
    NCHAIL & Hail number density in $\text{m}^{-3}$ \\ \hline 
    QR\_OUT & Sedimentation of rain mass density out of the air parcel in $\text{kg}\,\text{m}^{-3}$ \\ \hline 
    QS\_OUT & Sedimentation of snow mass density out of the air parcel in $\text{kg}\,\text{m}^{-3}$ \\ \hline 
    QI\_OUT & Sedimentation of ice mass density out of the air parcel in $\text{kg}\,\text{m}^{-3}$ \\ \hline 
    QG\_OUT & Sedimentation of graupel mass density out of the air parcel in $\text{kg}\,\text{m}^{-3}$ \\ \hline 
    QH\_OUT & Sedimentation of hail mass density out of the air parcel in $\text{kg}\,\text{m}^{-3}$ \\ \hline 
    NR\_OUT & Sedimentation of rain number density out of the air parcel in $\text{m}^{-3}$ \\ \hline 
    NS\_OUT & Sedimentation of snow number density out of the air parcel in $\text{m}^{-3}$ \\ \hline 
    NI\_OUT & Sedimentation of ice number density out of the air parcel in $\text{m}^{-3}$ \\ \hline 
    NG\_OUT & Sedimentation of graupel number density out of the air parcel in $\text{m}^{-3}$ \\ \hline 
    NH\_OUT & Sedimentation of hail number density out of the air parcel in $\text{m}^{-3}$ \\ \hline 
    latent\_heat & Latent heat released by cloud microphysical processes in $\text{J}\,\text{kg}^{-1}$ \\ \hline 
    latent\_cool & Latent heat absorbed by cloud microphysical processes in $\text{J}\,\text{kg}^{-1}$ \\ \hline 
    z & Height in m \\ \hline 
    Inactive & Number of nuclei that can not be activated for ice, snow, graupel or hail \\ \hline 
    deposition & Mass density of water vapor deposited in ice, snow, graupel and hail \\ \hline 
    sublimination & Mass density of water vapor from ice, snow, graupel and hail \\ \hline 
    time\_after\_ascent & Time centered to the start of the fastest ascent in a 2\,h time window \\ \hline 
    conv\_400 & Flag for a convective ascent of 400\,hPa \\ \hline 
    conv\_600 & Flag for a convective ascent of 600\,hPa \\ \hline 
    slan\_400 & Flag for a slantwise ascent of 400\,hPa \\ \hline 
    slan\_600 & Flag for a slantwise ascent of 600\,hPa \\ \hline 
    step & Simulation step \\ \hline 
    phase & Flag for different phases of the trajectory. 0: warm phase, 1: mixed phase, 2: ice phase, 3: neutral phase \\ \hline 
    %\vspace{-3.5cm}
    \end{tabu}
\end{table}

\begin{table}[t]
  \vspace{-1.5cm}
  \caption{Parameter Names in the Dataset}
  \label{tab:parameter_names}
  \centering
  \begin{tabu}{@{}l|X@{}}
  \toprule
  \textbf{Parameter} & \textbf{Description}  \\ 
  \midrule
    inv\_z & Inverse of air parcel size (height) used in explicit sedimentation) (cf. \cite{hieronymus_algorithmic_2022}) \\ \hline 
    rho\_vel & Exponent for density correction (cf. \cite{seifert_two-moment_2006}, Eq. (33)) \\ \hline 
    D\_rainfrz\_gh &  Size threshold for partitioning of freezing rain in the hail scheme (cf. \cite{seifert_two-moment_2006}) \\ \hline 
    p\_sat\_melt & Saturation pressure at $ \text{T}=273.15\,\text{K} $ (cf. \cite{seifert_two-moment_2006}) \\ \hline 
    a\_HET & Exponent for heterogeneous rain freezing with data of Barklie and Gokhale (cf. \cite{seifert_two-moment_2006}) \\ \hline 
    k\_r & Coefficient for accretion of QC to QR (cf. \cite{seifert_two-moment_2006}) \\ \hline 
    a\_ccn\_1 & Parameter for CCN concentration (cf. \cite{hande_parameterizing_2016}) \\ \hline 
    a\_ccn\_4 & Parameter for CCN concentration (cf. \cite{hande_parameterizing_2016}) \\ \hline 
    b\_ccn\_1 & Parameter for CCN concentration (cf. \cite{hande_parameterizing_2016}) \\ \hline 
    b\_ccn\_3 & Parameter for CCN concentration (cf. \cite{hande_parameterizing_2016}) \\ \hline 
    b\_ccn\_4 & Parameter for CCN concentration (cf. \cite{hande_parameterizing_2016}) \\ \hline 
    c\_ccn\_1 & Parameter for CCN concentration (cf. \cite{hande_parameterizing_2016}) \\ \hline 
    c\_ccn\_3 & Parameter for CCN concentration (cf. \cite{hande_parameterizing_2016}) \\ \hline 
    c\_ccn\_4 & Parameter for CCN concentration (cf. \cite{hande_parameterizing_2016}) \\ \hline 
    d\_ccn\_1 & Parameter for CCN concentration (cf. \cite{hande_parameterizing_2016}) \\ \hline 
    d\_ccn\_2 & Parameter for CCN concentration (cf. \cite{hande_parameterizing_2016}) \\ \hline 
    d\_ccn\_3 & Parameter for CCN concentration (cf. \cite{hande_parameterizing_2016}) \\ \hline 
    d\_ccn\_4 & Parameter for CCN concentration (cf. \cite{hande_parameterizing_2016}) \\ \hline 
    rain\_a\_geo & Coefficient for diameter size calculation (cf. \cite{seifert_two-moment_2006} Eq. (32)) \\ \hline 
    rain\_b\_geo & Exponent for diameter size calculation (cf. \cite{seifert_two-moment_2006} Eq. (32)) \\ \hline 
    rain\_min\_x & Minimum size of the particle used after the microphysics (cf. \cite{seifert_two-moment_2006}, Eqs. (94), (97)) \\ \hline 
    rain\_a\_vel & Coefficient for particle velocity (cf. \cite{seifert_two-moment_2006} Eq. (33)) \\ \hline 
    rain\_b\_vel & Exponent for particle velocity (cf. \cite{seifert_two-moment_2006} Eq. (33)) \\ \hline 
    rain\_alpha & Constant in rain sedimentation (cf. \cite{seifert_parameterization_2008}, Eq. (A10)) \\ \hline 
    rain\_beta & Coefficient for rain sedimentation (cf. \cite{seifert_parameterization_2008}, Eq. (A10)) \\ \hline 
    rain\_gamma & Exponent for rain sedimentation (cf. \cite{seifert_parameterization_2008}, Eq. (A10)) \\ \hline 
    rain\_nu & Parameter to calculate the shape of the generalized $ \Gamma $-distribution (cf. \cite{seifert_two-moment_2006}, Eq. (79)) \\ \hline 
    rain\_mu & Shape parameter of the generalized $ \Gamma $-distribution (cf. \cite{seifert_two-moment_2006}, Eq. (79)) \\ \hline 
    graupel\_a\_geo & Coefficient for diameter size calculation (cf. \cite{seifert_two-moment_2006} Eq. (32)) \\ \hline 
    graupel\_b\_geo & Exponent for diameter size calculation (cf. \cite{seifert_two-moment_2006} Eq. (32)) \\ \hline 
    graupel\_a\_vel & Coefficient for particle velocity (cf. \cite{seifert_two-moment_2006} Eq. (33)) \\ \hline 
    graupel\_b\_vel & Exponent for particle velocity (cf. \cite{seifert_two-moment_2006} Eq. (33)) \\ \hline 
    graupel\_vsedi\_max & Maximum sedimentation velocity parameter (cf. \cite{hieronymus_algorithmic_2022}) \\ \hline 
    ice\_a\_geo & Coefficient for diameter size calculation (cf. \cite{seifert_two-moment_2006} Eq. (32)) \\ \hline 
    ice\_b\_geo & Exponent for diameter size calculation (cf. \cite{seifert_two-moment_2006} Eq. (32)) \\ \hline 
    ice\_b\_vel & Exponent for particle velocity (cf. \cite{seifert_two-moment_2006} Eq. (33)) \\ \hline 
    ice\_vsedi\_max & Maximum sedimentation velocity parameter (cf. \cite{hieronymus_algorithmic_2022}) \\ \hline 
    snow\_b\_geo & Exponent for diameter size calculation (cf. \cite{seifert_two-moment_2006} Eq. (32)) \\ \hline 
    snow\_b\_vel & Exponent for particle velocity (cf. \cite{seifert_two-moment_2006} Eq. (33)) \\ \hline 
    snow\_vsedi\_max & Maximum sedimentation velocity parameter (cf. \cite{hieronymus_algorithmic_2022}) \\ \hline 
  \end{tabu}
\end{table}

\clearpage

\begin{figure*}[!b]
 \centering
 \includegraphics[width=\textwidth]{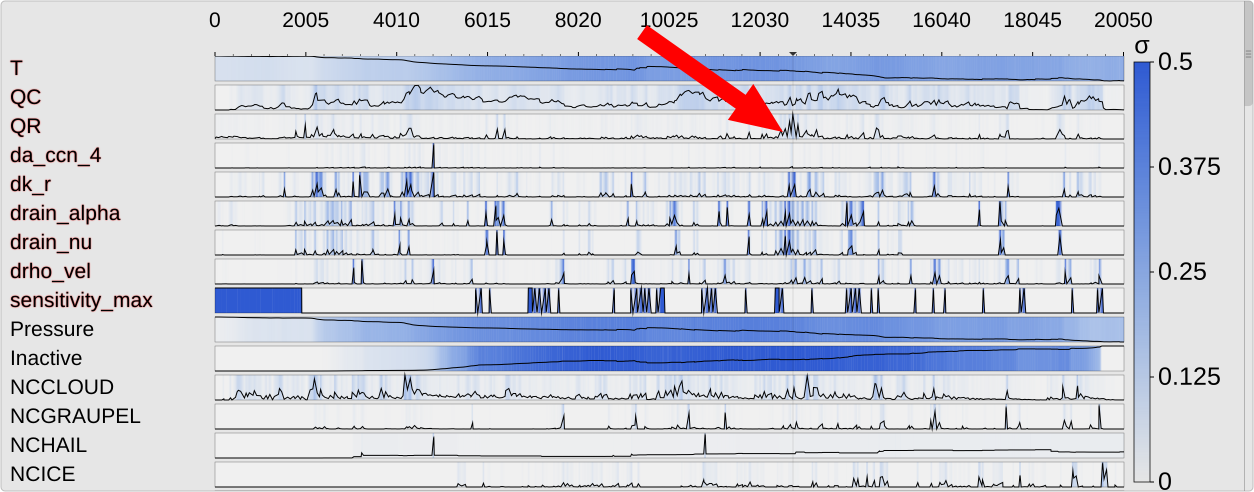}
 \caption{Concentrating on rain mass density (QR), we use the curve plot to look for time steps with extreme values worth further investigation. Such a selection can be made with multiple variables or sensitivities in mind.}
 \label{fig:regions-step-1}
\end{figure*}

\section{Regions and Times of Interest Selection for Case-Study ``Vladiana''}\label{sec:appendix-roi-toi}

%\christoph{Text proposal A: In Section 6 of our work, we demonstrate the value of our method by discussing first investigations of the sensitivity of the rain mass density (QR) simulated by the numerical model to microphysical parameters along WCB trajectories within ``Vladiana'' concerning the amount of rain mass density simulated by the numerical model. We are particularly interested in whether there are differences in the sensitivities along the ascent based on the location and the ascent rate. For this analysis, our domain scientists first employed the exploratory visual analysis capabilities of the proposed method to discover regions and times of interest. The different steps in this process are described below.}

%\christoph{Text proposal B: Section 6 of our work gives an example for investigating rain mass density along WCB trajectories within ``Vladiana'' focusing on the joint development of multiple sensitivities. In addition, we demonstrate in this section how our domain scientists identify regions and times of interest using our exploratory visual analysis capabilities. The different steps in this process are described below.}
Section 6 of our work gives an example for investigating the rain mass density (QR) along WCB trajectories within ``Vladiana'', focusing on the joint development of multiple model parameter sensitivities. In this section, we demonstrate how our domain scientists identify regions and times of interest using our exploratory visual analysis capabilities. The different steps in this process are described below.

\begin{figure*}[ht]
 \centering
 \includegraphics[width=\textwidth]{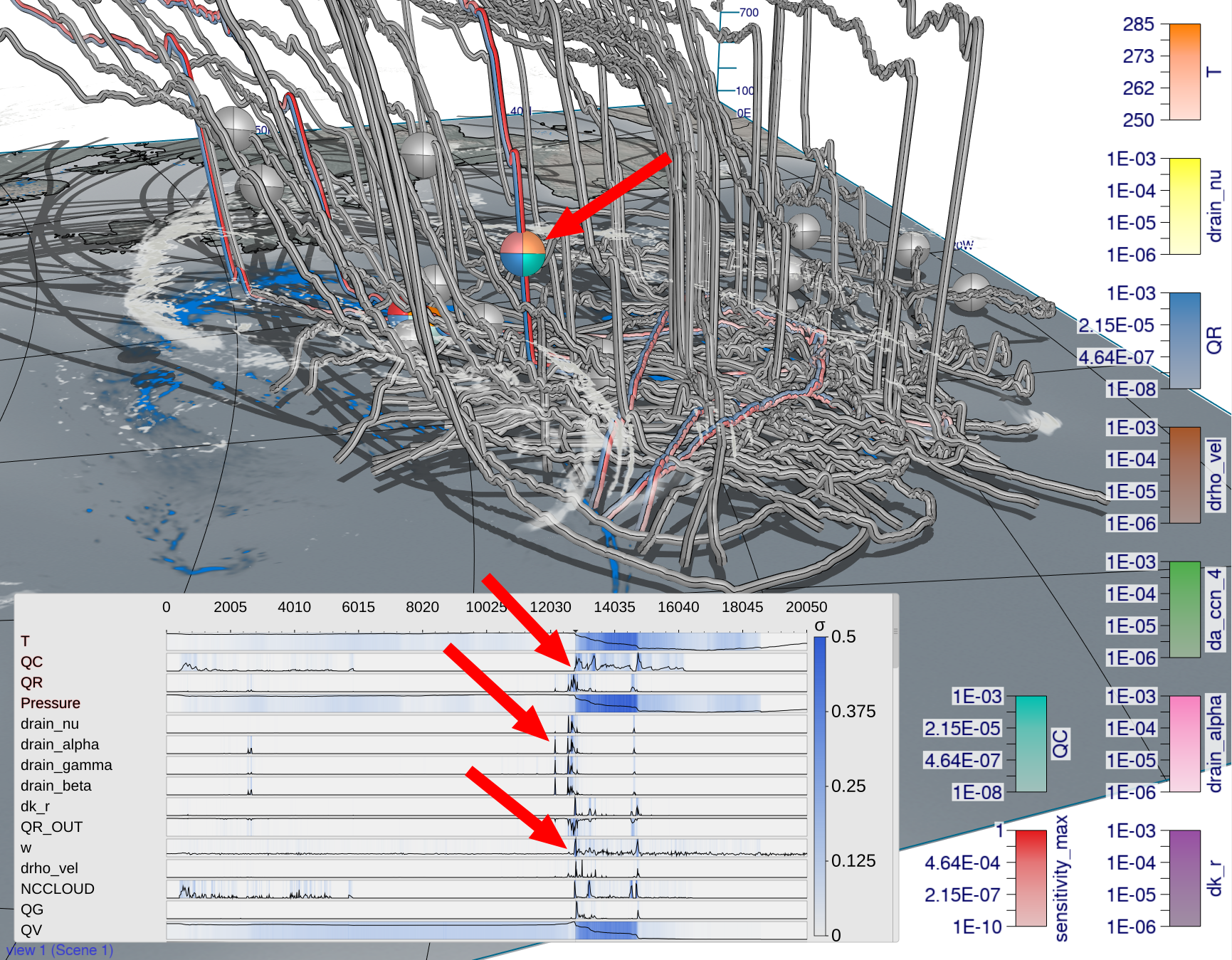}
 \caption{Using the spheres, we select the trajectories that contribute to the peak previously identified. The blue segment of the sphere corresponds to high rain mass density.
 Once all trajectories in question are selected, the curve plots are sorted by similarity to rain mass density. Comparing the location of the peaks in rain mass density with peaks in sensitivities reveals that activation of precipitation processes coincides with increased rain mass. In contrast, cloud formation indicated by QC starts afterwards and coincides with rising levels of the trajectories given by pressure.}
 \label{fig:regions-step-2}
 \afterpage{\clearpage}
\end{figure*}

\begin{figure*}[ht]
 \centering
 \includegraphics[width=\textwidth]{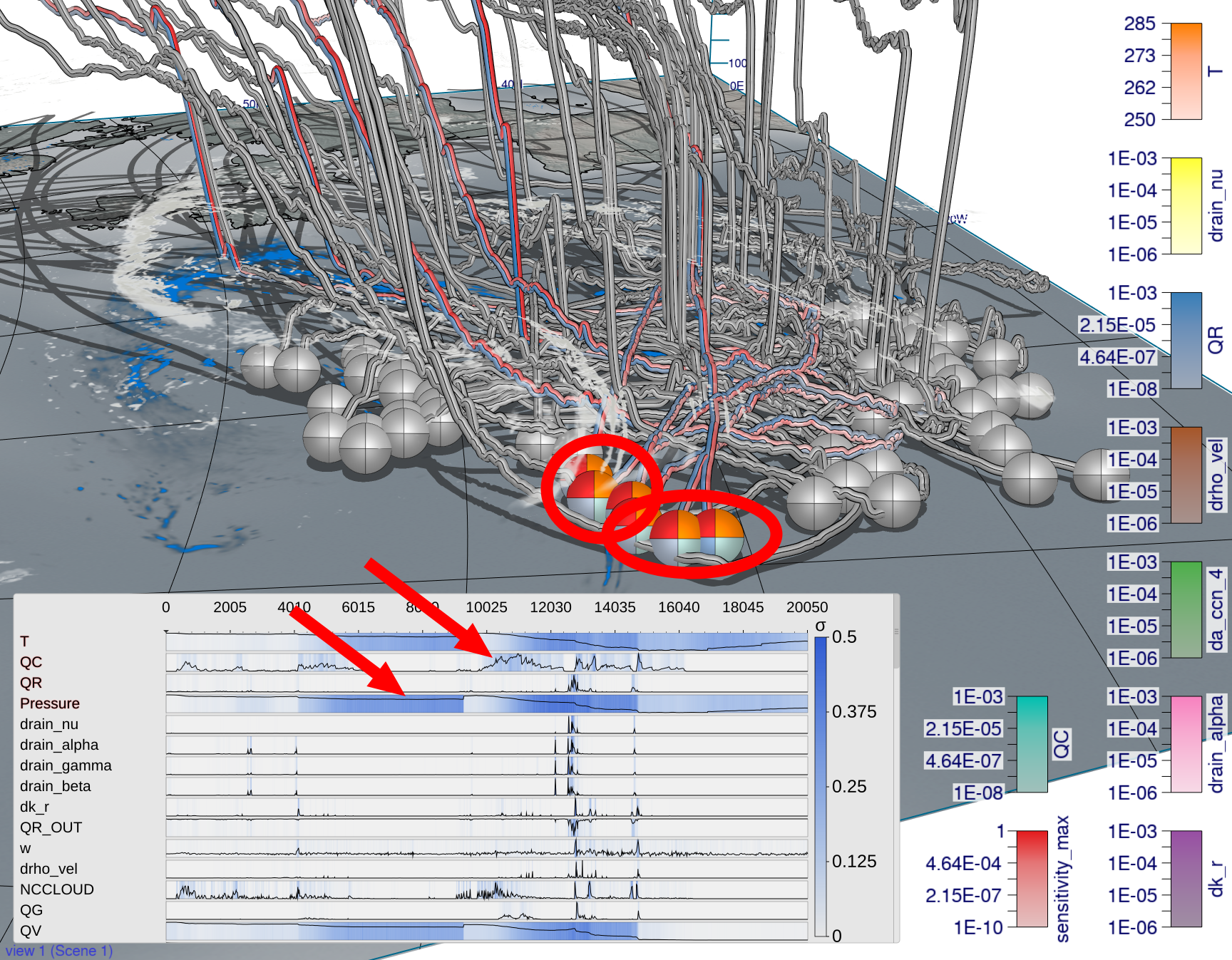}
 \caption{By changing the time step, we can trace the origin of the selected trajectories (left group). Several more trajectories in that area have not been selected previously (right group). We select the other trajectories to assess the difference between the initially chosen trajectories and the new ones with the same origin.
 The newly selected trajectories ascend earlier than before, albeit in a similar region. Furthermore, the curve plots clearly show an early cloud formation via QR and the large variance in ascent time given by the blue shade in pressure.
 }
 \label{fig:regions-step-3}
 \afterpage{\clearpage}
\end{figure*}

Trajectories initialized at similar time steps and regions may show different behavior responsible for diverse cloud structures and precipitation patterns. With Met.3D, one can identify such coherent clusters of trajectories and their characteristics. As a first step, we load a sample of 50 pre-selected trajectories from our trajectory dataset and select all to see the curve plot.
%In \autoref{fig:regions-step-1}, there is a clear peak of rain mass density (QR) that we want to investigate.
\autoref{fig:regions-step-1} shows a clear peak of rain mass density (QR) that we will investigate.
We select the time step of the peak in QR and select all trajectories that contribute to the peak around this time step. For this, we click all trajectories where the sphere shows no blue segment to deselect them, see \autoref{fig:regions-step-2}. To visualize the characteristics, we sort the similarity of the curve plot by QR, see \autoref{fig:regions-step-2}. These trajectories have similar times of ascent (compare pressure evolution), and precipitation occurs before cloud formation starts in these trajectories, i.e., rainfall sediments into the ascending WCB air from cloud layers above its current position.
The ascent of a trajectory is usually accompanied by cloud formation and then precipitation. One may further assume that trajectories started at similar areas and times have identical characteristics. To check this assumption, we go to the initial time step and select trajectories in a similar region. The right group in \autoref{fig:regions-step-3} are three trajectories we selected in addition. The curve plot reveals the different ascent times (early blue high-variance area in curve-plot pressure) and earlier cloud formation (new peaks in QC).
% To analyze if this behavior is to be expected from trajectories initialized in the same area, we go to the initial time step and select trajectories in a similar region, see \autoref{fig:regions-step-3}. We may increment the time step to see where and when these trajectories ascend. Furthermore, we can use the curve plot to see the different ascent times, the earlier cloud formation, and the lower precipitation rates, see \autoref{fig:regions-step-3}. 
We discovered two sets of trajectories: One involves large precipitation rates and another set with strong cloud formation at an earlier time and in a different area. Based on the identified characteristics, we can filter our dataset for more similar trajectories and investigate them as a whole.

\clearpage

\clearpage

%\FloatBarrier
\clearpage

\begin{figure*}[b]
 \centering
 \includegraphics[width=\textwidth]{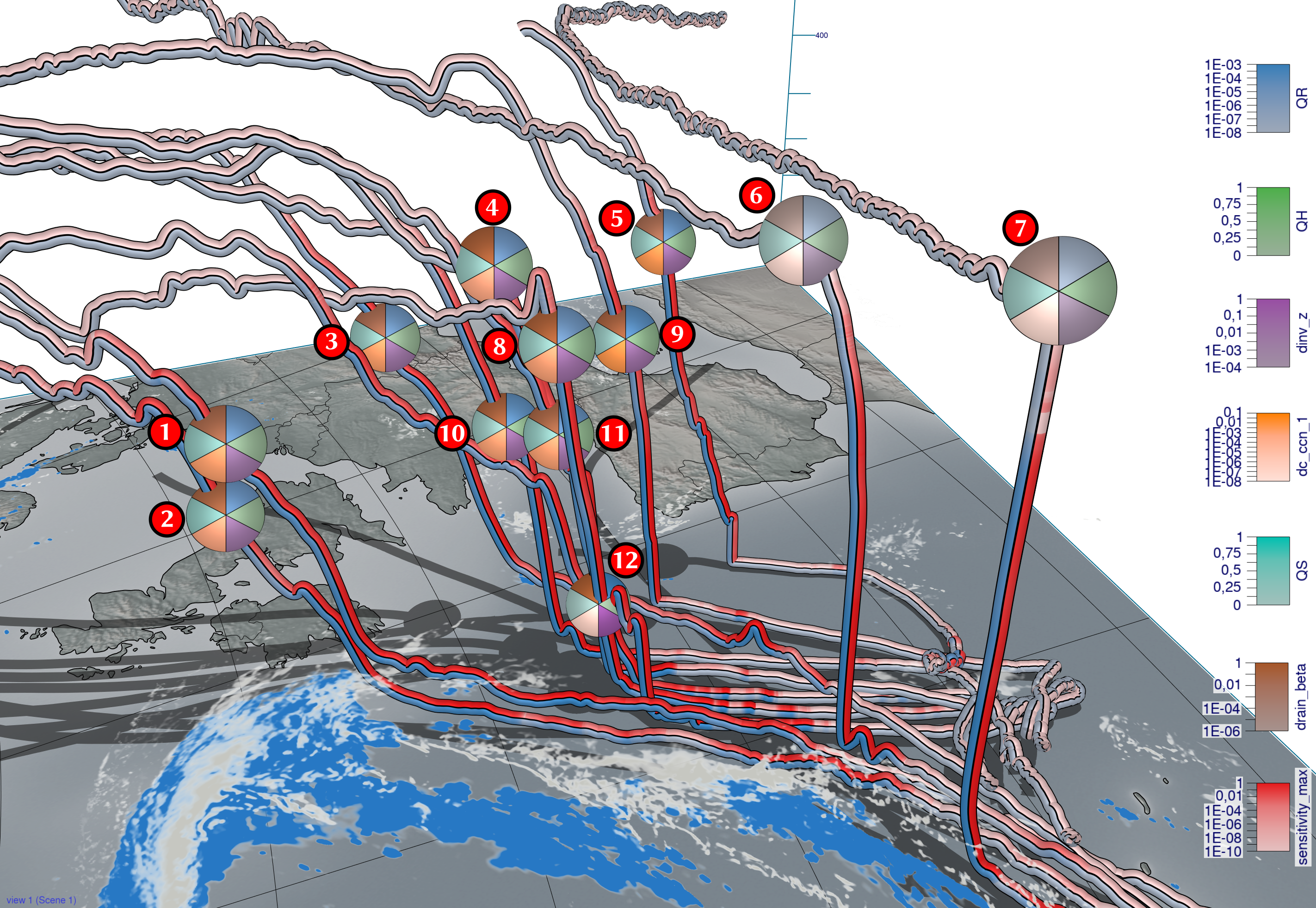}
 \caption{Representative image for user study task T1. Users needed to find trajectories which behave against the trend at the selected time step.}
 \label{fig:user-study-t1}
\end{figure*}

\begin{figure*}[b]
 \centering
 \includegraphics[width=\textwidth]{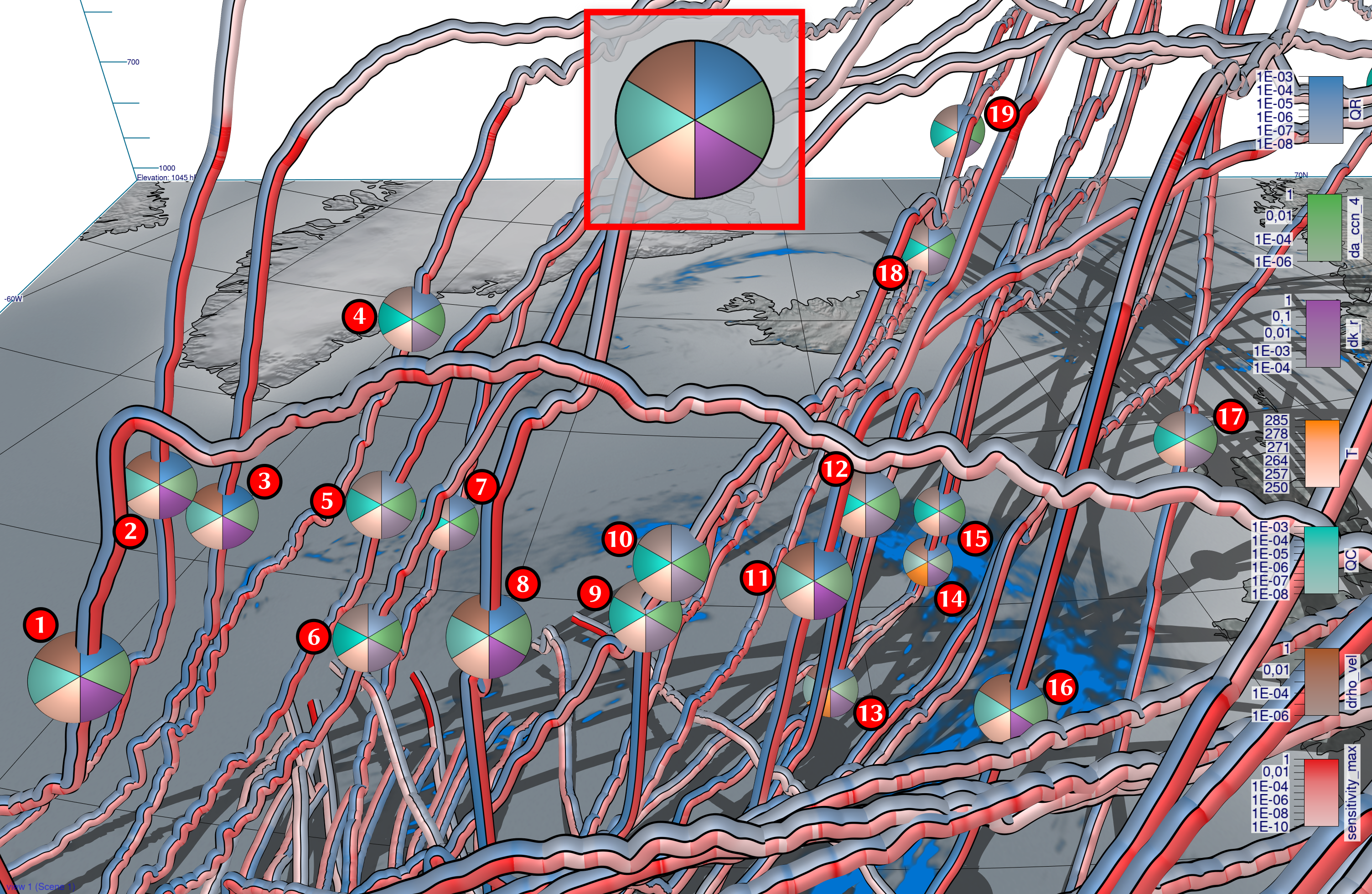}
 \caption{Representative image for user study task T2. Users needed to find trajectories with multi-parameter distributions similar to the reference distribution.}
 \label{fig:user-study-t2}
\end{figure*}

\begin{figure*}[b]
 \centering
 \includegraphics[width=\textwidth]{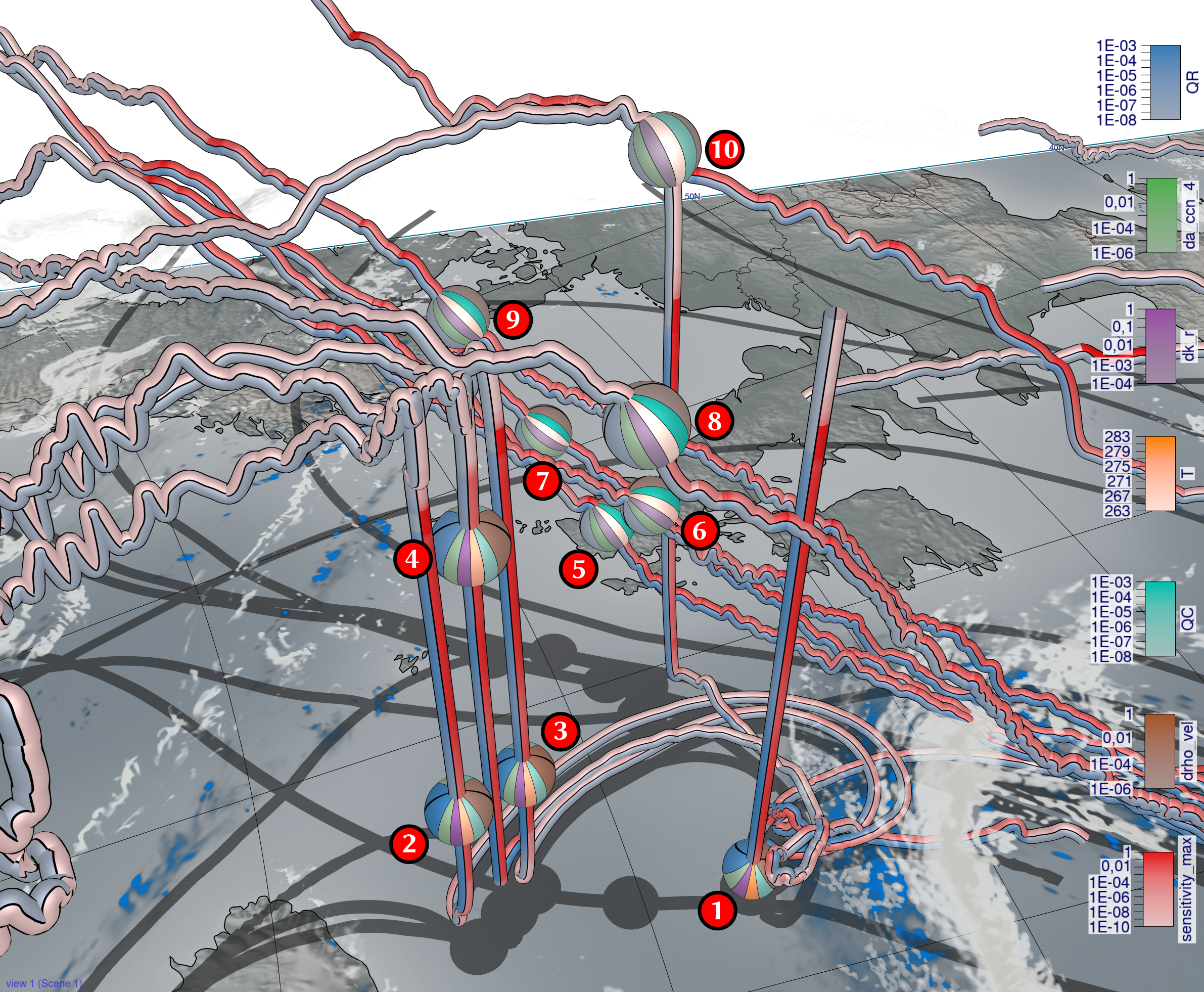}
 \caption{Representative image for user study task T3. Participants needed to assign trajectories with similar multi-parameter distributions to the same cluster.}
 \label{fig:user-study-t3}
\end{figure*}

\section{User Study Tasks}\label{sec:appendix-user-study}

In \autoref{fig:user-study-t1}, \autoref{fig:user-study-t2} and \autoref{fig:user-study-t3}, we show representative images from the user study for task T1, T2 and T3, respectively. The problem statements users were asked to solve with respect to these images were as follows.

\begin{enumerate}
    \item T1: ``Which trajectories are outliers with respect to their multi-parameter distribution at the selected time step?''
    \item T2: ``Which trajectories have the most similar distributions of quantities to the target distribution at the top?''
    \item T3: ``Group the trajectories into two clusters by the similarity of their multi-parameter distributions at the selected time step.''
\end{enumerate}

Whether participants were presented an image using bands-based or pie charts-based focus spheres was randomized. Each user got the same number of images for each visual mapping.

\clearpage

\begin{figure*}[b]
 \centering
 \includegraphics[width=\textwidth]{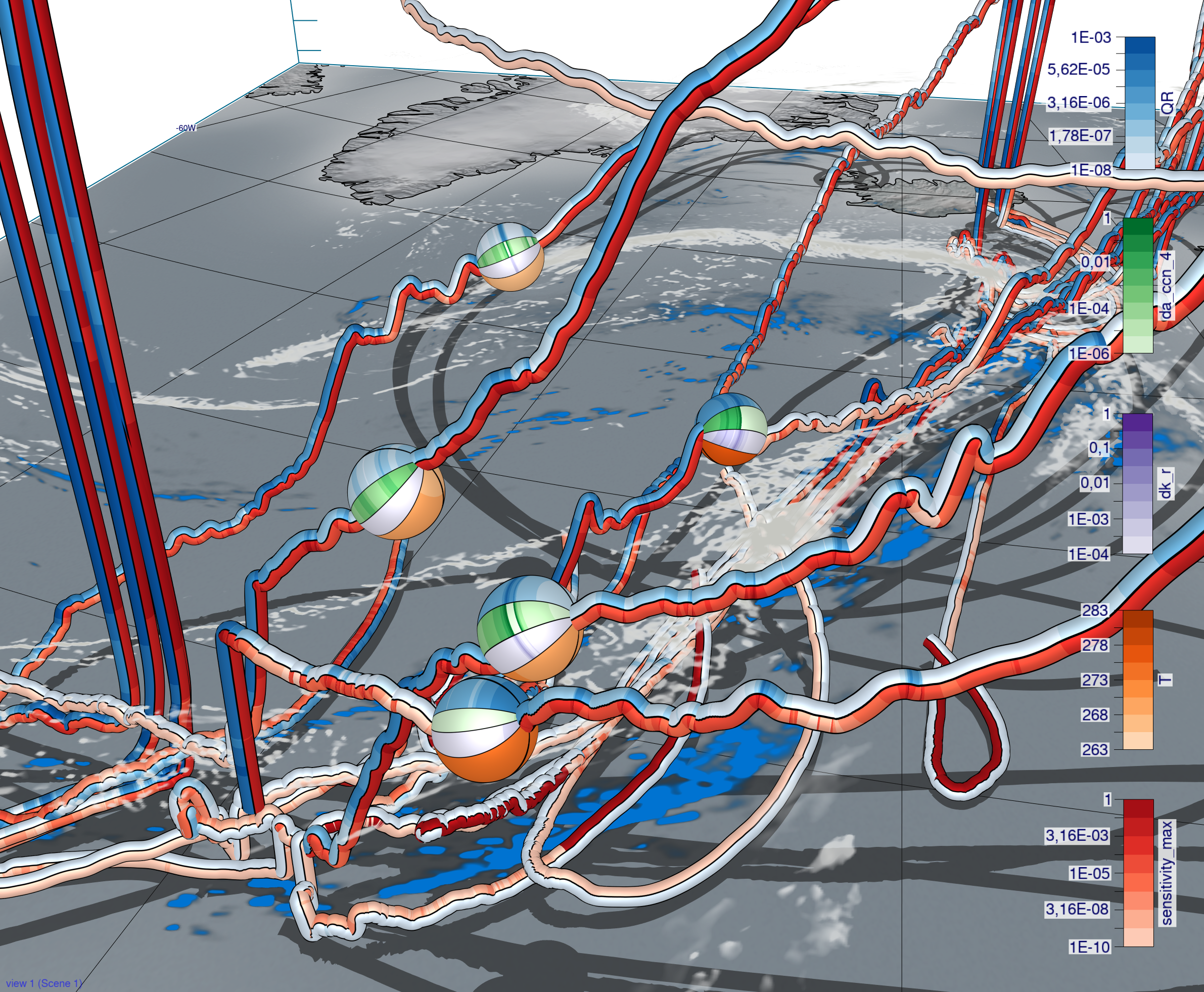}
 \caption{Visualization using discrete, quantized 8-class single hue color maps from ColorBrewer.
 }
 \label{fig:discrete}
 \afterpage{\clearpage}
\end{figure*}

\section{Visualization using Discrete Color Maps}\label{sec:appendix-discrete}

In \autoref{fig:discrete}, we demonstrate the use of discrete, quantized 8-class single hue color maps from ColorBrewer~\cite{brewer}. Using quantized color maps, it can be easier to assign colors on the trajectories and bands to individual values in the color maps.

\clearpage

\ifCLASSOPTIONcompsoc
  \section*{Acknowledgments}
\else
  \section*{Acknowledgment}
\fi

The authors acknowledge support by the Deutsche Forschungsgemeinschaft (DFG) within the Transregional Collaborative Research Centre TRR165 Waves to Weather, (www.wavestoweather.de), Projects A7, Z2 and B8 as well as funding from JGU Mainz.

\ifCLASSOPTIONcaptionsoff
  \newpage
\fi


%\begin{samepage}

\bibliographystyle{IEEEtran}
\bibliography{IEEEabrv,x_appendix}
%\begin{thebibliography}{1}


%\end{thebibliography}
%\nopagebreak


\begin{IEEEbiography}[{\includegraphics[width=1in,height=1.25in,clip,keepaspectratio]{figures/authors/christoph.jpg}}]{Christoph Neuhauser}
is a PhD candidate at the Computer Graphics and Visualization Group at the Technical University of Munich (TUM). He received his Bachelor's and Master's degrees in computer science from TUM in 2019 and 2020. Major interests in research comprise scientific visualization and real-time rendering.
\end{IEEEbiography}

% if you will not have a photo at all: IEEEbiographynophoto

%\newpage

\begin{IEEEbiography}[{\includegraphics[width=1in,height=1.25in,clip,keepaspectratio]{figures/authors/maicon.jpg}}]{Maicon Hieronymus}
is a PhD candidate at the Efficient Computing and Storage Group at the Johannes Gutenberg University Mainz (JGU) since 2019. He studied computer science (B.Sc.) and computer science with minor in physics (M.Sc.) at JGU. Research interests involve model analysis of cloud microphysical models and algorithmic differentiation in atmospheric physics.
\end{IEEEbiography}

\begin{IEEEbiography}[{\includegraphics[width=1in,height=1.25in,clip,keepaspectratio]{figures/authors/michael3.png}}]{Michael Kern}
is a research scientist and software engineer at Advanced Micro Devices (AMD) with major focus on real-time computer graphics and ray tracing. He studied computer science and received his Ph.D. from TUM in 2020. His thesis was concerned about feature detection and uncertainty visualization in meteorological data. Besides work, his main interests are scientific visualization and high-performance GPU computing. 
\end{IEEEbiography}

%\end{samepage}
%\end{minipage}

\vfill
\newpage

\begin{IEEEbiography}[{\includegraphics[width=1in,height=1.25in,clip,keepaspectratio]{figures/authors/marc.jpg}}]{Marc Rautenhaus}
received the M.Sc. degree in atmospheric science from the University of
British Columbia, Vancouver, in 2007, and the Ph.D. degree in computer science from TUM, Munich, in 2015. He currently leads the Visual Data Analysis Group at the Regional Computing Centre of Universit\"at Hamburg. Marc's research interests focus on the intersection of visualization, data analysis, and meteorology.
\end{IEEEbiography}

\begin{IEEEbiography}[{\includegraphics[width=1in,height=1.25in,clip,keepaspectratio]{figures/authors/annika.png}}]{Annika Oertel}
Annika Oertel studied atmospheric science and received a Ph.D. in atmospheric dynamics from ETH Zurich. Since 2020 she works as postdoctoral researcher in the Institute of Meteorology and Climate Research at Karlsruhe Institute of Technology. Her research focuses on the model representation of microphysical processes in warm conveyor belts and their interactions with the larger-scale extratropical circulation.
\end{IEEEbiography}

\begin{IEEEbiography}[{\includegraphics[width=1in,height=1.25in,clip,keepaspectratio]{figures/authors/ruediger2.jpg}}]{R\"udiger Westermann}
studied computer science at the Technical University Darmstadt and received his Ph.D. in computer science from the University of Dortmund, both in Germany. In 2002, he was appointed the chair of Computer Graphics and Visualization at TUM. His research interests include scalable data visualization and simulation algorithms, GPU computing, real-time rendering of large data, and uncertainty visualization.
\end{IEEEbiography}

\vfill


%\vfill

